\documentclass[aps,prd,reprint,amsmath,superscriptaddress,nofootinbib]{revtex4-2}
\usepackage[colorlinks,linkcolor=blue,hyperfootnotes=true]{hyperref}
\usepackage{amsmath}
\usepackage{graphicx}% Include figure files
\usepackage{dcolumn}% Align table columns on decimal point
\usepackage{bm}% bold math
\usepackage{footnote}
\usepackage{extarrows}
\usepackage{longtable}
\usepackage{threeparttable}
\usepackage{tabularx}
\usepackage{adjustbox}
\usepackage{amssymb}
\allowdisplaybreaks[2]

\begin{document}

%\preprint{}

\title{Geometric Approach to Light Rings in Axially Symmetric Spacetimes}


\author{Chenkai Qiao}
\email{Email: chenkaiqiao@cqut.edu.cn}
%\thanks{chenkaiqiao@cqut.edu.cn}
\affiliation{School of Physical Science and New Energy, Chongqing University of Technology, \\ Chongqing, 400054, The People's Republic of China}
\affiliation{School of Mathematical Science, Chongqing University of Technology, \\ Chongqing, 400054, The People's Republic of China}

\author{Ming Li}
\email{Email: mingli@cqut.edu.cn}
\affiliation{School of Mathematical Science, Chongqing University of Technology, \\ Chongqing, 400054, The People's Republic of China}
\affiliation{Mathematical Science Research Center, Chongqing University of Technology, \\  Chongqing, 400054, The People's Republic of China}

\author{Donghui Xie}
\email{Email: 202311998198@mail.bnu.edu.cn}
\affiliation{School of Physics and Astronomy, Beijing Normal University, \\ Beijing, 100875, The People's Republic of China}

\author{Minyong Guo}
\email{Email: minyongguo@bnu.edu.cn}
\affiliation{School of Physics and Astronomy, Beijing Normal University, \\ Beijing, 100875, The People's Republic of China}
\affiliation{Key Laboratory of Multiscale Spin Physics (Ministry of Education), Beijing Normal University, \\ Beijing, 100875, The People's Republic of China}


\date{\today}

\begin{abstract}
Circular photon orbits have become an attractive topic in recent years. They play extremely important roles in black hole shadows, gravitational lensings, quasi-normal modes, and spacetime topological properties. %The development of analytical methods for these circular orbits has also drawn extensive attention. 
In our recent work, \href{https://doi.org/10.1103/PhysRevD.106.L021501}{Phys. Rev. D \textbf{106}, L021501 (2022)}, a geometric approach to circular photon orbits was proposed for spherically symmetric spacetimes. In the present study, we extend this geometric approach from spherically symmetric spacetimes to axially symmetric spacetimes. In this geometric approach, light rings in the equatorial plane are determined by the intrinsic curvatures in the optical geometry of Lorentz spacetime, which gives rise to a Randers-Finsler geometry in axially symmetric cases. Specifically, light rings can be precisely determined by the vanishing of geodesic curvature, and the stability of light rings is classified using the intrinsic flag curvature in Randers-Finsler optical geometry. This geometric approach presented in this work is generally applicable to any stationary and axially symmetric spacetime, without imposing any restriction on the spacetime metric forms. Furthermore, we provide a rigorous demonstration to show that our geometric approach yields results that are completely equivalent to those derived from the conventional approach (based on the effective potential of photons). 
\end{abstract}


\maketitle

\section{Introduction \label{section1}}

The circular photon orbits (e.g., photon spheres and light rings) have emerged as highly attractive topics in the investigations of black holes and other ultra-compact astrophysical objects as black hole mimickers. Their significance is manifested through multiple avenues in theoretical and observational studies. Firstly, they are directly linked to the observed photon rings in black hole optical images \cite{EHT2019a,EHT2019b,EHT2022,Wald2019,Perlick2022,Vagnozzi2023,Roussille2025}. Secondly, they have dominant influences on other astrophysical observations, such as gravitational lensing in astrophysical systems \cite{Virbhadra2000,Bozza2002,Iyer2006,Tsukamoto2016}. Thirdly, circular photon orbits reveal nontrivial relationships among the Lyapunov exponent, the chaotic motion of photons, and quasi-normal modes in gravitational perturbations \cite{Cardoso2008,Giataganas2024}. The Lyapunov exponent of photons perturbed from circular orbits, which characterizes the chaotic behavior of photon motions \cite{Deich2023}, has a remarkable connection with the imaginary part of the quasi-normal mode frequency in the eikonal limit \cite{Cardoso2008,ChenYB2012,LiPC2021,Das2023}. Fourthly, the existence of stable circular photon orbits may serve as an indicator of spacetime instability. Some recent works suggested that stable circular photon orbits may induce nonlinear instabilities in the dynamical evolution of black holes coupled with matter fields, producing notable observational signatures in gravitational waves and quasi-normal modes \cite{Cardoso2014,Keir2016,GaoSJ2022,Cunha2023}. Finally, the circular photon orbits may also reveal the topological properties of spacetimes \cite{YinJ2023,Cunha2017,Cunha2018,Cunha2020,WeiSW2020}. Many studies have shown that the photon spheres and light rings in different classes of spacetimes (e.g., black hole spacetimes, horizon-less spacetimes produced by ultracompact objects, and naked singularity spacetimes) exhibit entirely different features \cite{Dolan2016,Cederbaum2016,Gibbons2016,JiaJJ2018a,JiaJJ2018b,Berry2020,Isomura2023,GaoSJ2021,Ghosh2021,GanQY2021,GuoGZ2023,Shaikh2019,JingJL2023,Murk2024,Cunha2017,Cunha2018,Cunha2020,WeiSW2020,Xavier2024}. The circular photon orbits in such spacetimes impose strong constraints on the event horizon, causal structure, and black hole hairs \cite{Cunha2020,Ghosh2023,Ghosh2025}.

\begin{figure*}
	\includegraphics[width=0.595\textwidth]{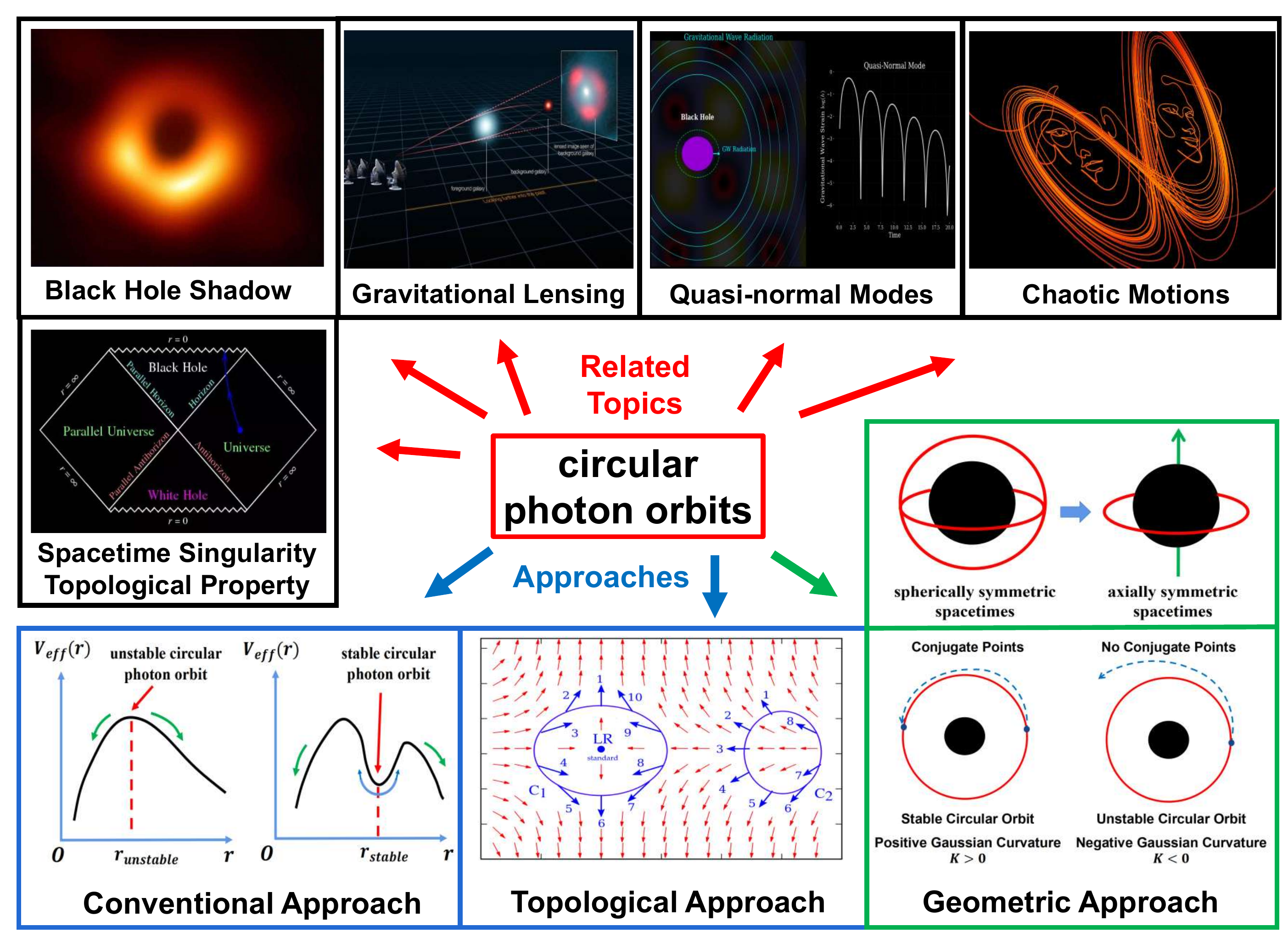}
	\caption{The research backgrounds and motivations of our present work. }
	\label{figure motivation}
\end{figure*}

There are several approaches for studying circular photon orbits (or other particles' circular orbits) in the gravitational fields. The conventional approach employs the effective potential of photons moving in gravitational fields \cite{Perlick2022,Hioki2008,Johannsen2013,Cunha2017b,Hartle2021,WangMZ2022,Bolokhov2025,Destounis2023,Vertogradov2024,Padhye2024,SongY2025}. This approach has proven highly valuable for analytically or numerically solving the photon orbit when a specific gravitational source is given (with explicit spacetime metric expressions). However, it may be less convenient when investigating general features of photon orbits that are universally applicable to arbitrary spacetime metrics, rather than dealing with a particular gravitational source with a known explicit metric form. To circumvent this limitation, topological approaches to circular photon orbits have emerged \cite{Cunha2017,Cunha2018,Cunha2020,WeiSW2020}. For each circular orbit, topological invariants of auxiliary vector fields can be assigned, such as the topological index of vector fields, Brouwer's degree of mapping, and topological charge of Duan's $\phi$ mapping. This kind of approach has inspired numerous studies on circular orbits and black hole topology \cite{Cunha2024,WeiSW2022a,WeiSW2022b,Afshar2023,Afshar2024a,Afshar2024b,WuD2024,WeiSW2023,WeiSW2024,BaiNC2023,SongY2025b}, offering us new insights and pathways for investigating the general characteristics of circular orbits, the topological structure of spacetimes, and the nature of gravitational fields. 

In addition to the methods mentioned above, we propose a novel geometric approach to circular photon orbits in our recent work \cite{QiaoCK2022a,QiaoCK2022b}, which relies on the mathematical construction of optical geometry and its intrinsic curvatures. This approach was initially developed for spherically symmetric spacetimes, where circular photon orbits become photon spheres. The intrinsic curvatures in optical geometry are crucial to determining photon spheres and their stability. Specifically, photon spheres are identified by the vanishing of geodesic curvature, and their stability is classified according to the sign of Gaussian curvature. Moreover, it has been demonstrated that this geometric approach yields results that are completely equivalent to those from the conventional approach based on local extrema of effective potentials \cite{QiaoCK2022a,QiaoCK2022b}. Our approach to photon spheres has enabled multiple applications and inspired several interesting works on related topics \cite{QiaoCK2024,QiaoCK2025,Cunha2022,Andino2024,Andino2025,Gallo2024,ZhangSH2025}. For instance, the number and distribution characteristics of stable and unstable photon spheres in various classes of spacetimes (including black hole spacetimes, ultra-compact objects' spacetime, regular spacetimes, and naked singularity spacetimes) have been obtained through the geometric analysis of Gaussian curvature and geodesic curvature \cite{QiaoCK2024,QiaoCK2025}. Additionally, recent works have suggested that similar geometric treatments can be used to study the circular orbits of massive particles and the massive particle surface \cite{Cunha2022,Andino2024,Andino2025}. Furthermore, it is also reported that the Gaussian curvature in optical geometry has a nontrivial connection with the Lyapunov exponent corresponding to unstable photon spheres \cite{Gallo2024,ZhangSH2025}. These studies provide solid evidence that the geometric quantities of auxiliary optical geometry can pioneer new directions in the study of gravitational fields and particle orbits.

Although our geometric approach has successfully characterized photon spheres in spherically symmetric spacetimes, its generalization to more general astrophysically relevant scenarios is necessary and compelling. The vast majority of gravitational systems in astrophysics are not spherically symmetric. Observationally, most of the supermassive black holes at the galaxy center typically exhibit rapid rotation, giving rise to axially symmetric spacetimes. Other ultra-compact astrophysical objects, such as pulsars, may exhibit substantial rotation. Consequently, a physically significant advancement of our geometric approach lies in its extension to rotating spacetimes. The backgrounds and motivations of the present work are illustrated in Figure \ref{figure motivation}. 

In this study, we provide an extension of our geometric approach to circular photon orbits to stationary and axially symmetric spacetimes. The generalization from spherically symmetric spacetimes to axially symmetric spacetimes requires additional mathematical concepts and techniques. Particularly, the constructed optical geometry for rotational spacetimes is no longer a Riemannian geometry, as in the spherically symmetric cases, and it becomes a Randers-Finsler geometry $dt = \sqrt{\alpha_{ij}dx^{i}dx^{j}} + \beta_{i}dx^{i}$ composed of a Riemannian part $\alpha$ and a non-Riemannian part $\beta$. In this work, it will be shown that intrinsic curvatures in Randers-Finsler geometry (geodesic curvature and flag curvature) provide a complete description of light rings in the equatorial plane. The light ring positions are determined by the vanishing Finslerian geodesic curvature condition $\kappa_{g}^{(F)} = 0$. The stability of light rings exhibits a nontrivial connection with the intrinsic flag curvature in Randers-Finsler optical geometry. The positive flag curvature indicates the light ring to be stable, while the negative flag curvature implies the light ring to be unstable. Furthermore, we can prove that this geometric description of stable and unstable light rings is fully equivalent to the conventional description based on the local maxima and local minima of effective potentials.

The organization of this work is as follows. Section \ref{section1} outlines the research backgrounds and motivations of our work. Section \ref{section2} provides a concise introduction to the optical geometry of Lorentz spacetimes, which plays a crucial role in our geometric approach to circular photon orbits. Section \ref{section3} presents a detailed formulation of our geometric approach to light rings in axially symmetric spacetimes. The curvature conditions to determine light rings and their stability are proposed in this section. Section \ref{section4} demonstrates the equivalence between our geometric approach and the conventional approach (based on the effective potential of photons). In Section \ref{section5}, representative examples are selected to show the validity and applicability of our approach. The conclusions and perspectives are summarized in section \ref{section6}. Furthermore, the mathematical preliminaries required for this work are presented in appendices, including the geodesic equation, the definition of geodesic curvature, and an introduction to the intrinsic flag curvature in 2-dimensional Randers-Finsler geometry.

\section{Optical Geometry of Stationary and Axially Symmetric Lorentz Spacetime \label{section2}}

The section briefly introduces the concept and properties of the optical geometry for axially symmetric (rotational asymmetric) spacetimes. The optical geometry serves as a powerful tool to study for studying the photon motion in gravitational fields. In the present work, our geometric approach to circular photon orbits (especially the light rings in axially symmetric spacetimes) is implemented using the intrinsic curvatures of optical geometry.

The underlying physical interpretation of the mathematical construction of optical geometry can be regarded as a generalization of Fermat's principle in curved spacetimes \cite{Abramowicz1988,Perlick2000,LinQ2008a,LinQ2008b,Piwnik2025}. The optical geometry emerges as a powerful framework for gravitational lensing studies, substantially reducing the computational complexity of the gravitational deflection angles for massless and massive particles \cite{Gibbons2008,Ono2019,Ishihara2016a,Ishihara2016b,Takizawa2020,Takizawa2023,LiZH2020a,LiZH2020b,LiZH2020c,ZhangZ2021,ZhangZ2024,HuangY2022,HuangY2023,Werner2012,Asida2017,Crisnejo2018}.
There are several equivalent ways to construct the optical geometry of stationary spacetimes \cite{Gibbons2008,Gibbons2009a,Gibbons2009b}. One of the most straightforward methods to construct the optical geometry is through a continuous mapping of spacetime geometry $ds^{2}=g_{\mu\nu}dx^{\mu}dx^{\nu}$ (which is a 4-dimensional Lorentz manifold) into a low-dimensional manifold with the null constraint $d\tau^{2}=-ds^{2}=0$ imposed \cite{Gibbons2008,Ono2019,Ishihara2016a} 
\begin{widetext}
\begin{equation}
	\underbrace{ds^{2} = g_{\mu\nu}dx^{\mu}dx^{\nu}}_{\text{Spacetime Geometry}}
	\ \ \ \overset{d\tau^{2}=-ds^{2}=0}{\Longrightarrow} \ \ \  
	\underbrace{dt^{2} = g^{\text{OP}}_{ij}dx^{i}dx^{j}}_{\text{Optical Geometry}}
	\ \ \ \text{or} \ \ \ 
	\underbrace{dt = \sqrt{\alpha^{\text{OP}}_{ij}dx^{i}dx^{j}} + \beta^{\text{OP}}_{i}dx^{i}}_{\text{Optical Geometry}}
\end{equation}
For static or stationary spacetimes, the photon orbits (which travel along lightlike/null geodesics in a 4-dimensional Lorentz manifold) become spatial geodesics when they are transformed into optical geometry. The stationary time coordinate $t$ plays the role of arc-length parameter (or spatial distance parameter) in optical geometry, which must be minimized along the photon orbits. Particularly, if we focus on the particle motions in the equatorial plane, a 2-dimensional optical geometry can be constructed.
\begin{equation}
	\underbrace{dt^{2} = g^{\text{OP}}_{ij}dx^{i}dx^{j}}_{\text{Optical Geometry}}
	\ \ \ \overset{\theta=\pi/2}{\Longrightarrow} \ \ \ 
	\underbrace{dt^{2}=\tilde{g}^{\text{OP-2d}}_{ij}dx^{i}dx^{j}}_{\text{Optical Geometry (Two Dimensional)}}
	\ \ \ \text{or} \ \ \ 
	\underbrace{dt = \sqrt{\alpha^{\text{OP-2d}}_{ij}dx^{i}dx^{j}} + \beta^{\text{OP-2d}}_{i}dx^{i}}_{\text{Optical Geometry (Two Dimensional)}}
\end{equation}
\end{widetext}
The properties of optical geometry strongly depend on the symmetries of gravitational fields and spacetime metrics. For a spherically symmetric spacetime, its optical geometry gives rise to a Riemannian manifold \cite{Gibbons2008,Ono2019,Ishihara2016a,Ishihara2016b}. For a stationary and axially symmetric spacetime, the corresponding optical geometry is described by a Randers-Finsler manifold \cite{Werner2012,Asida2017,Crisnejo2018,Jusufi2018a,Jusufi2018b,Gibbons2009b,LiZH2024}. 
%\end{widetext}

In the static and spherically symmetric gravitational field, the Riemannian geometry nature of optical geometry is easy to observe. Considering a general spherically symmetric spacetime with the metric 
\begin{equation}
	ds^{2} = g_{tt}dt^{2}+g_{rr}dr^{2}+g_{\theta\theta}d\theta^{2}+g_{\phi\phi}d\phi^{2} ,
\end{equation}
the corresponding optical geometry can be obtained from the null constraint $d\tau^{2}=-ds^{2}=0$, which eventually gives a 3-dimensional Riemannian manifold
\begin{equation}
	dt^{2} = g_{ij}^{\text{OP}}dx^{i}dx^{j}
	= -\frac{g_{rr}}{g_{tt}}\cdot dr^{2} - \frac{g_{\theta\theta}}{g_{tt}} \cdot d\theta^{2} - \frac{g_{\phi\phi}}{g_{tt}} d\phi^{2} .
	\label{optical geometry3}
\end{equation}
When analyzing the photon spheres in spherically symmetric spacetimes, one can always restrict this optical geometry to the equatorial plane $\theta=\frac{\pi}{2}$ without loss of generality. The explicit form of the 2-dimensional optical geometry is
\begin{equation}
	dt^{2}=\tilde{g}^{\text{OP-2d}}_{ij}dx^{i}dx^{j}
	= - \frac{g_{rr}}{g_{tt}} \cdot dr^{2} - \frac{\overline{g}_{\phi\phi}}{g_{tt}} \cdot d\phi^{2} , 
\end{equation}
where $\tilde{g}^{\text{OP-2d}}_{ij}$ denotes the 2-dimensional optical geometry metric, and the simplified notation $\overline{g}_{\phi\phi}$ represents the metric component restricted to the equatorial plane $\overline{g}_{\phi\phi}=g_{\phi\phi}(r,\theta=\frac{\pi}{2},\phi)$. 

For any stationary and axially symmetric gravitational systems, we now explain that the optical geometry eventually results in a Randers-Finsler manifold. Considering the standard rotational spacetime metric
\begin{equation}
	ds^{2} = g_{tt}dt^{2}+2g_{t\phi}dtd\phi+g_{rr}dr^{2}+g_{\theta\theta}d\theta^{2}+g_{\phi\phi}d\phi^{2} ,
\end{equation} 
the optical geometry can be obtained in a similar way by imposing the null constraint $d\tau^{2}=-ds^{2}=0$. Eventually, the arc-length parameter (or spatial distance parameter) in optical geometry becomes
\begin{equation}
	dt = \sqrt{-\frac{g_{rr}}{g_{tt}} \cdot dr^{2} - \frac{g_{\theta\theta}}{g_{tt}} \cdot d\theta^{2} + \frac{g_{t\phi}^{2}-g_{tt}g_{\phi\phi}}{g_{tt}^{2}} \cdot d\phi^{2}} 
	- \frac{g_{t\phi}}{g_{tt}}\cdot d\phi , \label{Optical geometry rotational}
\end{equation}
which exactly gives a Randers-Finsler manifold. Mathematically, the Randers-Finsler geometry is an extension of the Riemannian geometry \cite{ChernSS,ShenYB,ChenXY}, allowing the separation of arc-length / spatial distance into two parts
\begin{equation}
	dt = \sqrt{\alpha_{ij}(x)dx^{i}dx^{j}} + \beta_{i}(x)dx^{i} . \label{Renders geometry}
\end{equation}
The first part $\alpha_{ij}$ is a Riemannian metric, and the second part $\beta=\beta_{i}dx^{i}$ is a one-form that quantifies the departure of this Renders-Finsler geometry from the Riemannian geometry $dt^{2}=\alpha_{ij}dx^{i}dx^{j}$. The above Renders-Finsler geometry recovers the Riemannian geometry if and only if $\beta=0$.

In the present work, we mainly focus on circular photon orbits in axially symmetric spacetimes, whose optical geometry gives a Randers-Finsler geometry in expressions (\ref{Optical geometry rotational}) or (\ref{Renders geometry}). In such Randers-Finsler optical geometry, for any continuous curve $\gamma=\gamma(\lambda)$ parameterized by $\lambda$, the arc-length (or spatial distance) of this curve is calculated through the integration 
\begin{eqnarray}
	L_{AB} & = & \int_{s_{A}}^{s_{B}} ||T||_{(x,T)}^{(F)} \cdot d\lambda \nonumber
	             \\
	       & = & \int_{s_{A}}^{s_{B}} 
	             \bigg[ \sqrt{\alpha_{ij}(x) \frac{dx^{i}}{d\lambda} \frac{dx^{j}}{d\lambda}} + \beta_{i}(x) \frac{dx^{i}}{d\lambda} \bigg] d\lambda .
    \label{arc-length Randers-Finsler geometry}
\end{eqnarray}
where $||T||_{(x,T)}^{(F)} \equiv \sqrt{<T, T>_{(x,T)}^{(F)}}$ is the modulus of the tangent vector $T=\frac{dx}{d\lambda}=\frac{dx^{i}}{d\lambda}\cdot\frac{\partial}{\partial x^{i}}$ in a Finsler manifold. Mathematically, the modulus of a tangent vector in Finsler geometry can always be defined by a Finsler function $||T||_{(x,T)}^{(F)} = \sqrt{<T, T>_{(x,T)}^{(F)}} = F(x,T)$
\footnote{In the Finsler geometry, the norm of a vector $V$ depends not only on the position $x \in M$, but also on the tangent vector at this point $y \in T_{x}M$, so it is necessary to add a subscript to the vector norm $||V||_{(x,y)}^{(F)} \equiv \sqrt{<V, V>_{(x,y)}^{(F)}}$ or the inner product $<V, W>_{(x,y)}^{(F)}$. The norms of the same vector can be different if we take inner products along different directions, namely $<V,V>_{(x,y_{1})}^{(F)} \neq <V,V>_{(x,y_{2})}^{(F)}$ with $y_{1}, y_{2} \in T_{x}M$. This is quite different from the Riemannian geometry, where the product of vectors $<V, W>$ is totally determined by the Riemannian metric $g_{ij}(x)$. The metric component $g_{ij}(x)$ in Riemannian geometry depends only on the position in this manifold, regardless of the tangent direction. However, in the Finsler geometry, the proper definition of inner product is through the fundamental tensor $g_{ij}^{(F)}$, relying on potion $x$ and tangent vector $y$. The fundamental tensor in Finsler geometry plays a similar role as the Riemannian metric tensor when raising and lowering tensor indices, and its definition is provided in Appendix \ref{appendix3}. It is worth noting that the Finsler function measures the norm of the tangent vector $T$ of continuous curves along its own direction (with $y=T$), that is $F(x,T) = ||T||_{(x,T)}^{(F)} = <T,T>_{(x,T)}^{(F)}$.}. 
% 也可以选择插入一幅做内积的图，可放在附录里
The Randers-Finsler geometry is a special class of general Finsler geometry, where the Finsler function can always be decomposed into contributions from a Riemannian part $\alpha$ and a non-Riemannian part $\beta$.
\begin{eqnarray}
	F(x,T) 	%& = & \sqrt{\alpha_{ij}(x) \frac{dx^{i}}{d\lambda} \frac{dx^{j}}{d\lambda}} + \beta^{i}(x) \frac{dx^{i}}{d\lambda} \nonumber
	%\\
	= \sqrt{\alpha_{ij}(x) T^{i} T^{j}} + \beta_{i}(x) T^{i} .
\end{eqnarray}
Furthermore, if the affine parameter is chosen to be the arc-length parameter $\lambda = s$, then the modulus of the tangent vector for any continuous curve becomes unit, $||T||_{(x,T)}^{(F)} = 1$. 
The photon orbits, which are null geodesics in 4-dimensional spacetime geometry, become spatial geodesics when mapped to the Randers-Finsler optical geometry, with the stationary time coordinate $t$ to be the arc-length parameter in optical geometry (see expressions (\ref{Renders geometry})). 

% 下面说说两类内蕴曲率
In the Randers-Finsler geometry, several intrinsic geometric quantities are used to measure the geometric properties of this geometry. The geodesic curvature $\kappa_{g}^{(F)}$ and flag curvature $\mathcal{K}^{(F)}_{\text{flag}}$ are two crucial intrinsic curvatures in the 2-dimensional Randers-Finsler geometry, which play significant roles in our geometric approach to light rings in rotational spacetimes. The geodesic curvature is an intrinsic curvature of a continuous curve, which measures how far this curve is from being a geodesic curve in this Finsler manifold. Mathematically, the geodesic curvature in Randers-Finsler geometry can be rigorously defined in analogy with Riemannian geometry, and its definition is presented in Appendix \ref{appendix2}. In particular, for any geodesics in 2-dimensional Randers-Finsler geometry, their geodesic curvatures naturally vanish ($\kappa_{g}^{(F)} = 0$), and the arc-length defined in (\ref{arc-length Randers-Finsler geometry}) becomes extreme under a local variation ($\delta L_{AB} = 0$). The flag curvature can be viewed as the generalization of the Gaussian curvature to Finsler geometry, which quantifies whether a 2-dimensional surface (or 2-dimensional subsurface) is intrinsically flat. The sign of flag curvature provides a non-trivial constraint on the existence of conjugate points in Finsler geometry, which will be extremely helpful for determining the stability of light rings. The more detailed introductions and discussions on geodesics, geodesic curvature, and flag curvature are given in Appendices \ref{appendix1}-\ref{appendix3}.

\section{Geometric Approach to Determine Light Rings \label{section3}}

In this section, we present the geometric approach to light rings for arbitrary stationary and axially symmetric spacetimes, using the intrinsic curvatures of optical geometry to analyze light rings and their stability. Firstly, we recover a well-known equation for the angular velocity of photons traveling along light rings through the unit norm of tangent vectors in optical geometry, manifesting the equivalence between optical geometry and spacetime geometry when analyzing circular photon orbits. Secondly, we propose a geodesic curvature condition for light rings, which suggests that the locations of light rings are characterized by vanishing geodesic curvature in Randers-Finsler optical geometry. Finally, we establish a criterion that the stability of light rings can be determined through the intrinsic flag curvature, based on the Cartan-Hadamard theorem in optical geometry.

For any stationary and axially symmetric spacetime, the constructed optical geometry is a Randers-Finsler geometry given by expressions (\ref{Optical geometry rotational}) and (\ref{Renders geometry}). The stationary time $t$ plays the role of arc-length parameter in optical geometry, and the lengths of photon orbits are measured by the stationary time interval between the emission time $t_{A}$ and the received time $t_{B}$
\begin{equation}
	t_{AB} = \int_{A}^{B} dt
	       = \int_{t_{A}}^{t_{B}} 
	       \bigg[ \sqrt{\alpha_{ij}(x) \frac{dx^{i}}{dt} \frac{dx^{j}}{dt}} + \beta^{i}(x) \frac{dx^{i}}{dt} \bigg] dt  .
\end{equation}
Following the basic idea of Fermat's principle, which suggests that the optical path of a photon orbit starting and ending at two fixed points $A$ and $B$ must be an extreme, it is clear to see that the stationary time interval $t_{AB}$ can play the role of optical path when Fermat's principle is generalized to curved spacetimes. Therefore, photon trajectories always travel along geodesics in optical geometry such that the arc-length parameter (optical path) attains the extreme under a local variation $\delta t = 0$. Using the stationary time $t$ as the affine parameter, the tangent vectors of light orbits in the 3-dimensional optical geometry are $T^{\text{OP}}=(\frac{dr}{dt}, \frac{d\theta}{dt}, \frac{d\phi}{dt})$, and these tangent vectors must be unit vectors in Randers-Finsler optical geometry (since the stationary time is identically the arc-length parameter in optical geometry), which suggests 
\begin{equation}
	\big|\big| T^{\text{OP}} \big|\big|_{(x,T^{\text{OP}})}^{(F)}
	= \bigg[ \sqrt{\alpha_{ij}(x) \frac{dx^{i}}{dt} \frac{dx^{j}}{dt}} + \beta^{i}(x) \frac{dx^{i}}{dt} \bigg]
	= 1  .
	\label{unit tangent vector norm}
\end{equation}

\textbf{Unit Tangent Vector Norm Condition for Light Rings:} 
Particularly, if we focus on light rings restricted in the equatorial plane (with $\theta=\frac{\pi}{2}$), the radial and polar components of tangent vectors for these circular orbits are both zero, namely $(T^{\text{OP}})^{r} = \frac{dr}{dt} = 0$ and $(T^{\text{OP}})^{\theta} = \frac{d\theta}{dt} = 0$. The tangent vector of light rings has only one non-zero component, which is defined as the corresponding angular velocity of photon beams moving along these circular orbits $(T^{\text{OP}})^{\phi}=\frac{d\phi}{dt}=\Omega$. The unit norm of tangent vectors in Randers-Finsler optical geometry via equation (\ref{unit tangent vector norm}) leads to
%\begin{widetext}
\begin{eqnarray}
	&             &
	< T^{\text{OP}} \cdot T^{\text{OP}} >_{(x,T^{\text{OP}})}^{(F)} = 1  \nonumber
	\\
	& \Rightarrow & 
	F(x,T^{\text{OP}}) = \sqrt{\alpha_{ij}\frac{dx^{i}}{dt}\frac{dx^{j}}{dt}} + \beta_{i}\frac{dx^{i}}{dt} \nonumber
	\\
	&  &  \ \ \ \ \ \ \ \ \ \ \ \ \ \ 
	= \sqrt{ \alpha_{\phi\phi} \Omega^{2} } 
	+ \beta_{\phi} \Omega
	\nonumber
	\\
	&  &  \ \ \ \ \ \ \ \ \ \ \ \ \ \ 
	= \sqrt{ \frac{g_{t\phi}^{2}-g_{tt}g_{\phi\phi}}{g_{tt}^{2}} \cdot \Omega^{2} } 
	- \frac{g_{t\phi}}{g_{tt}} \cdot \Omega \nonumber
	\\
	&  &  \ \ \ \ \ \ \ \ \ \ \ \ \ \
	= 1 , \nonumber
	\\
	& \Rightarrow &
	g_{tt} + 2g_{t\phi} \Omega + g_{\phi\phi} \Omega^{2} = 0  .
\end{eqnarray}
The last line is exactly the well-known equation for the angular velocity of photon beams moving along light rings. Based on the aforementioned derivation, the following conclusion can be drawn from the Randers-Finsler optical geometry
\begin{widetext}
\begin{subequations}
\begin{eqnarray}
	< T^{\text{OP}} \cdot T^{\text{OP}} >_{(x,T^{\text{OP}})}^{(F)} = 1 
	\ \ \text{and} \ \ 
	(T^{\text{OP}})^{r} = (T^{\text{OP}})^{\theta} = 0 
	\ & \Rightarrow & \ 
	\bigg[ g_{tt} + 2g_{t\phi}\Omega + g_{\phi\phi} \Omega^{2} \bigg]_{r=r_{\text{LR}}} 
	= 0  ,
	\label{angular velocity equation}
	\\
	\bigg[ g_{tt} + 2g_{t\phi}\Omega + g_{\phi\phi} \Omega^{2} \bigg]_{r=r_{\text{LR}}} 
	= 0 
	\ & \Leftrightarrow & \
	\bigg[ \sqrt{ \alpha_{\phi\phi} \Omega^{2} } + \beta_{\phi} \Omega \bigg]_{r=r_{\text{LR}}} = 1  .
	\label{light ring unitarity condition}
\end{eqnarray}
\end{subequations}
It is interesting to note that equation (\ref{angular velocity equation}) can also be obtained from the conventional approach, simply by setting the effective potential of photons to zero. In the spacetime geometry, the effective potential of photons is usually defined through the reduced null geodesic equation, and the vanishing of the effective potential can be connected with a constraint on tangent vectors of null geodesics, via $u^{r} = \frac{dr}{d\lambda} = 0$ and $u^{\theta} = \frac{d\theta}{d\lambda} = 0$. 
\begin{equation}
	u \cdot u = 0 
	\ \ \text{and} \ \ 
	u^{r} = u^{\theta} = 0 
	\ \ \Rightarrow \ \ 
	V_{\text{eff}}(r) = 0  .
	%\ \ \Leftrightarrow \ \ 
	%g_{tt} + 2g_{t\phi}\Omega + g_{\phi\phi} \Omega^{2} = 0 \ .
	\label{constraint on tangent vector of null geodesics}
\end{equation}
Comparing the relations in (\ref{angular velocity equation}) and (\ref{constraint on tangent vector of null geodesics}), it is clearly manifested that the optical geometry and spacetime geometry play an equivalent role when analyzing circular photon orbits. Furthermore, an equivalent relationship can be established between our geometric approach and the conventional effective potential approach
%\begin{widetext}
\begin{subequations}
	\begin{eqnarray}
		< T^{\text{OP}} \cdot T^{\text{OP}} >_{(x,T_{\text{OP}})}^{(F)} = 1 
		\ \ \text{and} \ \ 
		(T^{\text{OP}})^{r} = (T^{\text{OP}})^{\theta} = 0 
		\ & \Leftrightarrow  & \  
		u \cdot u = 0 
		\ \ \text{and} \ \  
		u_{r} = u_{\theta} = 0  ,
		\\
		\bigg[ \sqrt{ \alpha_{\phi\phi} \Omega^{2} } + \beta_{\phi} \Omega \bigg]_{r=r_{\text{LR}}} = 1
		\ \ \Leftrightarrow \ \    
		\bigg[ g_{tt} + 2g_{t\phi}\Omega + g_{\phi\phi} \Omega^{2} \bigg]_{r=r_{\text{LR}}} = 0
		\ & \Leftrightarrow & \ 
		V_{\text{eff}}(r=r_{\text{LR}}) = 0  .
	\end{eqnarray}
\end{subequations}
The verification of this relation is presented in the next section.
\end{widetext}

At this stage, we have successfully re-derived an equation for the angular velocity of photons traveling along light rings through a pure geometric analysis. The derivation process is independent of any physical properties of gravitational sources, basically from the unit norm of tangent vectors in Randers-Finsler optical geometry. In the following part of this section, we shall provide the corresponding geometric conditions for light rings and their stability using intrinsic curvatures in optical geometry.

\textbf{Geodesic Curvature Condition for Light Rings:}
The light rings in the equatorial plane are spatial geodesics when they are transformed into the optical geometry, so the geodesic curvature of light rings in 2-dimensional Randers-Finsler automatically vanishes
\begin{equation}
	\text{Light Rings}
	\ \ \Leftrightarrow \ \ 
	\kappa_{g}^{(F)}(r=r_{\text{LR}}) = 0  .
	\label{light ring geodesic condition0}
\end{equation}	
Here, the radial variable $r$ in parentheses implies that we are dealing with a circular continuous curve with constant radius. Since the Randers-Finsler geometry can be decomposed into a Riemannian part $\alpha$ and a non-Riemannian part $\beta$ (see expression (\ref{Renders geometry})), it is natural to expect that the geodesic curvature in Randers-Finsler geometry also consists of contributions from two parts
\begin{equation} 
	\kappa_{g}^{(F)} = \kappa_{g}^{(\alpha)} + \kappa_{\beta}^{(\alpha)}  .
	\label{geodesic condition expression 1}
\end{equation}	
Here, the notation $\kappa_{g}^{(F)}$ labels the geodesic curvature for a continuous curve in Randers-Finsler geometry, $\kappa_{g}^{(\alpha)}$ denotes the geodesic curvature for the same continuous curve with respect to the Riemannian metric part $\alpha_{ij}$, and $\kappa_{\beta}^{(\alpha)}$ is the additional contribution acting on geodesic curvature $\kappa_{g}^{(\alpha)}$ due to the presence of non-Riemannian part $\beta$. Particularly, for any geodesic curves in optical geometry (\ref{Optical geometry rotational}), the additional contribution from $\beta$ part satisfies 
\begin{equation}
	\kappa_{\beta}^{(\alpha)}
	= \frac{\text{Sign}(\Omega)}{\sqrt{\alpha_{rr} \alpha_{\phi\phi}}} 
	\cdot \frac{\partial \beta_{\phi}}{\partial r}  .
	\label{geodesic condition expression 2}
\end{equation}
with the notation $\text{Sign}(\Omega)$ labels the sign of angular velocity. The same result has been derived in a recent work of Asida \emph{et al.} \cite{Asida2017}, but they interpreted it as the ``gravitomagnetic'' effect
\footnote{In reference \cite{Asida2017}, Asida \emph{et al.} give the expression $\kappa_{\beta}^{(\alpha)} = \frac{1}{\sqrt{\alpha_{rr}\alpha_{\phi\phi}}} \cdot \frac{\partial \beta_{\phi}}{\partial r}$. In this work, we have adopted a slightly different regularization from what Asida \emph{et al.} have used in reference, so that the retrograde motion (counter-rotating motion) of light can contribute to an additional minus sign in the calculation of $\kappa_{\beta}^{(\alpha)}$, compared with those for prograde motion (co-rotating motion) of light. This would lead to an additional factor $\text{Sign}(\Omega)$ in the geodesic curvature contribution $\kappa_{\beta}^{(\alpha)}$, see Appendix \ref{appendix2} for more detailed discussions.}.
Here we give a pure geometric interpretation on this term, arising from the non-Riemannian nature of optical geometry. On the other hand, for a light ring in the equatorial plane (with constant radius $r=r_{\text{LR}}$), its geodesic curvature with respect to the Riemannian metric part $\alpha$ can be calculated using the classical Liouville's relation \cite{Carmo1976}
\begin{equation}
	\kappa_{g}^{(\alpha)} (r=r_{\text{LR}})
	= \frac{1}{2\sqrt{\alpha_{rr}}} \frac{\partial \log(\alpha_{\phi\phi})}{\partial r} 
	\bigg|_{r=r_{\text{LR}}} .
	\label{geodesic condition expression 3}
\end{equation}
Combining the results in equations (\ref{geodesic condition expression 1})-(\ref{geodesic condition expression 3}), it can be demonstrated that the vanishing of geodesic curvature in Randers-Finsler optical geometry for light rings implies that
\begin{widetext}
\begin{equation}
	\kappa_{g}^{(F)} (r=r_{\text{LR}}) = 0 
	\ \ \Leftrightarrow \ \ 
	\bigg\{
	\frac{1}{2\sqrt{\alpha_{\phi\phi}}} 
	\frac{\partial \alpha_{\phi\phi}}{\partial r}
	+ \text{Sign}(\Omega) \cdot \frac{\partial \beta_{\phi}}{\partial r} 
	\bigg\}_{r=r_{\text{LR}}}
	= 0   .
	\label{light ring geodesic curvature condition}
\end{equation}	
This is the geodesic curvature condition for light rings in axially symmetric spacetimes. Furthermore, by comparing our geometric approach with the conventional effective potential approach, we establish a correspondence for light rings in such rotational spacetimes
\begin{equation}
	\kappa_{g}^{(F)}(r=r_{\text{LR}}) = 0 
	\ \ \Leftrightarrow \ \ 
	\frac{dV_{\text{eff}}(r)}{dr} \bigg|_{r=r_{\text{LR}}} = 0  .
	\label{equivalence relation for light rings}
\end{equation}
\end{widetext}
The demonstration of this equivalence relationship will be given in the next section.

\begin{figure*}[t]
	\includegraphics[width=0.65\textwidth]{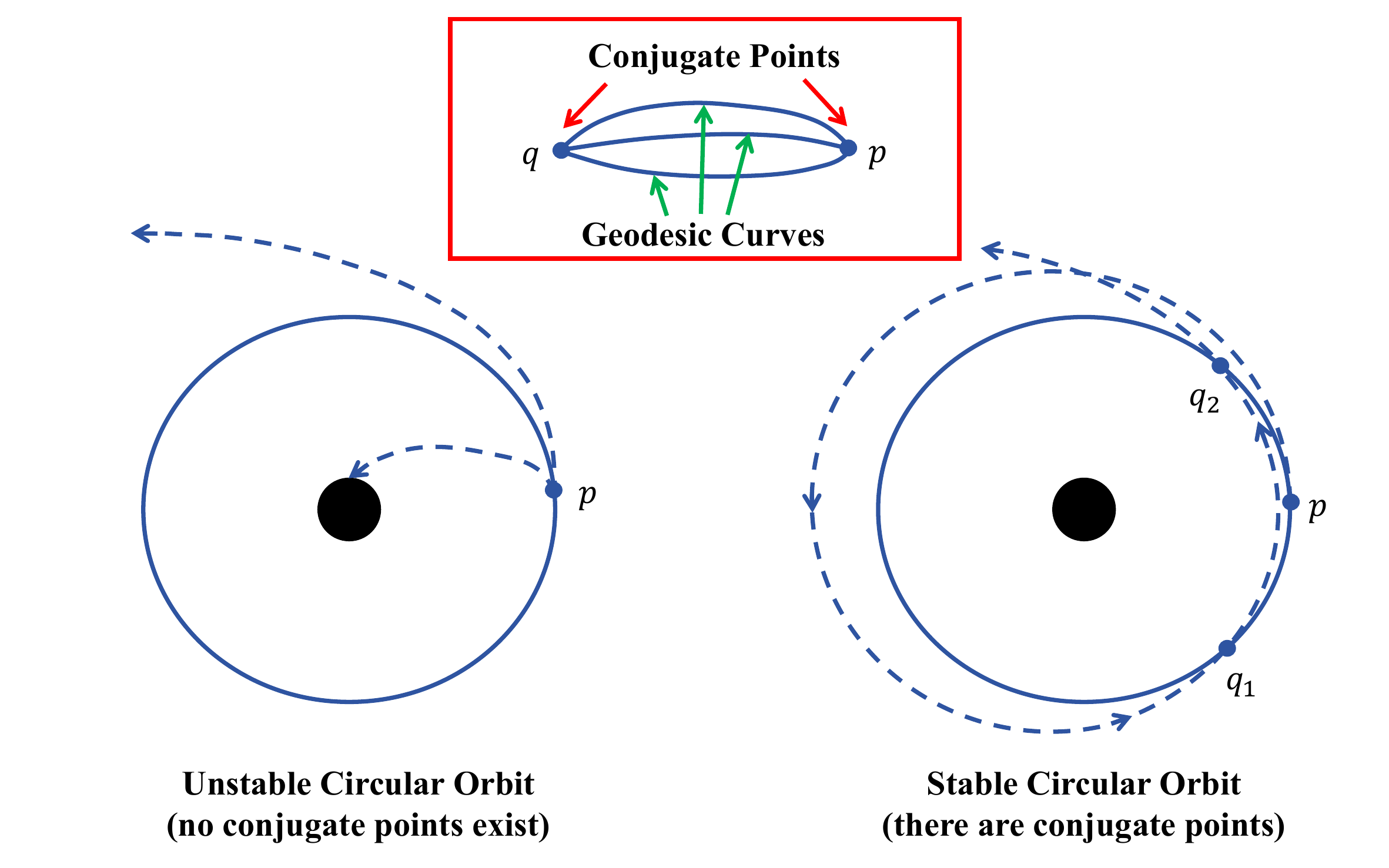}
	\caption{Illustration of the nontrivial connections between the stability of circular photon orbits and the existence of conjugate points in optical geometry. \textbf{(a)} The left panel illustrates the photon beams perturbed from an unstable circular photon orbit at a given point $p$. The perturbed photons would inevitably move away from this unstable circular orbit (escape to infinity or fall into the event horizon produced by black holes), and it is not possible to find another point $q$ conjugate to $p$ in this unstable circular photon orbit. \textbf{(b)} The right panel illustrates the photon beams perturbed from a stable circular photon orbit. In such cases, the perturbed photons may travel along other bound photon orbits near this stable circular photon orbit. There are conjugate points in this stable circular photon orbit (it is easy to observe that $p$ and $q_{1}$ are conjugate points; meanwhile, $q_{1}$ and $q_{2}$ are also conjugate points).}
	\label{figure light ring stability}
\end{figure*}

\textbf{Flag Curvatures Condition to the Stability of Light Rings, Conjugate Points, and Cartan-Hadamard Theorem:}
Having derived the geodesic curvature condition for light rings in equations (\ref{light ring geodesic condition0}) and (\ref{light ring geodesic curvature condition}), it is necessary to further explore the stability of these light rings in the equatorial plane. To carry out a geometric analysis of the light ring stability using geometric properties of optical geometry, we resort to the mathematical concept of conjugate points. The existence of conjugate points turns out to be crucial for distinguishing the stable and unstable light rings. Mathematically, two points $p$ and $q$ are defined to be conjugate points in a manifold, if a convergence of nearby geodesics starting from point $p$ can eventually converge to another point $q$ \cite{Carmo1976,Berger1988}. Conversely, if two different geodesic curves starting from the point $p$ can not converge at another point, then there are no points conjugate to $p$ in this manifold. In the optical geometry, the properties of stable and unstable circular photon orbits differ significantly, which have nontrivial connections with the conjugate points. Firstly, if photons are perturbed from an unstable circular photon orbit at a given point $p$, they would inevitably move away from this unstable circular orbit (the perturbed photons may eventually escape to infinity or fall into the event horizon produced by black holes). In such cases, the perturbed photons' orbit and the original unstable circular photon orbit, which belong to a geodesic congruence starting from the same point $p$, can not converge to each other at another point. Consequently, there are no conjugate points in the unstable circular photon orbit, as illustrated in the left part of Figure \ref{figure light ring stability}. On the other hand, if photons are perturbed from a stable photon orbit, instead of moving away completely, they may travel along other bound photon orbits in the vicinity of this stable circular photon orbit. In such cases, the original stable circular orbit and the nearby bound photon orbits may have intersect points (the points $q_{1}$ and $q_{2}$ in the right part of Figure \ref{figure light ring stability}, producing conjugate points in this stable circular photon orbit (for instance, $p$ and $q_{1}$, $q_{1}$ and $q_{2}$). Based on the above analysis, we conclude the following nontrivial relation between the stability of circular photon orbits and the existence of conjugate points:
\begin{widetext}  
\begin{eqnarray}
	\text{Stable Circular Photon Orbits} \ & \Rightarrow & \ \text{Conjugate points exist in this circular photon orbit} , \nonumber
	\\
	\text{Unstable Circular Photon Orbits} \ & \Rightarrow & \ \text{No conjugate points exist in this circular photon orbit}  . \nonumber
\end{eqnarray}
\end{widetext} 
Therefore, for light rings in axially symmetric spacetimes, their stability can be determined from an analysis of conjugate points in the equatorial plane of Randers-Finsler optical geometry. In differential geometry, the Cartan-Hadamard theorem strongly constrains the existence of conjugate points with the intrinsic geometric curvatures.

In the spherically symmetric cases, the Cartan-Hadamard theorem in optical geometry (which is a Riemannian geometry) constrains the existence of conjugate points with the sign (positive or negative) of the Gaussian curvature, yielding a Gaussian curvature condition for stable and unstable photon spheres \cite{QiaoCK2022a,QiaoCK2022b,QiaoCK2024}. However, in the stationary and axially symmetric cases, the corresponding optical geometry turns into a Randers-Finsler geometry, where non-Riemannian geometric effects from the $\beta$ part come into play, and intrinsic curvatures become more complex than those in Riemannian geometry. To carry out a similar geometric analysis on the stability of light rings, we require a Randers-Finsler geometry version of the Cartan-Hadamard theorem
\begin{quote}
    \textbf{Cartan-Hadamard Theorem (Finsler Geometry):} Let $(M, F)$ to be any connected and forward geodesically complete Finsler manifold with non-positive flag curvature. Then, geodesics in this Finsler manifold do not contain conjugate points \cite{ChernSS}. 
\end{quote}
This Cartan-Hadamard theorem in Randers-Finsler optical geometry gives a correspondence among the intrinsic flag curvature, the existence of conjugate points, and the stability of light rings. For any stable light ring, the existence of conjugate points in this light ring definitely violates the prerequisite of the Cartan-Hadamard theorem, which suggests that the flag curvature in the vicinity of stable light rings must be positive. Conversely, for the unstable light rings, the absence of conjugate points is consistent with the cases described in the Cartan-Hadamard theorem. This suggests that the flag curvature in the vicinity of unstable light rings is always negative 
\footnote{There are some subtleties in this conclusion, since our assumption is slightly different from the cases described in the Cartan-Hadamard theorem in mathematical literature. Strictly speaking, the Cartan-Hadamard theorem is a theorem relevant to global geometric and topological properties. The prerequisite of the Cartan-Hadamard theorem is that the flag curvature must be non-positive everywhere, and the conclusion is that no conjugate points exist in the entire Finsler manifold. In the present work, we do not care about whether conjugate points exist in the entire equatorial plane of optical geometry (especially the region very far from the light ring positions). What we are really concerned about is the conjugate points located in light rings (or near the light rings). So we slightly weaken the prerequisite condition of the Cartan-Hadamard theorem, assuming the flag curvature in the local region near light rings is negative (from this assumption, we also postulate that light ring position is not geometrically flat). However, the readers need not worry about this point, because an equivalence relation between our geometric approach and the conventional effective potential approach presented in the next section could provide a justification for our assumption.}. 
According to the Cartan-Hadamard theorem in 2-dimensional Randers-Finsler optical geometry, the following flag curvature condition for stable and unstable light rings can be proposed:
%\begin{widetext}
\begin{subequations}
\begin{eqnarray}
	\text{Stable Light Ring}   
	\ & \Leftrightarrow & \ 
	% \mathcal{K}^{(F)}_{\text{flag}}(r, T^{\text{OP}}, \partial^{\text{OP}}_{r}) > 0 ,
	\mathcal{K}^{(F)}_{\text{flag}} > 0 , 
	\ \ \ \ \ \ 
	\label{flag curvature condition for stable light ring}
	\\
	\text{Unstable Light Ring} 
	\ & \Leftrightarrow & \ 
	% \mathcal{K}^{(F)}_{\text{flag}}(r, T^{\text{OP}}, \partial^{\text{OP}}_{r}) < 0 . 
	\mathcal{K}^{(F)}_{\text{flag}} < 0 .
	\ \ \ \ \ \ 
	\label{flag curvature condition for unstable light ring}
\end{eqnarray}
\end{subequations}
%\end{widetext}
There are additional points on flag curvature that are worthy of notice. In the Riemannian geometry, the intrinsic Gaussian totally depends on the position in the manifold, via $\mathcal{K}_{\text{Gaussian}}=\mathcal{K}_{\text{Gaussian}}(x)$, while its successor in Finsler geometry --- flag curvature --- becomes more complicated and depends on more variables. 
In the Finsler geometry, the rigorous definition of flag curvature depends not only on the position $x \in M$, but also on two non-parallel vectors in the tangent space $T_{x}M$: a flagpole vector $y$ and a transverse edge vector $V$ (refer to Appendix \ref{appendix3} for more details), which suggests $\mathcal{K}_{\text{flag}} = \mathcal{K}_{\text{flag}}(x,y,V)$ \cite{ChernSS}. In the equatorial plane of optical geometry, it is convenient to choose the flagpole vector as the tangent vector of light rings $y = T^{\text{OP}} = (\frac{dr}{dt}, \frac{d\phi}{dt}) = (0, \Omega)$, and the transverse edge is selected as the radial frame vector $V = \partial^{\text{OP}}_{r}$ towards the outward radial direction. In this manner, the flag curvature in the 2-dimensional Randers-Finsler optical geometry at a given radius $r$ can be expressed as $\mathcal{K}^{(F)}_{\text{flag}}(x, y, V) = \mathcal{K}^{(F)}_{\text{flag}}(r, T^{\text{OP}}, \partial^{\text{OP}}_{r})$. 

\begin{widetext}
For general axially symmetric spacetimes with the optical geometry given by expression (\ref{optical geometry3}), the flag curvature in the equatorial plane with flag pole vector $y = T^{\text{OP}}$ and transverse edge vector $V=\partial^{\text{OP}}_{r}$ can be calculated through expression (\ref{flag curvature expression}) in Appendix \ref{appendix3}
\begin{eqnarray}
	\mathcal{K}^{(F)}_{\text{flag}}(r, T^{\text{OP}}, \partial^{\text{OP}}_{r}) 
	& = & 
	\frac{ 1 }{ \alpha_{rr} \big( 1+\frac{\beta_{\phi}}{\sqrt{\alpha_{\phi\phi}}} \big) } 
	\times
	\bigg\{ \
	\frac{ 3\Omega^{2} \sqrt{\alpha_{\phi\phi}} }{ 2\alpha_{rr} }  \frac{d\alpha_{rr}}{dr} 
	\cdot \bigg[ \frac{1}{2\sqrt{\alpha_{\phi\phi}}} \frac{d\alpha_{\phi\phi}}{dr} + \text{Sign}(\Omega) \frac{d\beta_{\phi}}{dr} \bigg]
	\nonumber
	\\
	&   &
	+ \frac{ 3\beta_{\phi}\Omega\sqrt{\Omega^{2}} }{ 2\alpha_{rr} } \frac{d\alpha_{rr}}{dr} \cdot \bigg[ \frac{1}{2\sqrt{\alpha_{\phi\phi}}} \frac{d\alpha_{\phi\phi}}{dr} + \text{Sign}(\Omega) \frac{d\beta_{\phi}}{dr} \bigg]
	\nonumber
	\\
	&   & 
	+ \frac{ \beta_{\phi}\Omega \sqrt{\alpha_{\phi\phi}\Omega^{2}} }{ \alpha_{\phi\phi} } 
	\cdot \bigg[ \frac{1}{2\alpha_{\phi\phi}} \frac{d\alpha_{\phi\phi}}{dr} + \text{Sign}(\Omega) \frac{d\beta_{\phi}}{dr} \bigg] 
	\cdot \bigg[ \frac{1}{2\alpha_{\phi\phi}} \frac{d\alpha_{\phi\phi}}{dr} - \text{Sign}(\Omega) \frac{d\beta_{\phi}}{dr} \bigg]
	\nonumber
	\\
	&   &
	- \frac{ \Omega \sqrt{\Omega^{2}} }{ 2 }  \frac{d\beta_{\phi}}{dr} \cdot \bigg[ \frac{1}{2\sqrt{\alpha_{\phi\phi}}} \frac{d\alpha_{\phi\phi}}{dr} + \text{Sign}(\Omega) \frac{d\beta_{\phi}}{dr} \bigg]
	+ \frac{ 3\beta_{\phi} \Omega^{2} }{ 2\sqrt{\alpha_{\phi\phi}} } \frac{d\beta_{\phi}}{dr} \cdot \bigg[ \frac{1}{2\sqrt{\alpha_{\phi\phi}}} \frac{d\alpha_{\phi\phi}}{dr} + \text{Sign}(\Omega) \frac{d\beta_{\phi}}{dr} \bigg]
	\nonumber
	\\
	&   & 
	+ \bigg( \sqrt{\alpha_{\phi\phi}\Omega^{2}} + \beta_{\phi}\Omega \bigg) 
	\cdot \bigg[ 
	\frac{ \sqrt{\alpha_{\phi\phi}\Omega^{2}} }{ \alpha_{\phi\phi} } 
	\cdot \bigg( \frac{d\beta_{\phi}}{dr} \bigg)^{2} 
	- \frac{ \sqrt{\alpha_{\phi\phi}\Omega^{2}} }{ 2\alpha_{\phi\phi} } 
	\cdot \frac{d^{2}\alpha_{\phi\phi}}{dr^{2}} 
	- \Omega \cdot \frac{d^{2}\beta_{\phi}}{dr^{2}} 
	\bigg]
	\ \bigg\} .
\end{eqnarray}
At the light ring position $r=r_{\text{LR}}$, the unit tangent vector condition $< T^{\text{OP}} \cdot T^{\text{OP}} >_{(x, T^{\text{OP}})}^{(F)} = 1$ in equation (\ref{light ring unitarity condition}) and geodesic curvature condition $\kappa_{g}^{(F)}(r=r_{\text{LR}}) = 0$ in equation (\ref{light ring geodesic curvature condition}) can be used, which leads to a simplified expression for the flag curvature
\begin{eqnarray}
	\mathcal{K}^{(F)}_{\text{flag}}(r, T^{\text{OP}}, \partial^{\text{OP}}_{r}) \bigg|_{r=r_{\text{LR}}}
	& = & 
	\bigg\{ \
	\frac{ 1 }{ \alpha_{rr} \big( 1+\frac{\beta_{\phi}}{\sqrt{\alpha_{\phi\phi}}} \big) } 
	\cdot
	\frac{\sqrt{\alpha_{\phi\phi}\Omega^{2}}}{\alpha_{\phi\phi}} 
	\cdot \bigg[ \bigg( \frac{d\beta_{\phi}}{dr} \bigg)^{2} 
	- \frac{1}{2} \frac{d^{2}\alpha_{\phi\phi}}{dr^{2}}
	- \frac{\alpha_{\phi\phi}\Omega}{\sqrt{\alpha_{\phi\phi}\Omega^{2}}} \cdot \frac{d^{2}\beta_{\phi}}{dr^{2}} \bigg]
	\ \bigg\}_{r=r_{\text{LR}}} .
	\label{flag curvature LR}
\end{eqnarray} 
%\end{widetext}
From this reduced flag curvature expression, we obtain the following flag curvature condition for stable and unstable light rings in axially symmetric spacetimes
\begin{subequations}
\begin{eqnarray}
	\text{Stable Light Ring}   
	\ & \Leftrightarrow & \ 
	\mathcal{K}^{(F)}_{\text{flag}}(r, T^{\text{OP}}, \partial^{\text{OP}}_{r}) \bigg|_{r=r_{\text{LR}}}
	%> 0
	\propto 
	\bigg[ \bigg( \frac{d\beta_{\phi}}{dr} \bigg)^{2} 
	- \frac{1}{2} \frac{d^{2}\alpha_{\phi\phi}}{dr^{2}}
	- \frac{\alpha_{\phi\phi}\Omega}{\sqrt{\alpha_{\phi\phi}\Omega^{2}}} \cdot \frac{d^{2}\beta_{\phi}}{dr^{2}} 
	\bigg]_{r=r_{\text{LR}}}
	> 0 ,
	\ \ \ \ \ \ \ \ 
	\label{flag curvature condition for stable light ring 2}
	\\
	\text{Unstable Light Ring}   
	\ & \Leftrightarrow & \ 
	\mathcal{K}^{(F)}_{\text{flag}}(r, T^{\text{OP}}, \partial^{\text{OP}}_{r}) \bigg|_{r=r_{\text{LR}}}
	%< 0
	\propto 
	\bigg[ \bigg( \frac{d\beta_{\phi}}{dr} \bigg)^{2} 
	- \frac{1}{2} \frac{d^{2}\alpha_{\phi\phi}}{dr^{2}}
	- \frac{\alpha_{\phi\phi}\Omega}{\sqrt{\alpha_{\phi\phi}\Omega^{2}}} \cdot \frac{d^{2}\beta_{\phi}}{dr^{2}} 
	\bigg]_{r=r_{\text{LR}}}
	< 0 .
	\ \ \ \ \ \ \ \ 
	\label{flag curvature condition for unstable light ring 2}
\end{eqnarray}
\end{subequations}
For physical spacetimes, $g_{tt}<0$, $g_{rr}>0$, $g_{\phi\phi}>0$ always hold for circular photon orbits, which suggests $\alpha_{rr} = -\frac{g_{rr}}{g_{tt}} > 0$ and $\alpha_{\phi\phi} = \frac{g_{t\phi}^{2}-g_{tt}g_{\phi\phi}}{g_{tt}^{2}} > \beta_{\phi}^{2} = \frac{g_{t\phi}^{2}}{g_{tt}^{2}} \ge 0$, so the first two factors outside the square brackets in the expression (\ref{flag curvature LR}) are always positive.
Furthermore, it is natural to think that our flag curvature conditions for stable and unstable light rings in equations (\ref{flag curvature condition for stable light ring}-\ref{flag curvature condition for unstable light ring}) or (\ref{flag curvature condition for stable light ring 2}-\ref{flag curvature condition for unstable light ring 2}) may be equivalent to the effective potential conditions (the local minima and local maxima of the effective potential correspond to stable and unstable light rings), since the different approaches determine the same light rings in a gravitational field. In the next section, we give a demonstration of this equivalence relation 
\begin{subequations}
	\begin{eqnarray}
		\mathcal{K}^{(F)}_{\text{flag}}(r, T^{\text{OP}}, \partial_{r}) \bigg|_{r=r_{\text{LR}}}
		> 0
		& \Leftrightarrow &
		\frac{d^{2}V_{\text{eff}}(r)}{dr^{2}} \bigg|_{r=r_{\text{LR}}} > 0 
		\ \ \ \ \
		\text{for stable light rings,} \ \ \ \ \ \ \ \ 
		\label{equivalence relation for stable light rings}
		\\
		\mathcal{K}^{(F)}_{\text{flag}}(r, T^{\text{OP}}, \partial_{r}) \bigg|_{r=r_{\text{LR}}}
		< 0
		& \Leftrightarrow &
		\frac{d^{2}V_{\text{eff}}(r)}{dr^{2}} \bigg|_{r=r_{\text{LR}}} < 0 
		\ \ \ \ \ 
		\text{for unstable light rings.} \ \ \ \ \ \ \ \ 
		\label{equivalence relation for unstable light rings}
	\end{eqnarray}
\end{subequations}

At the end of this section, it is worthwhile to consider a special case of rotational spacetimes, elaborating that our analysis presented in this work indeed serves as an extension of our geometric approach in our previous works \cite{QiaoCK2022a,QiaoCK2022b}. When the gravitational system is slowly rotating (where the non-Riemannian part $\beta$ is sufficiently small compared with the Riemannian part $\alpha$), we can approximate the flag curvature in 2-dimensional Randers-Finsler optical geometry as the Gaussian curvature for the Riemannian metric part $\alpha$. In the circular photon orbits, a simple reduction of the flag curvature in expression (\ref{flag curvature LR}) yields
\begin{equation}
	\mathcal{K}^{(F)}_{\text{flag}}(r, T^{\text{OP}}, \partial_{r}^{\text{OP}}) \bigg|_{r=r_{\text{ph}}}
	\approx \mathcal{K}_{\text{Gaussian}}^{(\alpha)}(r=r_{\text{ph}})
	= \bigg[ - \frac{1}{2\alpha_{rr}\alpha_{\phi\phi}} \cdot \frac{d^{2}\alpha_{\phi\phi}}{dr^{2}} \bigg]_{r=r_{\text{ph}}}
	\ \ \ \ \ 
	\text{for slowly rotating gravitational systems.}  
\end{equation}
This is precisely the Gaussian curvature in the vicinity of circular photon orbits calculated in our previous works \cite{QiaoCK2022a,QiaoCK2022b,QiaoCK2024,QiaoCK2025}. Therefore, under the consideration of slowly rotating gravitational spacetimes, it is clearly manifested that our geometric approach in this work successfully reproduces the geometric approach to photon spheres for spherically symmetric spacetimes developed in previous works. Moreover, in the next section, we also provide a demonstration of the equivalence relation between our geometric approach and the effective potential approach for slowly rotating gravitational systems.
\begin{subequations}
\begin{eqnarray}
	\frac{d^{2}V_{\text{eff}}(r)}{dr^{2}} \bigg|_{r=r_{\text{ph}}} > 0 
	& \Leftrightarrow &
	\mathcal{K}^{(F)}_{\text{flag}}(r, T^{\text{OP}}, \partial_{r}) \bigg|_{r=r_{\text{ph}}} 
	\approx 
	\mathcal{K}_{\text{Gaussian}}^{(\alpha)}(r=r_{\text{ph}}) > 0
	\ \ 
	\text{for stable orbits in slowly rotating spacetimes,} \ \ \ \ \ \ \ \ 
	\\
	\frac{d^{2}V_{\text{eff}}(r)}{dr^{2}} \bigg|_{r=r_{\text{ph}}} < 0 
	& \Leftrightarrow &
	\mathcal{K}^{(F)}_{\text{flag}}(r, T^{\text{OP}}, \partial_{r}) \bigg|_{r=r_{\text{ph}}} 
	\approx 
	\mathcal{K}_{\text{Gaussian}}^{(\alpha)}(r=r_{\text{ph}}) < 0
	\ \ 
	\text{for unstable orbits in slowly rotating spacetimes.} \ \ \ \ \ \ \ \ 
\end{eqnarray}
\end{subequations}
\end{widetext}

\section{Equivalence Between Our Geometric Approach and the Conventional Effective Potential Approach \label{section4}}

In this section, we give a demonstration of the equivalence between our geometric approach and the conventional effective potential approach. The demonstration process relies on the analysis and simplification of the effective potential, through the utility of conserved quantities in axially symmetric spacetimes. Consider the stationary and axially symmetric spacetime with the general metric form
\begin{eqnarray}
	ds^{2} & = & g_{\mu\nu} dx^{\mu}dx^{\nu} \nonumber
	\\
	& = & g_{tt}dt^{2} + 2g_{t\phi}dtd\phi + g_{rr}dr^{2} 
	      + g_{\theta\theta}d\theta^{2} + g_{\phi\phi}d\phi^{2} . %\nonumber
	\ \ \ \ \ \ 
\end{eqnarray}
For test particles moving in this rotational spacetime, the Killing vectors $K_{t}=\frac{\partial}{\partial t}$ and $K_{\phi}=\frac{\partial}{\partial \phi}$ determine the following conserved quantities along the particle orbit
\begin{subequations}
\begin{eqnarray}
	E & \equiv & - K_{t} \cdot u 
	= - g_{tt} \cdot \frac{dt}{d\lambda} - g_{t\phi} \cdot \frac{d\phi}{d\lambda} ,
	\\
	L & \equiv & K_{\phi} \cdot u
	= g_{t\phi} \cdot \frac{dt}{d\lambda} + g_{\phi\phi} \cdot \frac{d\phi}{d\lambda} .
\end{eqnarray}
\end{subequations}
where $u=(\frac{dt}{d\lambda}, \frac{dr}{d\lambda}, \frac{d\theta}{d\lambda}, \frac{d\phi}{d\lambda})$ is the tangent vector along the particle orbit 
\footnote{We use the notation $u$ to label the tangent vector along particle orbits in the 4-dimensional Lorentz spacetime, and the notation $T^{\text{OP}}$ represents the tangent vector in the optical geometry of spacetime.}. 
The $E$, $L$ are conserved energy and conserved angular momentum in a given particle orbit. Using these Killing vectors and conserved quantities in rotational spacetimes, the reduced equation of motion for test particles can be derived
\begin{eqnarray}
	&             &
	L_{\text{test particle}} 
	= \frac{1}{2} m \bigg( g_{\mu\nu} \frac{dx^{\mu}}{d\lambda} \frac{dx^{\nu}}{d\lambda} \bigg)
	= \frac{1}{2} m \epsilon \nonumber
	\\
	& \Rightarrow &
	g_{rr}  \bigg( \frac{dr}{d\lambda} \bigg)^{2}
	+ g_{\theta\theta}  \bigg( \frac{d\theta}{d\lambda} \bigg)^{2} 
	+ V_{\text{eff}}(r) = 0 .
	\label{reduced equation of motion}
\end{eqnarray}
In the second line, $V_{\text{eff}}(r)$ is the effective potential of test particles moving in the equatorial plane for an axially symmetric gravitational field. The explicit expression of the effective potential gives
\begin{eqnarray}
	V_{\text{eff}}(r) 
	& = & g_{tt} \bigg( \frac{dt}{d\lambda} \bigg)^{2} 
	+ 2g_{t\phi} \frac{dt}{d\lambda} \frac{d\phi}{d\lambda}
	+ g_{\phi\phi} \bigg( \frac{d\phi}{d\lambda} \bigg)^{2}
	- \epsilon \nonumber
	\\
	& = & - \frac{ E^{2}g_{\phi\phi}+2ELg_{t\phi}+L^{2}g_{tt} }{ g_{t\phi}^{2}-g_{tt}g_{\phi\phi} } 
	- \epsilon .
	\label{effective potential definition}
\end{eqnarray}
The parameter $\epsilon$ distinguishes the particle types in a gravitational field: $\epsilon=0$ characterizes massless photon orbits (e.g., photon orbits) and $\epsilon=1$ corresponds to massive particle orbits. It is worth noting that in many references, an additional minus sign is often absorbed into the definition of effective potential, $V_{\text{eff}}(r) \to - V_{\text{eff}}(r)$. However, our choice of effective potential in expression (\ref{effective potential definition}) is made such that the local minima of effective potential $\frac{d^{2}V_{\text{eff}}}{dr^{2}} > 0$ always correspond to stable particle orbits, and the local maxima of effective potential $\frac{d^{2}V_{\text{eff}}}{dr^{2}} < 0$ always correspond to unstable particle orbits.

For any light ring moving in the equatorial plane of rotational spacetimes, the photon's velocity components satisfy $u^{r} = \frac{dr}{d\lambda} = 0$, $u^{\theta} = \frac{d\theta}{d\lambda} = 0$. The equation of motion in (\ref{reduced equation of motion}) suggests that the effective potential for photons naturally vanishes
\begin{equation}
	V_{\text{eff}} (r=r_{\text{LR}}) = 0 .
\end{equation}
On the other hand, using the definition of conserved quantities $E$, $L$, and the angular velocity $\Omega=\frac{d\phi}{dt}$, a simple calculation on the effective potential gives
\begin{eqnarray}
	V_{\text{eff}}(r) 
	& = & g_{tt} \bigg( \frac{dt}{d\lambda} \bigg)^{2} 
	+ 2g_{t\phi} \frac{dt}{d\lambda} \frac{d\phi}{d\lambda}
	+ g_{\phi\phi} \bigg( \frac{d\phi}{d\lambda} \bigg)^{2} \nonumber
	\\
	& = & \bigg( \frac{dt}{d\lambda} \bigg)^{2} 
	\cdot \bigg( g_{tt} + 2g_{t\phi} \Omega + g_{\phi\phi} \Omega^{2} \bigg) .
\end{eqnarray}
The first part is a square, so it is non-negative for any particle orbit. In this way, the effective potential condition for light rings leads to
\begin{equation}
	V_{\text{eff}} (r=r_{\text{LR}}) = 0
	\ \ \Leftrightarrow \ \
	\bigg[ g_{tt} + 2g_{t\phi} \Omega + g_{\phi\phi} \Omega^{2} \bigg]_{r=r_{\text{LR}}} = 0 .
	\label{effective potential V=0}
\end{equation}
This equation precisely describes the angular velocity of photons moving along light rings. Following the derivation process from the reduced equation of motion in (\ref{reduced equation of motion}) to the angular velocity relation in (\ref{effective potential V=0}), we can summarize the following correspondences
\begin{widetext}
\begin{equation}
	u \cdot u = \epsilon = 0 
	\ \ \text{and} \ \  
	u^{r} = u^{\theta} = 0  
	\ \ \Rightarrow \ \ 
	V_{\text{eff}}(r) = 0  
	\ \ \Leftrightarrow \ \ 
	g_{tt} + 2g_{t\phi}\Omega + g_{\phi\phi} \Omega^{2} = 0 .
	\label{zero norm of tangent vector in equatorial plane}
\end{equation}
It is noteworthy that the above derivation process is very similar to what we have presented in expression (\ref{angular velocity equation}) using the unit norm of tangent vectors in optical geometry. Consequently, this implies that there are correspondences among the unit tangent vector norm condition given in expression (\ref{angular velocity equation}) for circular curves in 2-dimensional Randers-Finsler optical geometry, the null condition of tangent vector along light rings in expression (\ref{zero norm of tangent vector in equatorial plane}) in spacetime geometry, and the vanishing of effective potential $V_{\text{eff}}(r) = 0$ for photon beams confined to the equatorial plane of spacetime geometry. In conclusion, we can summarize the following equivalent relationships between optical geometry and spacetime geometry
\begin{equation}
	<T^{\text{OP}}, T^{\text{OP}}>_{(x,T^{\text{OP}})}^{(F)} = 1 
	\ \text{and} \ 
	(T^{\text{OP}})^{r} = (T^{\text{OP}})^{\theta} = 0
	\ \ \ \Leftrightarrow \ \ \ 
	u \cdot u = 0 \ \text{and} \ u^{r} = u^{\theta} = 0  
	\ \ \ \text{for light rings in the equatorial plane.}  
	\label{equivalence relation T}
\end{equation}
Furthermore, for photon orbits not confined to the equatorial plane, a similar equivalence (or correspondence) can also be obtained. Specifically, a routine calculation from the mathematical construction of Randers-Finsler optical geometry in expression (\ref{Optical geometry rotational}) would inevitably give rise to $u \cdot u = 0 \ \Leftrightarrow \ <T^{\text{OP}}, T^{\text{OP}}>_{(x,T^{\text{OP}})}^{(F)} = 1$, since the construction of optical geometry is achieved via the null constraint $d\tau^{2} = - ds^{2} = 0$ 
\footnote{A calculation of the invariant distance for a axially symmetric Lorentz spacetime through equations (\ref{Optical geometry rotational}) and (\ref{Renders geometry}) would give $ds^{2} = g_{\mu\nu} dx^{\mu} dx^{\nu} = (u \cdot u) \cdot d\lambda^{2} = g_{tt} dt^{2} \cdot \big[1 - \sqrt{ \alpha_{ij} \frac{dx^{i}}{dt} \frac{dx^{j}}{dt} } - \beta_{i} \frac{dx^{i}}{dt} \big] \cdot \big[ 1 + \sqrt{\alpha_{ij} \frac{dx^{i}}{dt} \frac{dx^{j}}{dt} } - \beta_{i} \frac{dx^{i}}{dt} \big] \propto g_{tt} dt^{2} \cdot  \big[ 1 - ||T^{\text{OP}}||^{(F)}_{(x,T^{\text{OP}})} \big]$, which eventually leads to $ds^{2} = 0  \Leftrightarrow u \cdot u = 0 \Leftrightarrow  ||T^{\text{OP}}||^{(F)}_{(x,T^{\text{OP}})} = \sqrt{<T^{\text{OP}}, T^{\text{OP}}>_{(x,T^{\text{OP}})}^{(F)}} = 1$.}. So we obtain the equivalence relation for general photon orbits that are not restricted in equatorial plane 
\begin{equation}
	<T^{\text{OP}}, T^{\text{OP}}>_{(x,T^{\text{OP}})}^{(F)} = 1
	\ \ \Leftrightarrow \ \ 
	u \cdot u = 0
	\ \ \ \ \ \ \ 
	\text{for photon orbits not restricted in the equatorial plane.} 
	\ \ \ \ \ \ \  
	\label{equivalence relation T 2}
\end{equation}
These correspondences in expressions (\ref{equivalence relation T}) and (\ref{equivalence relation T 2}) clearly show that the optical geometry and spacetime geometry play an equivalent role in the analysis of circular photon orbits. In particular, for light rings in the equatorial plane, combining results in (\ref{angular velocity equation}), (\ref{light ring unitarity condition}), and (\ref{effective potential V=0}), an additional equivalence relation can be derived
\begin{equation}
	\bigg[ \sqrt{ \alpha_{\phi\phi} \Omega^{2} } + \beta_{\phi} \Omega \bigg]_{r=r_{\text{LR}}} = 1
	\ \ \Leftrightarrow \ \   
	\bigg[ g_{tt} + 2g_{t\phi}\Omega + g_{\phi\phi} \Omega^{2} \bigg]_{r=r_{\text{LR}}} = 0
	\ \ \Leftrightarrow \ \ 
	V_{\text{eff}}(r=r_{\text{LR}}) = 0 .
	\label{equivalence relation V}
\end{equation}
Therefore, the unit tangent vector norm condition for circular curves in optical geometry can result in the vanishing of the effective potential for photon beams in the spacetime geometry. Eventually, we have proven that our geometric approach and the conventional effective potential approach yield the same well-known equation for angular velocity, which implies the equivalence of these two approaches when analyzing light rings in axially symmetric spacetimes.
\end{widetext}

Besides the above equation for angular velocity, it is necessary to prove that our geometric approach and the conventional effective potential approach can lead to all the equivalent results relevant to light rings. The most important information on light rings is their locations and stability. If we can demonstrate that the criteria for determining the location of light rings and their stability are equivalent in two approaches, as we have postulated in expression (\ref{equivalence relation for light rings}) and (\ref{equivalence relation for stable light rings}-\ref{equivalence relation for unstable light rings}), then all the derived conclusions associated with light rings must be identical. In order to further demonstrate the equivalence between our geometric approach and the conventional effective potential approach, it is convenient to reformulate the effective potential in terms of the impact parameter. For photon orbits confined to the equatorial plane, the impact parameter can be defined as the ratio of conserved angular momentum to conserved energy, via $b = \frac{L}{E}$. Using the impact parameter, the effective potential of photon beams moving in the equatorial plane can be rewritten as
\begin{eqnarray}
	V_{\text{eff}}(r) 
	& = & 
	- \frac{ E^{2} g_{\phi\phi} + 2EL g_{t\phi} + L^{2} g_{tt} }
	       { g_{t\phi}^{2}-g_{tt}g_{\phi\phi} } \nonumber
	\\
	& = &
	- \frac{ E^{2} }{ g_{tt} }  \cdot
	\frac{ \frac{g_{\phi\phi}}{g_{tt}} + 2 \frac{L}{E} \frac{g_{t\phi}}{g_{tt}} + \frac{L^{2}}{E^{2}} }{ \frac{g_{t\phi}^{2}-g_{tt}g_{\phi\phi}}{g_{tt}^{2}} } \nonumber
	\\ 
	& = &
	- \frac{ E^{2} }{ g_{tt} }  \cdot
	\frac{ - (\alpha_{\phi\phi} - \beta_{\phi}^{2}) - 2 b \beta_{\phi} + b^{2} }{ \alpha_{\phi\phi} } .
	\label{effective potential}
\end{eqnarray}
Furthermore, through a simple reduction, the impact parameter of photon orbits can be expressed using the Randers metric $\alpha$, $\beta$, and the angular velocity $\Omega$
\begin{eqnarray}
		b & = & \frac{L}{E} 
		= - \frac{ g_{t\phi}\cdot\frac{dt}{d\lambda} + g_{\phi\phi}\cdot\frac{d\phi}{d\lambda} }{ g_{tt}\cdot\frac{dt}{d\lambda} + g_{t\phi}\cdot\frac{d\phi}{d\lambda} } \nonumber
		%= - \frac{g_{t\phi} + g_{\phi\phi}\Omega }{ g_{tt} + g_{t\phi}\Omega } \nonumber
		\\ 
		& = &
		- \frac{ \frac{g_{t\phi}}{g_{tt}} + \frac{g_{\phi\phi}}{g_{tt}} \Omega }{ 1 + \frac{g_{t\phi}}{g_{tt}} \Omega } \nonumber
		\\ 
		& = &
		\frac{ \beta_{\phi} + (\alpha_{\phi\phi}-\beta_{\phi}^{2}) \cdot\Omega } 
		     { 1 - \beta_{\phi} \Omega }  .
		\label{expression-0}
\end{eqnarray}
Since the derivation process does not rely on the vanishing of the effective potential in equation (\ref{effective potential V=0}), the equality holds for all photon orbits traveling in the equatorial plane, not limited to light ring positions.

In the following, we give a demonstration of the equivalence between the geodesic curvature condition for light rings in expression (\ref{light ring geodesic curvature condition}) and the local extremum condition of effective potential for light rings. In axially symmetric spacetimes, light rings are always located at the local extrema of the effective potential, where the first-order derivative of the effective potential vanishes
\begin{equation}
	\frac{dV_{\text{eff}}(r)}{dr} \bigg|_{r=r_{\text{LR}}} = 0 .
\end{equation}
%\begin{widetext}
Using expression (\ref{effective potential}), the first-order derivative of the effective potential at the light ring position $r=r_{\text{LR}}$ can be calculated as
\begin{widetext}
\begin{eqnarray}
	\frac{dV_{\text{eff}}(r)}{dr} \bigg|_{r=r_{\text{LR}}} 
	& = & 
	\frac{d}{dr} 
	\bigg\{
	  - \frac{E^{2}}{g_{tt}} \cdot 
	    \frac{ b^{2} - 2\beta_{\phi} b - \big( \alpha_{\phi\phi}-\beta_{\phi}^{2} \big) }{ \alpha_{\phi\phi} }  
	\bigg\}_{r=r_{\text{LR}}} \nonumber
	\\
	& = & 
	\bigg\{
	  - \frac{d}{dr} \bigg( \frac{E^{2}}{g_{tt}} \bigg) \cdot 
	  \frac{ b^{2} - 2\beta_{\phi} b - \big( \alpha_{\phi\phi}-\beta_{\phi}^{2} \big) }{ \alpha_{\phi\phi} } 
	  - \frac{E^{2}}{g_{tt}} \cdot 
	  \frac{d}{dr} \bigg[ \frac{ b^{2} - 2\beta_{\phi} b - \big( \alpha_{\phi\phi}-\beta_{\phi}^{2} \big) }{ \alpha_{\phi\phi} } \bigg] 
	\bigg\}_{r=r_{\text{LR}}} .
	\label{effective potential first derivative}
\end{eqnarray}
At the light ring position, the vanishing of the effective potential $V_{\text{eff}}(r=r_{\text{LR}}) = 0$ implies that the first term in the brackets becomes zero. Meanwhile, the second term in the brackets can be simplified as
\begin{subequations}
\begin{eqnarray}
	\text{first term} 
	& = &
	\bigg\{
	  - \frac{d}{dr} \bigg( \frac{E^{2}}{g_{tt}} \bigg) \cdot 
	    \frac{ b^{2} - 2\beta_{\phi} b - \big( \alpha_{\phi\phi}-\beta_{\phi}^{2} \big) }{ \alpha_{\phi\phi} }
	\bigg\}_{r=r_{\text{LR}}}
	= 0  .
	\label{first term dV_dr}
    \\
	\text{second term}
	& = & 
	\bigg\{ 
	    - \frac{E^{2}}{g_{tt}} \cdot 
	      \frac{d}{dr} \bigg[ \frac{ b^{2} - 2\beta_{\phi} b - \big( \alpha_{\phi\phi}-\beta_{\phi}^{2} \big) }{ \alpha_{\phi\phi} } \bigg] 
	\bigg\}_{r=r_{\text{LR}}} \nonumber
	\\
	& = & 
	\bigg\{ 
	  \frac{E^{2}}{g_{tt}} \cdot
	  \frac{ b^{2} - 2\beta_{\phi} b - \big( \alpha_{\phi\phi}-\beta_{\phi}^{2} \big) }{ \alpha_{\phi\phi}^{2} } 
	  \cdot \frac{d\alpha_{\phi\phi}}{dr}
	  + \frac{E^{2}}{g_{tt}}  \frac{1}{\alpha_{\phi\phi}} \cdot
	    \bigg[ \frac{d\alpha_{\phi\phi}}{dr} + 2\big(b-\beta_{\phi}\big) \cdot \frac{d\beta_{\phi}}{dr} \bigg]
	\bigg\}_{r=r_{\text{LR}}} \nonumber
	\\
	& = & 
	\bigg\{ 
	  \frac{E^{2}}{g_{tt}}  \frac{1}{\alpha_{\phi\phi}} \cdot
	  \bigg[ \frac{d\alpha_{\phi\phi}}{dr} + 2\bigg( \frac{\beta_{\phi} + ( \alpha_{\phi\phi}-\beta_{\phi}^{2} ) \Omega}{1-\beta_{\phi}\Omega} - \beta_{\phi} \bigg) \cdot \frac{d\beta_{\phi}}{dr} \bigg]
	\bigg\}_{r=r_{\text{LR}}} \nonumber
	\\
	& = & 
	\bigg\{ 
	  \frac{E^{2}}{g_{tt}}  \frac{1}{\alpha_{\phi\phi}} \cdot
	  \bigg[ \frac{d\alpha_{\phi\phi}}{dr} +  \frac{2\alpha_{\phi\phi}\Omega}{1-\beta_{\phi}\Omega} \cdot \frac{d\beta_{\phi}}{dr} \bigg]
	\bigg\}_{r=r_{\text{LR}}} \nonumber
	\\
	& = & 
	\bigg\{ 
	  \frac{E^{2}}{g_{tt}}  \frac{2}{\sqrt{\alpha_{\phi\phi}}} \cdot
	  \bigg[ \frac{1}{2\sqrt{\alpha_{\phi\phi}}} \frac{d\alpha_{\phi\phi}}{dr} +  \text{Sign}(\Omega) \cdot \frac{d\beta_{\phi}}{dr} \bigg]
	\bigg\}_{r=r_{\text{LR}}} .
	\label{second term dV_dr}
\end{eqnarray}
\end{subequations}
Physically, the effective potential $V_{\text{eff}}(r)$ and its derivative $\frac{dV_{\text{eff}}(r)}{dr}$ are conventionally defined for a photon orbit in the equatorial plane, with $L$ and $E$ to be conserved quantities along this orbit. The impact parameter, which is defined as $b=\frac{L}{E}$, is also conserved. Consequently, $\frac{db}{dr}=0$ holds along the photon orbit. In the third line of (\ref{second term dV_dr}), the vanishing of the effective potential $V_{\text{eff}}(r=r_{\text{LR}}) = 0$ and the expression (\ref{expression-0}) for the impact parameter are utilized. In the fourth line, we have imposed the unit tangent vector condition in expression (\ref{light ring unitarity condition}) for any light ring position, which suggests $\big[1-\beta_{\phi}\Omega\big]_{r=r_{\text{LR}}} = \big[\sqrt{\alpha_{\phi\phi}\Omega^{2}}\big]_{r=r_{\text{LR}}}$. Substituting (\ref{first term dV_dr}) and (\ref{second term dV_dr}) into expression (\ref{effective potential first derivative}), the first-order derivative of the effective potential brings about
\begin{equation}
	\frac{dV_{\text{eff}}(r)}{dr} \bigg|_{r=r_{\text{LR}}}
	= 
	\bigg\{ 
	  \frac{E^{2}}{g_{tt}} \frac{2}{\sqrt{\alpha_{\phi\phi}}} \cdot
	  \bigg[
	    \frac{1}{2\sqrt{\alpha_{\phi\phi}}}  \frac{d\alpha_{\phi\phi}}{dr}
	    + \text{Sign}(\Omega) \cdot \frac{d\beta_{\phi}}{dr} 
	  \bigg] 
	\bigg\}_{r=r_{\text{LR}}}  .
	\label{dV_dr expr}
\end{equation} 
In any light ring position $r=r_{\text{LR}}$, the metric components maintain $g_{tt} < 0$ and $\alpha_{\phi\phi} > 0$. The vanishing of the first-order derivative of effective potential indicates that
\begin{equation}
	\frac{dV_{\text{eff}}(r)}{dr} \bigg|_{r=r_{\text{LR}}} = 0
	\ \ \Leftrightarrow \ \ 
	\bigg[
	    \frac{1}{2\sqrt{\alpha_{\phi\phi}}}  \frac{d\alpha_{\phi\phi}}{dr}
	    + \text{Sign}(\Omega) \cdot \frac{d\beta_{\phi}}{dr} 
	\bigg]_{r=r_{\text{LR}}}
	= 0
	\ \ \Leftrightarrow \ \
	\kappa_{g}^{(F)}(r=r_{\text{LR}}) 
	%= \bigg[ \kappa_{g}^{(\alpha)} + \kappa_{\beta}^{(\alpha)} \bigg]_{r=r_{\text{LR}}}
	= 0  .
	\label{effective potential dV_dr=0}
\end{equation}
In this way, we have demonstrated the equivalence between our geodesic curvature condition for light rings and the local extremum condition of the effective potential in the conventional approach. 
%\end{widetext}

Then we present an analysis of the light ring's stability, proving the equivalence between our flag curvature criterion in optical geometry and the local maxima (local minima) criterion of effective potential for stable (unstable) light rings. To achieve this point, it is necessary to compare the analytical expressions for the second-order derivative of effective potential and the flag curvature in Randers-Finsler optical geometry. The second-order derivative of the effective potential at the light ring positions can be calculated as
\begin{eqnarray}
	\frac{d^{2}V_{\text{eff}}(r)}{dr^{2}} \bigg|_{r=r_{\text{LR}}} 
	& = & 
	\frac{d^{2}}{dr^{2}} 
	\bigg\{ 
	  - \frac{E^{2}}{g_{tt}} \cdot 
	     \frac{ b^{2} - 2\beta_{\phi} b - \big( \alpha_{\phi\phi}-\beta_{\phi}^{2} \big) }{ \alpha_{\phi\phi} } 
	\bigg\}_{r=r_{\text{LR}}} \nonumber
	\\
	& = & 
	\bigg\{ 
	  \frac{d^{2}}{dr^{2}} 
	  \bigg( - \frac{E^{2}}{g_{tt}} \bigg) \cdot 
	  \frac{ b^{2} - 2\beta_{\phi} b - \big( \alpha_{\phi\phi}-\beta_{\phi}^{2} \big) }{ \alpha_{\phi\phi} }  
	\bigg\}_{r=r_{\text{LR}}} 
	+ 2 \bigg\{ 
	  \frac{d}{dr} \bigg( - \frac{E^{2}}{g_{tt}} \bigg) \cdot
	  \frac{d}{dr} \bigg[ \frac{ b^{2} - 2\beta_{\phi} b - \big( \alpha_{\phi\phi}-\beta_{\phi}^{2} \big) }{ \alpha_{\phi\phi} } \bigg] 
	\bigg\}_{r=r_{\text{LR}}} \nonumber
	\\
	&   & 
	+ \bigg\{ 
	    - \frac{E^{2}}{g_{tt}} \cdot
	      \frac{d^{2}}{dr^{2}} \bigg[ \frac{ b^{2} - 2\beta_{\phi} b - \big( \alpha_{\phi\phi}-\beta_{\phi}^{2} \big) }{ \alpha_{\phi\phi} } \bigg] \bigg\}_{r=r_{\text{LR}}} \nonumber
	\\
	& = & 
	\bigg\{  
	  - \frac{E^{2}}{g_{tt}} \cdot
	    \frac{d^{2}}{dr^{2}} \bigg[ \frac{ b^{2} - 2\beta_{\phi} b - \big( \alpha_{\phi\phi}-\beta_{\phi}^{2} \big) }{ \alpha_{\phi\phi} } \bigg]
	\bigg\}_{r=r_{\text{LR}}} \nonumber
	\\
	& = & 
	\bigg\{ 
	  - \frac{E^{2}}{g_{tt}} \cdot
	    \frac{ b^{2} - 2\beta_{\phi} b - \big( \alpha_{\phi\phi}-\beta_{\phi}^{2} \big) }{ \alpha_{\phi\phi}^{3} }
	    \cdot 2 \bigg( \frac{d\alpha_{\phi\phi}}{dr} \bigg)^{2}
	  - \frac{E^{2}}{g_{tt}}  \frac{2}{\alpha_{\phi\phi}^{2}}
	    \cdot \frac{d\alpha_{\phi\phi}}{dr} \cdot
	    \bigg[ \frac{d\alpha_{\phi\phi}}{dr} + 2 \big(b-\beta_{\phi}\big) \cdot \frac{d\beta_{\phi}}{dr} \bigg] \nonumber
	\\
	&   &
	  + \frac{E^{2}}{g_{tt}} \cdot
	    \frac{ b^{2} - 2\beta_{\phi} b - \big( \alpha_{\phi\phi}-\beta_{\phi}^{2} \big) }{ \alpha_{\phi\phi}^{2} }
	    \cdot \frac{d^{2}\alpha_{\phi\phi}}{dr^{2}}
	  - \frac{E^{2}}{g_{tt}}  \frac{2}{\alpha_{\phi\phi}} \cdot
	    \bigg[ \bigg( \frac{d\beta_{\phi}}{dr} \bigg)^{2} - \frac{1}{2} \frac{d^{2}\alpha_{\phi\phi}}{dr^{2}} - (b-\beta_{\phi}) \cdot \frac{d^{2}\beta_{\phi}}{dr^{2}} \bigg]
	\bigg\}_{r=r_{\text{LR}}} .
	\label{effective potential second order derivative}
\end{eqnarray}
In the third equal sign, we have used the $V_{\text{eff}}(r=r_{\text{LR}})=0$ and $\frac{dV_{\text{eff}}(r)}{dr} \big|_{r=r_{\text{LR}}} = 0$ for light rings. Similar to the reduction in the first-order derivative of the effective potential, the various terms in the brackets can be simplified as 
\begin{subequations}
\begin{eqnarray}
	\text{first term}
	& = & 
	\bigg\{ 
	- \frac{E^{2}}{g_{tt}} \cdot
	\frac{ b^{2} - 2\beta_{\phi} b - \big( \alpha_{\phi\phi}-\beta_{\phi}^{2} \big) }{ \alpha_{\phi\phi}^{3} }
	\cdot 2 \bigg( \frac{d\alpha_{\phi\phi}}{dr} \bigg)^{2} 
	\bigg\}_{r=r_{\text{LR}}}
	= 0 ,
	\label{first term d2V_dr2}
	\\
	\text{second term}
	& = &
	\bigg\{
	  - \frac{E^{2}}{g_{tt}} \frac{2}{\alpha_{\phi\phi}^{2}}
	    \frac{d\alpha_{\phi\phi}}{dr} \cdot
	    \bigg[ \frac{d\alpha_{\phi\phi}}{dr} + 2 \big(b-\beta_{\phi}\big) \cdot \frac{d\beta_{\phi}}{dr} \bigg]
	\bigg\}_{r=r_{\text{LR}}} \nonumber
	\\
	& = &
	\bigg\{
	  - \frac{E^{2}}{g_{tt}} \frac{2}{\alpha_{\phi\phi}^{2}} 
	    \frac{d\alpha_{\phi\phi}}{dr} \cdot 
	    \bigg[ 
	      \frac{d\alpha_{\phi\phi}}{dr}
	      + 2 \bigg( \frac{\beta_{\phi} + ( \alpha_{\phi\phi}-\beta_{\phi}^{2} ) \Omega }{ 1-\beta_{\phi}\Omega } - \beta_{\phi} \bigg) \cdot \frac{d\beta_{\phi}}{dr}
	    \bigg]
	\bigg\}_{r=r_{\text{LR}}} \nonumber
	\\
	& = &
	\bigg\{
	- \frac{E^{2}}{g_{tt}} \frac{4\sqrt{\alpha_{\phi\phi}}}{\alpha_{\phi\phi}^{2}}
	\frac{d\alpha_{\phi\phi}}{dr} \cdot 
	\bigg[
	  \frac{1}{2\sqrt{\alpha_{\phi\phi}}} \frac{d\alpha_{\phi\phi}}{dr}
	  + \text{Sign}(\Omega) \cdot \frac{d\beta_{\phi}}{dr}
	\bigg]
	\bigg\}_{r=r_{\text{LR}}} 
	= 0 ,
	\label{second term d2V_dr2}
    \\
    \text{third term}
    & = &
    \bigg\{
      \frac{E^{2}}{g_{tt}} \cdot
      \frac{ b^{2} - 2\beta_{\phi} b - \big( \alpha_{\phi\phi}-\beta_{\phi}^{2} \big) }{ \alpha_{\phi\phi}^{2} }
      \cdot \frac{d^{2}\alpha_{\phi\phi}}{dr^{2}}
    \bigg\}_{r=r_{\text{LR}}}
    = 0 ,
    \label{third term d2V_dr2}
    \\
    \text{fourth term}
    & = &
    \bigg\{
      - \frac{E^{2}}{g_{tt}}  \frac{2}{\alpha_{\phi\phi}} \cdot
        \bigg[ \bigg( \frac{d\beta_{\phi}}{dr} \bigg)^{2} - \frac{1}{2} \frac{d^{2}\alpha_{\phi\phi}}{dr^{2}} - (b-\beta_{\phi}) \cdot \frac{d^{2}\beta_{\phi}}{dr^{2}} \bigg]
    \bigg\}_{r=r_{\text{LR}}} \nonumber
    \\
    & = &
    \bigg\{
       - \frac{E^{2}}{g_{tt}}  \frac{2}{\alpha_{\phi\phi}} \cdot
         \bigg[ 
           \bigg( \frac{d\beta_{\phi}}{dr} \bigg)^{2} 
           - \frac{1}{2} \frac{d^{2}\alpha_{\phi\phi}}{dr^{2}} 
           - \bigg( \frac{\beta_{\phi} + ( \alpha_{\phi\phi}-\beta_{\phi}^{2} ) \Omega }{ 1-\beta_{\phi}\Omega } - \beta_{\phi} \bigg) 
           \cdot \frac{d^{2}\beta_{\phi}}{dr^{2}} \bigg]
    \bigg\}_{r=r_{\text{LR}}} \nonumber
    \\
    & = &
    \bigg\{
    - \frac{E^{2}}{g_{tt}}  \frac{2}{\alpha_{\phi\phi}} \cdot
    \bigg[ 
    \bigg( \frac{d\beta_{\phi}}{dr} \bigg)^{2} 
    - \frac{1}{2} \frac{d^{2}\alpha_{\phi\phi}}{dr^{2}} 
    - \frac{\alpha_{\phi\phi}\Omega }{ \sqrt{\alpha_{\phi\phi}\Omega^{2}} }  
    \cdot \frac{d^{2}\beta_{\phi}}{dr^{2}} \bigg]
    \bigg\}_{r=r_{\text{LR}}} .
\end{eqnarray}
\end{subequations}
Combining these terms, the second-order derivative of the effective potential becomes
\begin{equation}
	\frac{d^{2}V_{\text{eff}}(r)}{dr^{2}} \bigg|_{r=r_{\text{LR}}}
	= \bigg\{
	- \frac{E^{2}}{g_{tt}}  \frac{2}{\alpha_{\phi\phi}} \cdot
	\bigg[ 
	\bigg( \frac{d\beta_{\phi}}{dr} \bigg)^{2} 
	- \frac{1}{2} \frac{d^{2}\alpha_{\phi\phi}}{dr^{2}} 
	- \frac{\alpha_{\phi\phi}\Omega }{ \sqrt{\alpha_{\phi\phi}\Omega^{2}} }  
	\cdot \frac{d^{2}\beta_{\phi}}{dr^{2}} \bigg]
	\bigg\}_{r=r_{\text{LR}}}	.
\end{equation}
Notably, it is clearly evident that this result closely resembles the flag curvature expression in (\ref{flag curvature LR}). Since the signs of spacetime metric components are not changed outside the ergosphere ($g_{tt}<0$ and $g_{\phi\phi}>0$, which suggests $\alpha_{\phi\phi}>0$), the second-order derivative of the effective potential and the flag curvature in Randers-Finsler optical geometry always have the same sign at light ring positions
\begin{equation}
	\mathcal{K}^{(F)}_{\text{flag}}(r, T^{\text{OP}}, \partial^{\text{OP}}_{r}) \bigg|_{r=r_{\text{LR}}}
	\ \ \propto \ \
	\bigg[
	\bigg( \frac{d\beta_{\phi}}{dr} \bigg)^{2} 
	- \frac{1}{2} \frac{d^{2}\alpha_{\phi\phi}}{dr^{2}} 
	- \frac{\alpha_{\phi\phi}\Omega }{ \sqrt{\alpha_{\phi\phi}\Omega^{2}} }  
	\cdot \frac{d^{2}\beta_{\phi}}{dr^{2}}
	\bigg]_{r=r_{\text{LR}}} 
	\ \ \propto \ \
	\frac{d^{2}V_{\text{eff}}(r)}{dr^{2}} \bigg|_{r=r_{\text{LR}}} .
\end{equation} 
From the above derivations, we have proved the following equivalence relationship on the stability of light rings
\begin{subequations}
\begin{eqnarray}
	\mathcal{K}^{(F)}_{\text{flag}}(r, T^{\text{OP}}, \partial^{\text{OP}}_{r}) \bigg|_{r=r_{\text{LR}}} > 0 
	\ \ \Leftrightarrow \ \ 
	& \text{stable light rings} &
	\ \ \Leftrightarrow \ \ 
	\frac{d^{2}V_{\text{eff}}(r)}{dr^{2}} \bigg|_{r=r_{\text{LR}}} > 0  ,
	\label{equivelence relation 3a}
	\\
	\mathcal{K}^{(F)}_{\text{flag}}(r, T^{\text{OP}}, \partial^{\text{OP}}_{r}) \bigg|_{r=r_{\text{LR}}}  < 0 
	\ \ \Leftrightarrow \ \ 
	& \text{unstable light rings} &
	\ \ \Leftrightarrow \ \ 
	\frac{d^{2}V_{\text{eff}}(r)}{dr^{2}} \bigg|_{r=r_{\text{LR}}} < 0  .
	\label{equivelence relation 3b}
\end{eqnarray}
\end{subequations}
In conclusion, when the flag curvature is negative, there are no conjugate points in light rings, suggesting the corresponding light rings to be unstable, and the effective potential naturally reaches a local maximum. Conversely, when the flag curvature becomes positive, conjugate points may exist in such light rings, indicating the corresponding light rings to be stable, and the effective potential automatically reaches a local minimum. 

At the end of this section, it is important to see how the geometric approach of light rings presented in this work reproduces the geometric approach to photon sphere for spherically symmetric spacetimes proposed in previous studies \cite{QiaoCK2022a,QiaoCK2022b}. For a static and spherically symmetric spacetime (or the slowly rotating spacetime), the contributions from the non-Riemannian part $\beta$ in Randers geometry become negligible. As a result, the optical geometry (restricted to the equatorial plane) reduces to a Riemannian geometry 
\begin{equation}
	dt = \sqrt{\alpha_{ij}dx^{i}dx^{j}} + \beta_{i} dx^{i}
	\approx \sqrt{\alpha_{rr}dr^{2} + \alpha_{\phi\phi}d\phi^{2}}
	\ \ \ \Rightarrow \ \ \ 
	dt^{2} = \alpha_{ij}dx^{i}dx^{j}
	= \alpha_{rr}dr^{2} + \alpha_{\phi\phi}d\phi^{2}
\end{equation}
In such cases, the geodesic curvature of a circular curve only contains contributions from Riemannian metric $\alpha$. The vanishing of geodesic curvature at the photon sphere radius leads to
\begin{equation}
	%\kappa^{(F)} = \kappa^{(\alpha)} , 
	%\ \ \ \ \ \ \ \ 
	\kappa^{(F)}(r=r_{\text{ph}}) 
	\approx \kappa^{(\alpha)}(r=r_{\text{ph}})
	= \bigg[ \frac{1}{2\sqrt{\alpha_{rr}}} \frac{\partial \log(\alpha_{\phi\phi})}{\partial r} \bigg]_{r=r_{\text{ph}}}
	= 0  
	\ \ \ \Rightarrow \ \ \ 
	\bigg[ \frac{1}{2\sqrt{\alpha_{rr}}} 
	\cdot \frac{1}{\alpha_{\phi\phi}} \frac{\partial \alpha_{\phi\phi}}{\partial r} \bigg]_{r=r_{\text{ph}}} = 0  . 
\end{equation}
Additionally, the flag curvature in the 2-dimensional Randers-Finsler optical geometry simply recovers the Gaussian curvature of the Riemannian geometry $\alpha$
\begin{equation}
	\mathcal{K}^{(F)}_{\text{flag}}(r, T^{\text{OP}}, \partial_{r}) \bigg|_{r=r_{\text{ph}}} 
	\approx  
	\bigg[ 
	    - \frac{1}{2\alpha_{rr}\alpha_{\phi\phi}} \cdot \frac{d^{2}\alpha_{\phi\phi}}{dr^{2}} 
	\bigg]_{r=r_{\text{ph}}}
	= \mathcal{K}_{\text{Gaussian}}^{(\alpha)} (r=r_{\text{ph}}) .
\end{equation}
On the other hand, the effective potential of photons moving in static and spherically symmetric spacetimes (or slowly rotating spacetimes) can be simplified as
\begin{equation}
	V_{\text{eff}}(r)
	= - \frac{ E^{2} }{ g_{tt} }  \cdot
	\frac{ b^{2} - 2 b \beta_{\phi} - (\alpha_{\phi\phi} - \beta_{\phi}^{2})  }{ \alpha_{\phi\phi} } 
	\approx - \frac{E^{2}}{g_{tt}} \bigg( \frac{b^{2}}{\alpha_{\phi\phi}} - 1 \bigg) .
\end{equation}
A straightforward simplification of expression (\ref{dV_dr expr}) indicates that extremum condition of effective potential at the photon sphere radius is equivalent to the vanishing of geodesic curvature in the optical geometry
\begin{equation} 
	\frac{dV_{\text{eff}}(r)}{dr} \bigg|_{r=r_{\text{ph}}}
	\approx 
	\bigg[ 
	  \frac{E^{2}}{g_{tt}}  \frac{1}{\alpha_{\phi\phi}}
	  \cdot \frac{d\alpha_{\phi\phi}}{dr}
	\bigg]_{r=r_{\text{ph}}}
	= 0 
	\ \ \Leftrightarrow \ \
	\kappa^{(\alpha)}(r=r_{\text{ph}}) = 0 .
\end{equation}
It can also be easily verified that the unit tangent vector norm condition in optical geometry implies that $<T^{\text{OP}}, T^{\text{OP}}>^{(F)}_{(x,T^{\text{OP}})} \approx <T^{\text{OP}}, T^{\text{OP}}>^{(\alpha)} = \alpha_{\phi\phi}\Omega^{2} = 1$ holds for circular photon orbits in the equatorial plane, with $T^{\text{OP}}=(\frac{dr}{dt},\frac{d\theta}{dt},\frac{d\phi}{dt})=(0,0,\Omega)$. The impact parameter along the circular orbits becomes $b = \frac{L}{E} \approx - \frac{ g_{\phi\phi}\cdot\frac{d\phi}{d\lambda} }{ g_{tt}\cdot\frac{dt}{d\lambda} } = \alpha_{\phi\phi}\Omega$ (since $g_{t\phi} \approx 0$), indicating that $b^{2} \approx \alpha_{\phi\phi}^{2} \Omega^{2} = \alpha_{\phi\phi}$. Based on the above analysis, we reach the following conclusion
\begin{equation}
	<T^{\text{OP}}, T^{\text{OP}}>^{(\alpha)} = 1 
	\ \ \text{and} \ \ 
	T^{\text{OP}}_{r} = T^{\text{OP}}_{\theta} = 0
	\ \ \Rightarrow \ \ 
	\big[ \alpha_{\phi\phi}\Omega^{2} \big]_{r=r_{\text{ph}}} = 1
	\ \ \Leftrightarrow \ \ 
	V_{\text{eff}}(r=r_{\text{ph}})
	\approx \bigg[ - \frac{E^{2}}{g_{tt}}\bigg(\frac{b^{2}}{\alpha_{\phi\phi}}-1\bigg) \bigg]_{r=r_{\text{ph}}}
	= 0  .
\end{equation}
Furthermore, the local maxima (or local minima) criterion for stable (or unstable) photon spheres in spherically symmetric spacetimes is equivalent to the Gaussian curvature criterion for stable (or unstable) photon spheres 
\begin{subequations}
\begin{eqnarray}
	\frac{d^{2}V_{\text{eff}}(r)}{dr^{2}} \bigg|_{r=r_{\text{ph}}} 
	\approx 
	\bigg[ 
	  - \frac{E^{2}}{g_{tt}} \cdot 
	    \bigg( - \frac{1}{\alpha_{\phi\phi}} \frac{d^{2}\alpha_{\phi\phi}}{dr^{2}} \bigg)
	\bigg]_{r=r_{\text{ph}}}
	> 0
	& \Leftrightarrow &
	\mathcal{K}_{\text{Gaussian}}^{(\alpha)} (r=r_{\text{ph}}) > 0 ,
	\label{equivelence relation 4a}
	\\
	\frac{d^{2}V_{\text{eff}}(r)}{dr^{2}} \bigg|_{r=r_{\text{ph}}} 
	\approx 
	\bigg[
	  - \frac{E^{2}}{g_{tt}} \cdot 
	    \bigg( - \frac{1}{\alpha_{\phi\phi}} \frac{d^{2}\alpha_{\phi\phi}}{dr^{2}} \bigg)
	\bigg]_{r=r_{\text{ph}}}
	< 0
	& \Leftrightarrow &
	\mathcal{K}_{\text{Gaussian}}^{(\alpha)} (r=r_{\text{ph}}) < 0 .
	\label{equivelence relation 4b}
\end{eqnarray}
\end{subequations}
It has been explicitly shown that the geometric approach of circular photon orbits presented in this work exactly reproduces the geometric approach to photon spheres for spherically symmetric spacetimes proposed in our previous studies. 

\begin{table*}
\caption{Comparisons of our geometric approach to photon spheres in spherical symmetric spacetimes, our geometric approach to light rings in axially symmetric spacetimes, and the conventional effective potential approach to circular photon orbits.}
\label{table1}
\vspace{1mm}
\begin{adjustbox}{width=1.06\linewidth,center}
\footnotesize
%\scriptsize
\begin{ruledtabular}
	\begin{tabularx}{1.1\textwidth}{lccccc}
		\\ [-7pt]
		Approach & Geometric Approach & Geometric Approach & Conventional Approach &
		\\
		         & (Spherically Symmetric Spacetimes) & (Axially Symmetric Spacetimes) & (All Stationary Spacetimes)
		\\ [3pt]
		\hline
		\\ [-7pt]
		Geometry & Optical Geometry      & Optical Geometry   & Spacetime Geometry &
		\\
		         & (Riemannian Geometry) & (Randers-Finsler Geometry) & (Lorentz Geometry)
		\\ [3pt]
		\hline
		\\ [-7pt]
		Basic quantities & Geodesic Curvature $\kappa_{g}^{(\alpha)}(r)$ & Geodesic Curvature $\kappa_{g}^{(F)}(r)$ & Effective Potential $V_{\text{eff}}(r)$ &
		\\ 
		& Gaussian Curvature $\mathcal{K}_{\text{Gaussian}}^{(\alpha)}(r)$ & Flag Curvature $\mathcal{K}^{(F)}_{\text{flag}}(r,T^{\text{OP}},\partial^{\text{OP}}_{r})$ & (and its derivatives) &     
		\\ [3pt]
		\hline
		\\ [-7pt]
		Circular Photon Orbit   & Photon Sphere & Light Ring & Photon Sphere or Light Ring
		\\ [3pt]
		\hline
		\\ [-7pt]
		Tangent Vector's Condition   & $<T^{\text{OP}}, T^{\text{OP}}>^{(\alpha)} = 1$ \ & \ $< T^{\text{OP}} \cdot T^{\text{OP}} >_{(x,T^{\text{OP}})}^{(F)} = 1$  & $u \cdot u = 0$ 
		\\ [3pt]
		& and $(T^{\text{OP}})^{r} = (T^{\text{OP}})^{\theta} = 0$ \ & \ and $(T^{\text{OP}})^{r} = (T^{\text{OP}})^{\theta} = 0$  & and $u^{r}=u^{\theta}=0$
		\\ [3pt]
		\hline
		\\ [-7pt]
		Circular Orbits's Condition   & $\alpha_{\phi\phi}\Omega^{2} = 1$ & $\sqrt{\alpha_{\phi\phi}\Omega^{2}} +\beta_{\phi}\Omega = 1$ & $V_{\text{eff}}(r)=0$
		\\ [3pt] 
		& $\kappa_{g}^{(\alpha)}(r)=0$ & $\kappa_{g}^{(F)}(r)=0$ & $\frac{dV_{\text{eff}}(r)}{dr}=0$
		\\ [3pt]
		\hline
		\\ [-7pt]
		Unstable Circular Orbit & \ $\kappa_{g}^{(\alpha)}(r)=0$ and $\mathcal{K}_{\text{Gaussian}}^{(\alpha)}(r)<0$ \ & \ $\kappa_{g}^{(F)}(r)=0$ and $\mathcal{K}^{(F)}_{\text{flag}}(r,T^{\text{OP}},\partial^{\text{OP}}_{r})<0$ \ & \ $\frac{dV_{\text{eff}}(r)}{dr}=0$ and $\frac{d^{2}V_{\text{eff}}(r)}{dr^{2}}<0$ &
		\\ [3pt]
		Stable Circular Orbit   & \ $\kappa_{g}^{(\alpha)}(r)=0$ and $\mathcal{K}_{\text{Gaussian}}^{(\alpha)}(r)>0$ \ & \ $\kappa_{g}^{(F)}(r)=0$ and  $\mathcal{K}^{(F)}_{\text{flag}}(r,T^{\text{OP}},\partial^{\text{OP}}_{r})>0$ \ & \ $\frac{dV_{\text{eff}}(r)}{dr}=0$ and $\frac{d^{2}V_{\text{eff}}(r)}{dr^{2}}>0$ &
		\\ [3pt]
	\end{tabularx}
\end{ruledtabular}
\end{adjustbox}
\end{table*}

Through the relations presented in expressions (\ref{equivalence relation T}-\ref{equivalence relation T 2}), (\ref{equivalence relation V}), (\ref{effective potential dV_dr=0}), (\ref{equivelence relation 3a}-\ref{equivelence relation 3b}), and (\ref{equivelence relation 4a}-\ref{equivelence relation 4b}), we successfully demonstrate the equivalence between our geometric approach (based on intrinsic curvatures in Randers-Finsler optical geometry) and the conventional approach (based on the effective potential of photons). Given this equivalence, it is reasonable to apply this geometric approach to an arbitrary axially symmetric spacetime. Our geometric approach, which provides a mathematically self-contained framework in the studies of photon spheres and light rings, serves as an alternative and complementary approach to the conventional effective potential approach. A systematic comparison of our geometric approach to photon spheres in spherically symmetric spacetimes, our geometric approach to light rings in axially symmetric spacetimes, and the conventional effective potential approach is summarized in Table \ref{table1}.

\section{Some Examples \label{section5}}

In this section, we present two representative examples to illustrate how to obtain light rings in axially symmetric spacetimes using our geometric approach. In gravity theories, the most fundamental axially symmetric spacetimes are the Kerr spacetime and Kerr-Newman spacetime. They have profound significance in both theoretical and observational studies. The application of our geometric approach to such spacetimes effectively demonstrates its practical utility and theoretical validity. 

%\begin{widetext}
\textbf{Kerr Spacetime:} The metric for Kerr spacetime in general relativity is given by
\begin{equation}
	ds^{2}  
	= - \bigg( 1 - \frac{2Mr}{\Sigma} \bigg) dt^{2}
	  - \frac{4aMr\sin^{2}\theta}{\Sigma} dt d\phi  
	  + \frac{\Sigma}{\Delta} dr^{2} 
	  + \Sigma d\theta^{2} 
	  + \bigg( r^{2} + a^{2} + \frac{2a^{2}Mr\sin^{2}\theta}{\Sigma} \bigg) \sin^{2}\theta d\phi^{2} .
\end{equation}
where $\Sigma$ and $\Delta$ are two functions defined as
\begin{equation}
	\Sigma = r^{2} + a^{2} \cos^{2}\theta ,
	\ \ \ \ \ \ 
	\Delta = r^{2} - 2Mr + a^{2} .
\end{equation}
The mathematical construction of optical geometry for this Kerr spacetime gives
\begin{equation}
	dt = \sqrt{-\frac{g_{rr}}{g_{tt}} dr^{2} - \frac{g_{\theta\theta}}{g_{tt}} d\theta^{2} + \frac{g_{t\phi}^{2}-g_{tt}g_{\phi\phi}}{g_{tt}^{2}} d\phi^{2}} 
	- \frac{g_{t\phi}}{g_{tt}} d\phi %\nonumber
	%\\
	= \sqrt{ \frac{\Sigma^{2}}{\Sigma-2Mr} \bigg( \frac{1}{\Delta} dr^{2} + d\theta^{2} + \frac{\Delta \sin^{2}\theta}{\Sigma-2Mr} d\phi^{2} \bigg) }  
	- \frac{2aMr\sin^{2}\theta}{\Sigma-2Mr} d\phi  .
\end{equation}
When confined to the equatorial plane ($\theta=\pi/2$), the optical geometry becomes
\begin{equation}
	dt = \sqrt{ \frac{r^{4}}{r^{2}-2Mr} \bigg( \frac{1}{\Delta} dr^{2} + \frac{\Delta \sin^{2}\theta}{r^{2}-2Mr} d\phi^{2} \bigg) }
	- \frac{2aMr}{r^{2}-2Mr} d\phi  .
\end{equation}
From our geometric approach, the light ring naturally vanishes its geodesic curvature in Randers-Finsler optical geometry. The geodesic curvature condition to light ring in Kerr spacetime leads to
\begin{eqnarray}
	\kappa_{g}^{(F)}(r=r_{\text{LR}}) = 0 
	& \Rightarrow &
	\bigg[
	\frac{1}{2\sqrt{\alpha_{\phi\phi}}} \frac{d\alpha_{\phi\phi}}{dr}
	+ \text{Sign}(\Omega) \cdot \frac{d\beta_{\phi}}{dr} 
	\bigg]_{r=r_{\text{LR}}}
	= 0 \nonumber
	\\
	& \Rightarrow & 
	\frac{1}{(r_{\text{LR}}-2M)^{2}} \cdot \bigg[ \frac{ r_{\text{LR}}^{3} - 5Mr_{\text{LR}}^{2} + 6M^{2}r_{\text{LR}} - 2a^{2}M }{ \text{Sign}(r_{\text{LR}}-2M) \cdot \sqrt{\Delta\big|_{r=r_{\text{LR}}}} } + \text{Sign}(\Omega) \cdot 2 a M \bigg] 
	= 0 \nonumber
	\\
	& \Rightarrow & r_{\text{LR}} (r_{\text{LR}}-3M)^{2} - 4Ma^{2} = 0 \nonumber
	\\
	& \Rightarrow & r_{LR} = 2M \cdot \bigg\{ 1 + \cos\bigg[ \frac{2}{3} \arccos\bigg(-\text{Sign}(\Omega)\cdot\frac{a}{M}\bigg) \bigg] \bigg\} 
	= 2M \cdot \bigg\{ 1 + \cos\bigg[ \frac{2}{3}\arccos\bigg(\mp\frac{a}{M}\bigg) \bigg] \bigg\} .
	\ \ \ \ \ \ \ \ 
	\label{light ring Kerr}
\end{eqnarray}
This gives the correct positions of light rings in Kerr spacetime, derived solely from the geodesic curvature condition in Randers-Finsler optical geometry. It is in full agreement with the well-known analytical results for light rings in Kerr spacetime \cite{Bardeen1973,Bambi2018}. Here, we have conveniently defined the direction of increasing azimuthal angle $\phi$ such that a positive angular velocity $\Omega > 0$ describes the prograde motion of light, and a negative angular velocity $\Omega < 0$ describes the retrograde motion of light. In this convention, the spin parameter of the Kerr spacetime is assumed to be nonnegative $a \ge 0$, with the minus and plus signs in the last equality of expression (\ref{light ring Kerr}) corresponding to light rings for prograde and retrograde photon motions. Remarkably, the second last line in expression (\ref{light ring Kerr}) is precisely the constraint that can be obtained from the vanishing of the Carter constant in the equatorial plane of Kerr spacetime \cite{Tavlayan2020,LiuYX2019} 
\begin{equation}
	\eta = \frac{\mathcal{Q}}{E^{2}} 
	= \frac{ r_{\text{LR}}^{3} }{ a^{2} (r_{\text{LR}}-M)^{2} } 
	\cdot \bigg[ 4a^{2}M - r_{\text{LR}} (r_{\text{LR}}-3M)^{2} \bigg]
	= 0
\end{equation}
Furthermore, numerical calculations suggest that the flag curvature calculated using (\ref{flag curvature LR}) at these light ring positions is negative, indicating that the light rings for prograde and retrograde photon motions in Kerr spacetime are unstable.

\textbf{Kerr-Newman Spacetime (with Electric and Magnetic Charge):} The metric for Kerr-Newman spacetime is expressed as
\begin{eqnarray}
	ds^{2}  
	& = & - \bigg( 1 - \frac{2Mr-Q_{e}^{2}-Q_{m}^{2}}{\Sigma} \bigg) dt^{2}
	- \frac{2a(2Mr-Q_{e}^{2}-Q_{m}^{2})\sin^{2}\theta}{\Sigma} dt d\phi  
	+ \frac{\Sigma}{\Delta_{\text{KN}}} dr^{2} \nonumber
	\\
	&   & + \Sigma d\theta^{2} 
	+ \bigg( r^{2} + a^{2} + \frac{(2Mr-Q_{e}^{2}-Q_{m}^{2})a^{2}\sin^{2}\theta}{\Sigma} \bigg) \sin^{2}\theta d\phi^{2} .
\end{eqnarray}
with $Q_{e}$ and $Q_{m}$ to be the electric and magnetic charge of the rotating black hole, and the functions $\Sigma$ and $\Delta_{\text{KN}}$ are defined as
\begin{equation}
	\Sigma = r^{2} + a^{2} \cos^{2}\theta , 
	\ \ \ \ \ \ 
	\Delta_{\text{KN}} = r^{2} - 2Mr + a^{2} + Q_{e}^{2} + Q_{m}^{2} .
\end{equation}
To study the photon orbits, the construction of optical geometry for this Kerr-Newman spacetime gives
\begin{eqnarray}
	dt & = & \sqrt{-\frac{g_{rr}}{g_{tt}} dr^{2} - \frac{g_{\theta\theta}}{g_{tt}} d\theta^{2} + \frac{g_{t\phi}^{2}-g_{tt}g_{\phi\phi}}{g_{tt}^{2}} d\phi^{2}} 
	- \frac{g_{t\phi}}{g_{tt}} d\phi \nonumber
	\\
	& = & \sqrt{ \frac{\Sigma^{2}}{\Sigma-2Mr+Q_{e}^{2}+Q_{m}^{2}} \bigg( \frac{1}{\Delta_{\text{KN}}} dr^{2} + d\theta^{2} + \frac{\Delta_{\text{KN}} \sin^{2}\theta}{\Sigma-2Mr+Q_{e}^{2}+Q_{m}^{2}} d\phi^{2} \bigg) }  
	- \frac{a(2Mr-Q_{e}^{2}-Q_{m}^{2})\sin^{2}\theta}{\Sigma-2Mr+Q_{e}^{2}+Q_{m}^{2}} d\phi .
	\ \ 
\end{eqnarray}
Similarly, for photons moving in the equatorial plane $\theta=\pi/2$, the optical geometry reduces to
\begin{equation}
	dt = \sqrt{ \frac{r^{4}}{r^{2}-2Mr+Q_{e}^{2}+Q_{m}^{2}} \bigg( \frac{1}{\Delta_{\text{KN}}} dr^{2} + \frac{\Delta_{\text{KN}} \sin^{2}\theta}{r^{2}-2Mr+Q_{e}^{2}+Q_{m}^{2}} d\phi^{2} \bigg) }
	- \frac{a(2Mr-Q_{e}^{2}-Q_{m}^{2})}{r^{2}-2Mr+Q_{e}^{2}+Q_{m}^{2}} d\phi .
\end{equation}
The geodesic curvature condition for light rings leads to
\begin{eqnarray}
	\kappa_{g}^{(F)}(r=r_{\text{LR}}) = 0 
	& \Rightarrow &
	\bigg[
	\frac{1}{2\sqrt{\alpha_{\phi\phi}}} \frac{d\alpha_{\phi\phi}}{dr}
	+ \text{Sign}(\Omega) \cdot \frac{d\beta_{\phi}}{dr} 
	\bigg]_{r=r_{\text{LR}}}
	= 0 \nonumber
	\\
	& \Rightarrow & \bigg[ \frac{r_{\text{LR}}^{4} - 5Mr_{\text{LR}}^{3} + (6M^{2}+3Q_{e}^{2}+3Q_{m}^{2})r_{\text{LR}}^{2} + 2(Q_{e}^{2}+Q_{m}^{2})^{2} - 2a^{2}(Mr_{\text{LR}}-Q_{e}^{2}-Q_{m}^{2}) }{ \sqrt{\Delta_{\text{KN}}\big|_{r=r_{\text{LR}}}} } \nonumber
	\\
	& & + \text{Sign}(\Omega) \cdot 2a(Mr-Q_{e}^{2}-Q_{m}^{2}) \bigg] \cdot \frac{r}{(r^{2}-2Mr_{\text{LR}}+Q_{e}^{2}+Q_{m}^{2})^{2}} 
	= 0 \nonumber
	\\
	& \Rightarrow & 
	r_{\text{LR}}^{4} - 6Mr_{\text{LR}}^{3} 
	+ ( 9M^{2} + 4Q_{e}^{2} + 4Q_{m}^{2} ) r_{\text{LR}}^{2} 
	- ( 12MQ_{e}^{2} + 12MQ_{m}^{2} + 4Ma^{2} ) r_{\text{LR}} \nonumber
	\\
	&             &
	+ 4(Q_{e}^{2}+Q_{m}^{2})^{2} + 4a^{2}(Q_{e}^{2}+Q_{m}^{2}) = 0 
	%\\
	%& \Rightarrow & r_{LR} = ?
\end{eqnarray}
The light rings calculated using our geometric approach are identical to those obtained through the effective potential of photons. Actually, the last line is exactly what we can obtain from the vanishing of the Carter constant in the equatorial plane \cite{ChenYX2022}
\begin{eqnarray}
	\eta & = & \frac{\mathcal{Q}}{E^{2}}
	= - \frac{r_{\text{LR}}^{2}}{a^{2}(r_{\text{LR}}-M)^{2}} \nonumber
	\times 
	\bigg[
	r_{\text{LR}}^{4} - 6Mr_{\text{LR}}^{3} 
	+ 9M^{2}r_{\text{LR}}^{2} 
	- 4Ma^{2} r_{\text{LR}} 
	\\
	&   &
	\ \ \ \ \ \ \ \
	  + 4 ( Q_{e}^{2} + 4Q_{m}^{2} ) r_{\text{LR}}^{2}
	  - 12M ( Q_{e}^{2} + Q_{m}^{2} ) r_{\text{LR}}
	  + 4(Q_{e}^{2}+Q_{m}^{2})^{2} + 4a^{2}(Q_{e}^{2}+Q_{m}^{2})
	\bigg] \nonumber
	\\
	&   & \ \ \ \ \ 
	= 0 
\end{eqnarray}
The analytical expression of light ring radii in the Kerr-Newman spacetime is much more complicated than that in Kerr spacetime, which is not presented here. The detailed analytical results on light ring radii in the Kerr-Newman spacetime can be consulted in references \cite{ChenYX2022,Hsiao2019}.

To summarize, the two illustrative examples (Kerr spacetime and Kerr-Newman spacetime) presented in this section demonstrate the efficacy of our geometric approach in such axially symmetric spacetimes. The results on the radii of light rings obtained from our approach are in complete agreement with those calculated from the conventional effective potential approach. In principle, our approach can be universally applied to any stationary and axially symmetric spacetime (for the study of light rings in the equatorial plane), regardless of the specified metric form. Due to space limitations, other examples are not given in this section.
\end{widetext}

\section{Conclusion and Perspectives \label{section6}}

Circular photon orbits are of paramount significance in astrophysical observations of black holes and other ultra-compact objects, and they also bring valuable insights into understanding the topological properties of spacetimes, gravitational waves, and quasi-normal modes. 

In this work, we present a novel geometric approach to light rings in axially symmetric spacetimes, studying light rings and their stability through intrinsic curvatures in optical geometry. This work provides a natural extension of our geometric approach to circular photon orbits (proposed in \href{https://doi.org/10.1103/PhysRevD.106.L021501}{Phys. Rev. D \textbf{106}, L021501 (2022)}) from spherically symmetric spacetimes to axially symmetric spacetimes. 

In axially symmetric spacetimes, the construction of optical geometry gives rise to a Randers-Finsler geometry $dt=\sqrt{\alpha_{ij}dx^{i}dx^{j}}+\beta_{i}dx^{i}$. Our analysis shows that the location of light rings can be determined by vanishing the geodesic curvature in the equatorial plane of Randers-Finsler optical geometry. Moreover, the stability of light rings can be distinguished through the existence of conjugate points, which are dominantly controlled by the sign of intrinsic flag curvature in Randers-Finsler optical geometry. Specifically, the positive flag curvature indicates a stable light ring, while the negative flag curvature implies an unstable light ring. To validate the effectiveness of our geometric approach, we have chosen two representative examples: Kerr spacetime and Kerr-Newman spacetime. In both instances, our method completely reproduces the correct light ring positions. Generally speaking, our approach can be applied to an arbitrary stationary spacetime, without any restrictions on the specific forms of the spacetime metrics. Furthermore, we provide a rigorous demonstration of the equivalence between our geometric approach and the conventional approach based on the effective potential of photons, showing the broad applicability and robustness of our geometric approach. This equivalence relationship, as summarized in Table \ref{table1}, not only provides novel perspectives on light rings but also builds bridges between different theoretical methods.

% 下面谈谈扩展 
The present work can inspire several research topics. For instance, our geometric approach can be extended to study the circular orbits of massive particles in axially symmetric spacetimes, particularly the innermost stable circular orbits (ISCO). To achieve this point, the mathematical construction of Jacobi geometry (rather than the optical geometry) is necessary \cite{Chanda2017,Chanda2019,Chanda2024,Andino2021,LiZH2021,LiZH2025}. Furthermore, it is highly significant to conduct a comprehensive analysis of the general properties of light rings through our geometric approach. This exploration could examine whether a geometric analysis from optical geometry can rederive some well-established theorems on circular photon orbits for axially symmetric spacetimes reported in literature \cite{Cunha2017,Cunha2018,Cunha2020,WeiSW2020,GaoSJ2021}, especially those regarding the existence and number of stable and unstable light rings in various classes of rotational spacetimes.

Our present work, combined with a series of investigations by the authors and other researchers in recent years, reveals a nontrivial relationship between circular particle orbits and the intrinsic curvatures of the low-dimensional geometry constructed from Lorentz spacetime (e.g., optical geometry and Jacobi geometry). The explorations in this direction are likely to uncover deeper and intricate connections between particle orbits in gravitational fields and the geometric properties of these low-dimensional constructed geometries, showing the significant potential of such geometrization methods in gravitational theory and mathematical physics.


\appendix

\section*{Appendices}

The appendices present the mathematical preliminaries required for our geometric analysis of circular photon orbits in axially symmetric spacetimes. Appendix \ref{appendix1} offers a brief introduction to the geodesic equation in Randers-Finsler geometry. Appendix \ref{appendix2} discusses the mathematical definition of geodesic curvature for continuous curves in Randers-Finsler optical geometry. The decomposition of geodesic curvature in Randers-Finsler geometry $\kappa_{g}^{(F)} = \kappa_{g}^{(\alpha)} + \kappa_{\beta}^{(\alpha)}$, which consists of the geodesic curvature with Riemannian part $\alpha$ and the additional contribution from the non-Riemannian part $\beta$, is presented in this section. Appendix \ref{appendix3} introduces one of the most important intrinsic curvatures in Randers-Finsler optical geometry --- the flag curvature.

\section{The Geodesics in Randers-Finsler Geometry \label{appendix1}}

The optical geometry of a stationary and axially symmetric gravitational spacetime is described by a Randers-Finsler geometry. In this framework, null geodesics in the spacetime geometry correspond to spatial geodesics in optical geometry. This appendix presents a brief introduction to geodesics in Randers-Finsler geometry and a concise overview of their properties.

Mathematically, the arc-length of a continuous curve in Finsler geometry is defined as
\begin{equation}
	s_{AB} \equiv \int_{\lambda_{A}}^{\lambda_{B}} F(x,T) \cdot d\lambda 
	= \int_{\lambda_{A}}^{\lambda_{B}} ||T||_{(x,T)}^{(F)} \cdot d\lambda  .
	\label{Finsler length}
\end{equation} 
In this expression, the $F(x,y)$ is a Finsler function, which is generally defined on the tangent bundle $(x,y) \in TM$. Analogous to the metric tensor $g_{ij}(x)$ in Riemannian geometry, the Finsler function $F(x,y)$ serves as a fundamental quantity in Finsler geometry. It determines the spatial distance between two distinct points, the arc length of continuous curves, the volume of local regions, and a number of intrinsic curvatures in Finsler geometry. Specifically, to define the arc-length of continuous curves in expression (\ref{Finsler length}), we choose the vector $y$ to be the tangent vector $y=T=\frac{dx}{d\lambda}$ along this continuous curve. Under these circumstances, the Finsler function equals the norm of tangent vector $T$ along its own direction, via $F(x,T) = ||T||_{(x,T)}^{(F)} = \sqrt{<T,T>_{(x,T)}^{(F)}}$.

Particularly, the Randers geometry is a specific subclass of Finsler geometry, in which the Finsler function can be decomposed into two parts
\begin{equation}
	F(x,y) = \sqrt{\alpha_{ij}(x) \cdot y^{i}y^{j}} + \beta_{i}(x) \cdot y^{i} .
	\label{Randers-Finsler geometry}
\end{equation}
Here, $\alpha_{ij}(x)$ represents a Riemannian metric, and the coefficient $\beta_{i}$ can compose a 1-form $\beta=\beta_{i}(x) dx^{i}$ that quantifies the deviation of this Finsler geometry from the Riemannian geometry $ds^{2} = \alpha_{ij}(x)dx^{i}dx^{j}$. Notably, when the one-form vanishes (i.e., $\beta=0$), the Randers-Finsler geometry defined in (\ref{Finsler length}) and (\ref{Randers-Finsler geometry}) reduces to the Riemannian geometry.

Similar to the Riemannian geometry, the spatial geodesics in Finsler geometry make the arc-length of continuous curves be extremum \cite{ChernSS,ShenYB}
\begin{equation}
	\delta s_{AB} = \delta \bigg[ \int_{A}^{B} F(x,T) \cdot d\lambda \bigg] = 0 .
\end{equation} 
This variational condition eventually leads to the following geodesic equation in Finsler geometry \cite{ChernSS}
\begin{equation}
	D_{T}^{(F)} \tilde{T} \bigg|_{y=T}
	= D_{T}^{(F)} \bigg[ \frac{T}{F(x,T)} \bigg] \bigg|_{y=T}
	= 0 .
	\label{Finsler geodesic equation}
\end{equation}
Therefore, a geodesic curve makes the local extremum of arc-length is simultaneously an auto-parallel curve along its tangent direction in Finsler geometry. In expression (\ref{Finsler geodesic equation}), the notation $D_{T}^{(F)}$ represents the Finslerian covariant derivative operator along the tangent vector $T=\frac{dx}{d\lambda}$ of a continuous curve $\gamma=\gamma(\lambda)$. We use the simplified notation $\tilde{T}$ to label the unit tangent vector along its own direction $\tilde{T} = T/\sqrt{<T,T>_{(x,T)}^{(F)}}=T/F(x,T)$. Notably, if we choose the arc-length parameter as the affine parameter (i.e., $\lambda=s$) , the tangent vector along any continuous curve becomes a unit tangent vector (namely $T = \tilde{T}$). In such circumstances, the Finslerian geodesic equation in (\ref{Finsler geodesic equation}) becomes
\begin{equation}
	D^{(F)}_{T} T \bigg|_{y=T} = 0
	\ \Leftrightarrow \
	\bigg[
	    \frac{d^{2}x^{i}}{ds^{2}} 
	  + \Gamma^{i}_{jk}(x,y) \cdot \frac{dx^{j}}{ds} \frac{dx^{k}}{ds}
	\bigg]_{y=T}
	= 0  .
	\label{Finsler geodesic equation2}
\end{equation} 
Here, $\Gamma^{i}_{jk}(x,y)$ is the coefficient of the Chern-Rund connection in Finsler geometry, whose expression is determined by the Finsler function $F(x,y)$. The explicit form of these coefficients is given in Appendix \ref{appendix3}. Furthermore, one of the distinctions of Finsler geometry from Riemannian geometry lies in the connection coefficients. In Finsler geometry, the coefficient of the Chern-Rund connection depends not only on base point $x \in M$, but also on tangent vector $y \in T_{x}M$. This is different from the scenario in Riemannian geometry, where the Levi-Civita connection depends only on the base point $x$. Therefore, when we use the covariant derivative to formulate the geodesic equation in Finsler geometry, it is necessary to specify the reference vector $y=T$ in the subscript, which fulfills the concept of the auto-parallel curve. In the general definition of covariant derivative in Finsler geometry, the reference vector may differ from the tangent vector along the geodesic curve
\begin{equation}
	D_{T}^{(F)} V \bigg|_{y} 
	\equiv
	\bigg[ 
	    \frac{dV^{i}}{d\lambda} 
	  + \Gamma^{i}_{jk}(x,y) \cdot V^{j} \cdot \frac{dx^{k}}{ds}
	\bigg]
	\cdot \frac{\partial}{\partial x^{i}} .
\end{equation}

Although the Finslerian geodesic equation presented in expression (\ref{Finsler geodesic equation2}) has an elegant form and can be formally analogous to the geodesic equation in Riemannian geometry (by replacing the Chern-Rund connection in Finsler geometry with the Levi-Civita connection), its direct calculation in Randers-Finsler geometry is computationally challenging, especially when we directly calculate the Chern-Rund connection using $\alpha_{ij}$ and $\beta_{i}$ described in (\ref{Randers-Finsler geometry}). In practical calculations, we prefer to find a more convenient form of the geodesic equation that utilizes $\alpha_{ij}$ and $\beta_{i}$. Fortunately, it turns out that the Finslerian geodesic equation (\ref{Finsler geodesic equation}) can be transformed into a relatively simple form in the Randers-Finsler geometry when a ``constant Riemannian speed'' affine parameter $l$ is adopted \cite{ChernSS}
\begin{equation}
	\frac{dx^{i}}{dl^{2}} + (\gamma^{\alpha})^{i}_{jk} \cdot \frac{dx^{j}}{dl} \frac{dx^{k}}{dl}
	+ \alpha^{ij} \bigg[ D_{k}^{(\alpha)} \beta_{j} - D_{j}^{(\alpha)} \beta_{k} \bigg] \cdot \frac{dx^{k}}{dl}
	= 0
	\label{geodesic equation in Randers-Finsler geometry}
\end{equation}
In this expression, $D_{k}^{(\alpha)} \beta_{j} = \frac{\partial \beta_{j}}{\partial x^{k}} - (\gamma^{(\alpha)})^{i}_{jk} \cdot \beta_{i}$ labels the covariant derivative of 1-form $\beta=\beta_{i}dx^{i}$ with respect to the Riemannian metric $\alpha_{ij}$, and their Levi-Civita connection is denoted as $(\gamma^{(\alpha)})^{i}_{jk}$ 
\footnote{One should not be confused with $(\gamma^{(\alpha)})^{i}_{jk}$ and $\Gamma^{i}_{jk}(x,y)$. The former is the coefficient of Levi-Civita connection in Riemannian geometry $ds^{2}=\alpha_{ij}dx^{i}dx^{j}$. The latter is the connection coefficient in the Randers-Finsler geometry $ds = \sqrt{\alpha_{ij}dx^{i}dx^{j}}+\beta_{i}dx^{i}$, which contains the contributions from 1-form $\beta$.}.
The affine parameter $l$ is chosen such that 
\begin{equation}
	<T \cdot T>^{(\alpha)} = 1 
	\ \ \Leftrightarrow \ \  
	\alpha_{ij} \cdot \frac{dx^{i}}{dl}\frac{dx^{j}}{dl} = 1
\end{equation}
which is called the affine parameter corresponding to ``constant Riemannian speed'' with respect to metric $\alpha_{ij}$ in references. Furthermore, a simple calculation suggests that 
\begin{equation}
	D_{k}^{(\alpha)}\beta_{j} - D_{j}^{(\alpha)}\beta_{k} 
	= \frac{\partial \beta_{j}}{\partial x^{k}} - \frac{\partial \beta_{k}}{\partial x^{j}}
\end{equation}
This geodesic equation in (\ref{geodesic equation in Randers-Finsler geometry}) eventually gives rise to the ``covariant acceleration'' of the tangent vector $T$ associated with the Riemannian metric part $\alpha$
\begin{eqnarray}
	a_{(\alpha)}^{i} 
	& \equiv & 
	\frac{dx^{i}}{dl^{2}} + (\gamma^{(\alpha)})^{i}_{jk} \cdot \frac{dx^{j}}{dl} \frac{dx^{k}}{dl} \nonumber
	\\
	& = &
	- \alpha^{ij}
	\bigg[ D_{k}^{(\alpha)} \beta_{k} - D_{j}^{(\alpha)} \beta_{j} \bigg] \cdot \frac{dx^{k}}{dl} \nonumber
	\\
	& = & 
	- \alpha^{ij} \bigg( \frac{\partial \beta_{j}}{\partial x^{k}} - \frac{\partial \beta_{k}}{\partial x^{j}} \bigg) \cdot \frac{dx^{k}}{dl}
	\label{covariant acceleration}
\end{eqnarray}
This is precisely the result that has been reported in references \cite{Asida2017,ChernSS}

\section{Geodesic Curvature in Randers-Finsler Optical Geometry \label{appendix2}}

Conventionally, geodesic curvature is the intrinsic curvature of a continuous curve in 2-dimensional Riemannian geometry, which quantifies the deviation of this continuous curve from being geodesic. However, the concept of geodesic curvature can be generalized into Finsler geometry. This appendix presents the formal definition of geodesic curvature for continuous curves in Randers-Finsler optical geometry. 

\begin{figure}
	\includegraphics[width=0.525\textwidth]{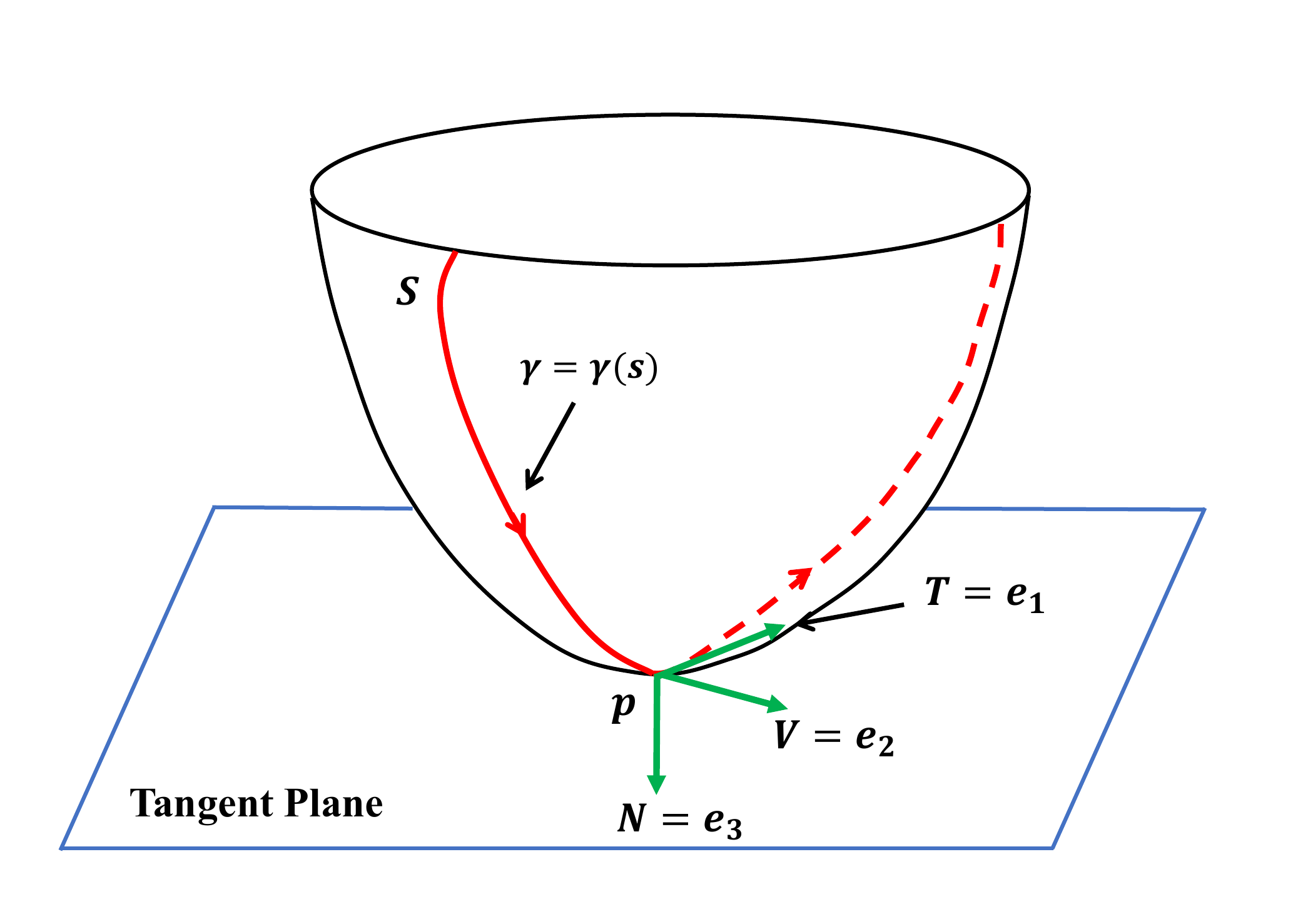}
	\caption{Illustration of the embedding of a curved surface $S$ (which is a 2-dimensional Riemannian geometry) into a 3-dimensional background Euclidean space. The $\gamma=\gamma(s)$ is a continuous curve residing on this surface $S$, and $p$ is an arbitrary point on this curve. The $\{ \boldsymbol{e_{1}},\boldsymbol{e_{2}},\boldsymbol{e_{3}}\}$ compose an orthonormal frame field in 3-dimensional Euclidean space such that $\boldsymbol{e_{1}}=\boldsymbol{T}$ is the unit tangent vector of this curve, $\boldsymbol{e_{2}}=\boldsymbol{V}$ is a unit vector in the tangent space of surface $S$ that is orthogonal to $\boldsymbol{T}$, and $\boldsymbol{e_{3}}=\boldsymbol{N}$ is the unit normal vector of surface $S$ at point $p$. The tangent plane of the curved surface $S$ at point $p$ is spanned by frame vectors $\boldsymbol{e_{1}}$ and $\boldsymbol{e_{2}}$.}
	\label{figure Embedding}
\end{figure}

In the context of 2-dimensional Riemannian geometry, the geodesic curvature of a curve admits a variety of equivalent definitions. Here, we present an instructive way to give the geodesic curvature, which utilizes the embedding of this 2-dimensional Riemannian geometry into a higher-dimensional background space. Consider a curved surface $S$ embedded into a 3-dimensional background Euclidean space, which is illustrated in Figure \ref{figure Embedding}. At each point $p$ along the curve $\gamma=\gamma(s)$, we can construct an orthonormal frame $\{ \boldsymbol{e_{1}}, \boldsymbol{e_{2}}, \boldsymbol{e_{3}} \}$ to the point $p$. Specifically, $\boldsymbol{e_{1}}=\boldsymbol{T}$ is the unit tangent vector of this curve $\gamma$, $\boldsymbol{e_{2}}=\boldsymbol{V}$ is a unit vector in the tangent space of surface $S$ that is orthogonal to $\boldsymbol{T}$, and $\boldsymbol{e_{3}}=\boldsymbol{N}$ represents the unit normal vector of surface $S$ at this point $p$. The tangent space of the curved surface $S$ at point $p$ is spanned by frame vectors $\boldsymbol{e_{1}}=\boldsymbol{T}$ and $\boldsymbol{e_{2}}=\boldsymbol{V}$. The geodesic curvature $\kappa_{g}$ of this continuous curve $\gamma$ is defined as follows \cite{ChenWH} 
\begin{equation}
	\kappa_{g} \equiv \frac{d\boldsymbol{T}}{ds} \cdot \boldsymbol{V} 
	= \frac{d\boldsymbol{e_{1}}}{ds} \cdot \boldsymbol{e_{2}}  .
	\label{geodesic curvature surface theory} 
\end{equation}
Furthermore, the geodesic curvature defined in this manner is an intrinsic geometric quantity, independent of the embedding of this surface $S$ into a higher-dimensional background Euclidean space. This intrinsic nature implies that $\kappa_{g}$ can be defined and computed solely using the intrinsic geometry of the surface $S$. Particularly, in the framework of the intrinsic Riemannian geometry of surface $S$, the mathematical definition of the geodesic curvature becomes
\begin{equation}
	\kappa_{g} 
	= \bigg< \frac{D\boldsymbol{T}}{ds}, \boldsymbol{V} \bigg>
	= \bigg< \frac{d\boldsymbol{e_{1}}}{ds} \cdot \boldsymbol{e_{2}} \bigg> ,
	\label{geodesic curvature Riemannian geometry} 
\end{equation}
with the inner product in background Euclidean space replaced by the inner product induced by the Riemannian metric, and the ordinary derivative is substituted with the covariant derivative along this curve.

The ideas and treatments for defining a geodesic curvature illustrated above can be extended to the Finsler geometry. For a continuous curve $\gamma$ in 2-dimensional Finsler geometry, its geodesic curvature can be defined in a similar manner. Specifically, the geodesic curvature for a curve is given by the inner product of the covariant derivative of the unit tangent vector $\tilde{T}=T/\sqrt{<T,T>_{(x,T)}^{(F)}}=T/F(x,T)$ and a unit vector $\tilde{V}$ that is orthogonal to $\tilde{T}$, via \cite{Shimada2010,Shimada2007}
\footnote{In reference \cite{Shimada2007}, the authors prefer to use an alternative definition of geodesic curvature $\kappa_{g}^{(F)} = \bigg|\bigg| D_{T}^{(F)}  \tilde{T} \big|_{y=T} \bigg|\bigg|_{(x,T)}^{(F)} = \sqrt{ \big< D_{T}^{(F)}  \tilde{T} \big|_{y=T} , D_{T}^{(F)}  \tilde{T} \big|_{y=T} \big>^{(F)}_{(x,T)} }$, and the quantity defined in expression (\ref{geodesic curvature Finsler}) is named the signed curvature. In this work, we adopt the quantity in (\ref{geodesic curvature Finsler}) as the definition of geodesic curvature, since it is consistent with the geodesic curvature in expressions (\ref{geodesic curvature surface theory}), (\ref{geodesic curvature Riemannian geometry}), and (\ref{geodesic curvature alpha}). Regardless of the nomenclature, the geometric quantity defined in expression (\ref{geodesic curvature Finsler}) vanishes for a geodesic curve in 2-dimensional Finsler geometry.}
\begin{eqnarray}
	\kappa_{g}^{(F)} 
	& \equiv & 
	\big< D_{T}^{(F)} \tilde{T} \big|_{y=T}, \tilde{V} \big>^{(F)}_{(x,T)} \nonumber
	\\
	& = & 
	\bigg< D_{T}^{(F)}  \bigg[ \frac{T}{F(x,T)} \bigg] \bigg|_{y=T} , \tilde{V} \bigg>^{(F)}_{(x,T)} ,
	\label{geodesic curvature Finsler}
\end{eqnarray}
Since $\tilde{V}$ is a unit vector in the 2-dimensional Finsler surface that is orthogonal to the tangent vector $\tilde{T}$, the relations $<\tilde{T}, \tilde{V}>_{(x,T)}^{(F)} = 0$ and $<\tilde{V}, \tilde{V}>_{(x,T)}^{(F)} = 1$ always hold along the continuous curve $\gamma$ (where the reference vector for taking inner vector products in Finsler geometry is chosen as $y=T$). 
Moreover, it can be easily noticed that for any geodesic curve in 2-dimensional Finsler geometry, its geodesic curvature defined in expression (\ref{geodesic curvature Finsler}) must vanish
\begin{equation}
	\text{geodesics}
	\ \ \Leftrightarrow \ \ 
	D_{T}^{(F)} \bigg[ \frac{T}{F(x,T)} \bigg] \bigg|_{y=T} = 0
	\ \ \Leftrightarrow \ \ 
	\kappa_{g}^{(F)} = 0 .
\end{equation}

Then we are able to apply the aforementioned Finslerian geodesic curvature $\kappa_{g}^{(F)}$ to the optical geometry of axially symmetric spacetimes, which gives a Randers-Finsler geometry in expression (\ref{Optical geometry rotational}). Since the 2-dimensional Randers-Finsler geometry can be decomposed into a Riemannian geometry part $\alpha$ and a non-Riemannian part $\beta$, it reasonable to anticipate that the Finslerian geodesic curvature $\kappa_{g}^{(F)}$ for a continuous curve is composed of two parts: the geodesic curvature with respect to the Riemannian metric $\alpha_{ij}$, and the additional contribution due to the existence of the non-Riemannian part $\beta$. Therefore, the Finslerian geodesic curvature is given by
\begin{equation}
	\kappa_{g}^{(F)} = \kappa_{g}^{(\alpha)} + \kappa_{\beta}^{(\alpha)} .
\end{equation}
Here, $\kappa_{\beta}^{(\alpha)}$ represents the additional contribution acting on the Riemannian geodesic curvature $\kappa_{g}^{(\alpha)}$ due to the existence of the non-Riemannian part $\beta$. For photon orbits restricted in the equatorial plane, which become spatial geodesics in 2-dimensional Randers-Finsler geometry, their corresponding Finslerian geodesic curvature should be zero
\begin{equation}
	\kappa_{g}^{(F)} = 0 
	\ \ \Leftrightarrow \ \ 
	\kappa_{g}^{(\alpha)} = - \kappa_{\beta}^{(\alpha)} .
\end{equation}
Here, it is clearly shown that the presence of $\beta$ leads to the emergence of nonzero ``covariant acceleration'' of photons with respect to the Riemannian metric $\alpha_{ij}$ (see Appendix \ref{appendix1} and expression (\ref{covariant acceleration})). Particularly, when the non-Riemannian one-form $\beta$ in 2-dimensional Randers-Finsler geometry vanishes (i.e. $\beta=0$), the additional contribution $\kappa_{\beta}$ becomes zero, and the Finslerian geodesic curvature $\kappa_{g}^{(F)}$ reduces to the Riemannian geodesic curvature $\kappa_{g}^{(\alpha)}$. The ``covariant acceleration'' of photon beams in equation (\ref{covariant acceleration}) vanishes in this case.

\begin{figure*}
	\includegraphics[width=0.615\textwidth]{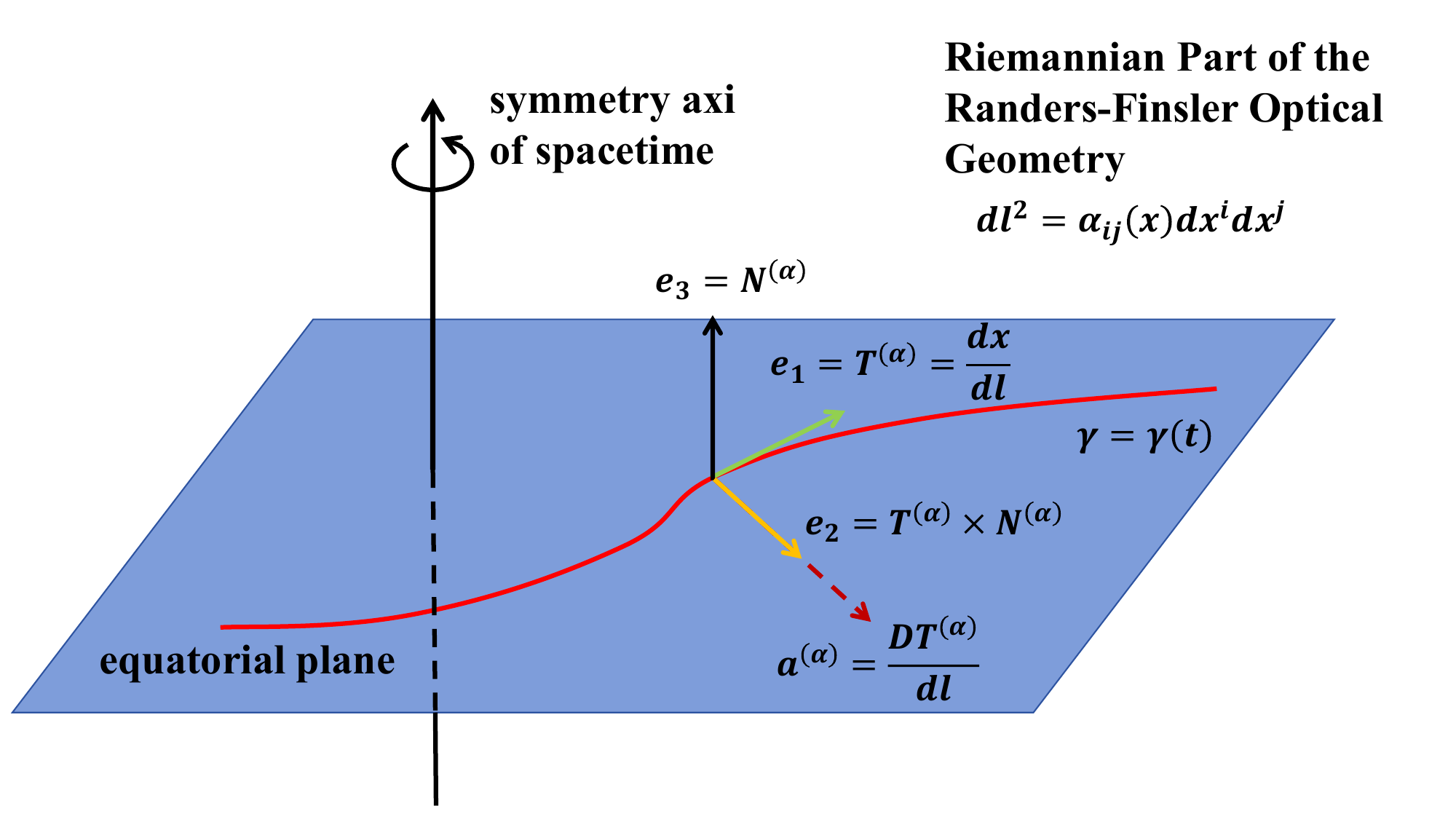}
	\caption{This figure illustrates the left-handed frame $\{ e_{1}^{(\alpha)}, e_{2}^{(\alpha)}, e_{3}^{(\alpha)} \} = \{ T^{\alpha}, T^{\alpha}\times N^{\alpha}, N^{\alpha} \}$ used in the geodesic curvature formula in (\ref{geodesic curvature alpha}). The handedness (left-handed or right-handed) of the frame $\{ e_{1}^{(\alpha)}, e_{2}^{(\alpha)}, e_{3}^{(\alpha)} \}$ or $\{ \boldsymbol{e_{1}}, \boldsymbol{e_{2}}, \boldsymbol{e_{3}} \}$ depends on the choice of normal vector orientation. Here we follow the convention given by Ono \emph{et. al} in reference \cite{Asida2017}, where the unit normal vector $N^{\alpha}$ is assumed in the upward direction. This is different from the right-handed frame presented in Figure \ref{figure Embedding}, where the normal vector $\boldsymbol{e_{2}} = \boldsymbol{N}$ is oriented downward. However, the vanishing of Finslerian geodesic curvature $\kappa_{g}^{(\alpha)} + \kappa_{\beta} = 0$ is independent of the choice of left-handed and right-handed frames. Furthermore, for any photon orbit parametrized by arc-length parameter $l$ in the Riemannian geometry $dl^{2}=\alpha_{ij}dx^{i}dx^{j}$, the tangent vector $T^{(\alpha)}=\frac{dx}{dl}$ always has a constant unit norm. In this case, the``covariant acceleration'' vector $a^{(\alpha)} = \frac{DT^{(\alpha)}}{dl}$ of photons is automatically orthogonal to the vector $T^{(\alpha)}$ in the tangent space.}
	\label{figure A2}
\end{figure*}

To determine the additional contribution $\kappa_{\beta}^{(\alpha)}$ arising from the non-Riemannian part, it is convenient to resort to the ``covariant acceleration'' of photons with respect to the Riemannian metric $\alpha_{ij}$, which is given in (\ref{covariant acceleration}). The existence of nonzero ``covariant acceleration'' vector $a^{(\alpha)}$ of photons with respect to Riemannian metric $\alpha_{ij}$ is analogous to the nonzero covariant derivative of tangent vector $\frac{D \boldsymbol{T}}{ds}$ encountered in the Riemannian geodesic curvature expression in (\ref{geodesic curvature Riemannian geometry}). Following the standard definition of geodesic curvature in Riemannian geometry, the geodesic curvature $\kappa_{g}^{(\alpha)}$ of a continuous photon orbit can be written as
\begin{equation}
	\kappa_{g}^{(\alpha)} 
	= \big< a^{(\alpha)}, ( T^{(\alpha)} \times N^{(\alpha)} ) \big>^{(\alpha)} 
	= - \kappa_{\beta}^{(\alpha)}.
	\label{geodesic curvature alpha}
\end{equation}
Here, $T^{(\alpha)} = \frac{dx}{dl}$ denotes the unit tangent vector to the photon orbit (parametrized by arc-length parameter $l$ in the Riemannian geometry part $dl^{2}=\alpha_{ij}dx^{i}dx^{j}$). Additionally, $N^{(\alpha)}$ represents the unit normal vector of the equatorial plane in 3-dimensional Riemannian geometry $\alpha$. In this formulation, the unit tangent vector $T^{(\alpha)}$ is analogous to the frame vector $\boldsymbol{e_{1}}=\boldsymbol{T}$ in expression (\ref{geodesic curvature Riemannian geometry}), while the unit normal vector $N^{(\alpha)}$ for the equatorial plane is analogous to the frame vector $\boldsymbol{e_{3}}=\boldsymbol{N}$. Moreover, once the frame vectors $\boldsymbol{e_{1}}$ and $\boldsymbol{e_{3}}$ are specified, another frame vector $\boldsymbol{e_{2}}$ can be constructed through the vector product of the unit tangent vector and unit normal vector, via $\boldsymbol{e_{2}} = \boldsymbol{e_{1}} \times \boldsymbol{e_{3}} = \boldsymbol{T} \times \boldsymbol{N}$. This choice of frame vectors ensures that $\{ e_{1}^{(\alpha)}, e_{2}^{(\alpha)}, e_{3}^{(\alpha)} \} = \{ T^{\alpha}, T^{\alpha}\times N^{\alpha}, N^{\alpha} \}$ forms a left-handed frame. 

In the 3-dimensional Riemannian geometry, the vector product can be formulated in terms of the  anti-symmetric Levi-Civita tensor
\begin{subequations}
\begin{eqnarray}
	(A \times B)_{i} & = & \epsilon_{ijk}^{(\alpha)} A^{j} B^{k} 
	= \sqrt{\alpha^{(3d)}} \varepsilon_{ijk} A^{j} B^{k} , 
	\\
	(A \times B)^{i} & = & \epsilon_{(\alpha)}^{ijk} A_{j} B_{k} 
	= \frac{\varepsilon^{ijk}}{\sqrt{\alpha^{(3d)}}} A_{j} B_{k} ,
\end{eqnarray}
\end{subequations}
In these expressions, $\varepsilon_{ijk}$ and $\varepsilon^{ijk}$ are called the anti-symmetric Levi-Civita symbol (which takes $+1$ for even permutation of indices and $-1$ for odd permutation of indices), $\epsilon_{ijk}^{(\alpha)}=\sqrt{\alpha^{(3d)}} \varepsilon_{ijk}$ and $\epsilon_{(\alpha)}^{ijk}=\frac{\varepsilon^{ijk}}{\sqrt{\alpha^{(3d)}}}$ are the anti-symmetric Levi-Civita tensor in Riemannian geometry $dl^{2}=\alpha_{ij}dx^{i}dx^{j}$, and the $\alpha^{(3d)}=\text{det}(\alpha^{(3d)}_{ij})=\alpha_{rr}\alpha_{\theta\theta}\alpha_{\phi\phi}$ represents the determinant of the Riemannian metric $\alpha_{ij}$. In this way, the geodesic curvature for this photon orbit with respect to the Riemannian metric $\alpha$ can be expressed using the Levi-Civita tensor
\begin{eqnarray}
	\kappa_{g}^{(\alpha)} = -\kappa_{\beta}^{(\alpha)}
	& = & \big< a^{(\alpha)}, ( T^{(\alpha)} \times N^{(\alpha)} ) \big>^{\alpha} \nonumber
	\\
	& = & \epsilon_{(\alpha)}^{ijk} \cdot a_{i}^{(\alpha)} T^{(\alpha)}_{j} N^{(\alpha)}_{k}
	\label{geodesic curvature alpha2}
\end{eqnarray}
Substituting the ``covariant acceleration'' of photons from expression (\ref{covariant acceleration}) into the geodesic curvature formula, the geodesic curvature $\kappa_{g}^{(\alpha)}$ can be calculated as
\begin{eqnarray}
	\kappa_{g}^{(\alpha)} 
	& = & - \epsilon_{(\alpha)}^{ijk} \cdot T^{(\alpha)}_{j} N^{(\alpha)}_{k}
	\cdot \bigg( \frac{\partial \beta_{i}}{\partial x^{m}} - \frac{\partial \beta_{m}}{\partial x^{i}} \bigg) \cdot (T^{(\alpha)})^{m}  \nonumber
	\\
	& = &
    - \epsilon_{(\alpha)}^{ijk} \cdot T^{(\alpha)}_{j} N^{(\alpha)}_{k} 
    \cdot \omega_{im} (T^{(\alpha)})^{m} \nonumber
	\\
	& = &
    - \epsilon_{(\alpha)}^{ijk} \cdot T^{(\alpha)}_{j} N^{(\alpha)}_{k}
	\cdot (**\omega)_{im} (T^{(\alpha)})^{m} \nonumber
	\\
	& = &
	- \frac{1}{2} \epsilon_{(\alpha)}^{ijk} \cdot T^{(\alpha)}_{j} N^{(\alpha)}_{k} \cdot
	\epsilon^{(\alpha)}_{imn} \epsilon_{(\alpha)}^{abn} \cdot \omega_{ab} (T^{(\alpha)})^{m}  \nonumber
	\\
	& = &
	- \big( \delta^{j}_{m}\delta^{k}_{n} - \delta^{k}_{m}\delta^{j}_{n} \big) \cdot T^{(\alpha)}_{j} N^{(\alpha)}_{k} \epsilon_{(\alpha)}^{abn} \cdot \omega_{ab} (T^{(\alpha)})^{m} \nonumber
	\\ 
	& = &
	- \epsilon_{(\alpha)}^{abk} \cdot \omega_{ab} \times \nonumber
	\\
	&   &  
	\bigg[ 
	    N^{(\alpha)}_{k} \cdot \big< T^{(\alpha)}, T^{(\alpha)} \big>^{(\alpha)}
	  - T^{(\alpha)}_{k} \cdot \big< N^{(\alpha)}, T^{(\alpha)} \big>^{(\alpha)} 
	\bigg] \nonumber
	\\
	& = &
	- \epsilon_{(\alpha)}^{abk} \cdot \omega_{ab} N^{(\alpha)}_{k} \nonumber
	\\
	& = &
	- \frac{1}{2}  
	\frac{\varepsilon^{abk}}{\sqrt{\alpha^{(3d)}}} 
	\bigg( \frac{\partial \beta_{a}}{\partial x^{b}} - \frac{\partial \beta_{b}}{\partial x^{a}} \bigg) N^{(\alpha)}_{k}  .
	\label{geodesic curvature calculation}
\end{eqnarray}
In the calculation, we employ the following anti-symmetric two-form $\omega$ in the Riemannian geometry $dl^{2}=\alpha_{ij}dx^{i}dx^{j}$ and its Hodge dual form
\begin{subequations}
\begin{eqnarray}
	\omega & = & d\beta 
	= \frac{1}{2} \bigg( \frac{\partial \beta_{i}}{\partial x^{j}} - \frac{\partial \beta_{j}}{\partial x^{i}} \bigg) \cdot dx^{i} \wedge dx^{j} 
	= \frac{1}{2} \omega_{ij} \cdot dx^{i} \wedge dx^{j} \nonumber
	\\
	\\
	*\omega & = & *\omega_{k} \cdot dx^{k} = (\varepsilon^{(\alpha)})^{ij}_{\ \ k} \cdot \omega_{ij} \cdot dx^{k}
\end{eqnarray}
\end{subequations}
Furthermore, in the above derivation process, we also utilize the contraction rule of the Levi-Civita tensor in 3-dimensional Riemannian geometry \cite{Carroll}
\begin{equation}
	\epsilon_{(\alpha)}^{ijk} \cdot \epsilon^{(\alpha)}_{imn}
	= 2! \cdot \big( \delta^{j}_{m}\delta^{k}_{n} - \delta^{k}_{m}\delta^{j}_{n} \big) .
\end{equation}
and the important relation on the Hodge star operator
\begin{equation}
	**\omega = (-1)^{2} \cdot \omega 
\end{equation}

\begin{widetext}
For the axially symmetric spacetime, its optical geometry takes the form
\begin{eqnarray}
	dt = \sqrt{\alpha_{ij}dx^{i}dx^{j}} + \beta_{i}dx^{i}  
	= \sqrt{-\frac{g_{rr}}{g_{tt}} \cdot dr^{2} - \frac{g_{\theta\theta}}{g_{tt}} \cdot d\theta^{2} + \frac{g_{t\phi}^{2}-g_{tt}g_{\phi\phi}}{g_{tt}^{2}} \cdot d\phi^{2}} 
	- \frac{g_{t\phi}}{g_{tt}}\cdot d\phi .
	\label{optical geometry rotational3}
\end{eqnarray}
There is only one nonzero derivative for the one-form $\beta$, which is $\frac{\partial \beta_{\phi}}{\partial r}$. Additionally, the unit normal vector $N^{(\alpha)}$ for the equatorial plane in 3-dimensional Riemannian geometry also has one nonzero component 
\begin{eqnarray}
	<N^{(\alpha)} \cdot N^{(\alpha)}>^{(\alpha)} 
	= \alpha^{\theta\theta} \cdot N^{(\alpha)}_{\theta} N^{(\alpha)}_{\theta} = 1 
	& \Rightarrow &
	N^{(\alpha)}_{\theta}=\frac{1}{\sqrt{\alpha^{\theta\theta}}} .
	\label{normal vector N}
\end{eqnarray}
\end{widetext}
Substituting the normal vector component in expression (\ref{normal vector N}) into the geodesic curvature $\kappa_{g}^{(\alpha)}$ in expression (\ref{geodesic curvature calculation}), the additional contribution $\kappa_{\beta}$ arising from $\beta$ can be obtained
\begin{eqnarray}
	\kappa_{g}^{(\alpha)} = - \kappa_{\beta}^{(\alpha)}
	& = & 
	- \frac{1}{2} \bigg( \frac{\partial \beta_{a}}{\partial x^{b}} - \frac{\partial \beta_{b}}{\partial x^{a}} \bigg) 
	\cdot \frac{\varepsilon^{abk}}{\sqrt{\alpha^{(3d)}}} N^{(\alpha)}_{k} \nonumber
	\\
	& = & - \frac{\partial \beta_{\phi}}{\partial r} \frac{\varepsilon^{\phi r\theta}}{\sqrt{\alpha_{rr}\alpha_{\theta\theta}\alpha_{\phi\phi}}}
	\frac{1}{\sqrt{\alpha^{\theta\theta}}} \nonumber
	\\
	& = & - \frac{1}{\sqrt{\alpha_{rr}\alpha_{\phi\phi}}} 
	\frac{\partial \beta_{\phi}}{\partial r} 
	\label{geodesic curvature contribution from beta}
\end{eqnarray}
This expression is completely identical to the one proposed by Ono \emph{et. al} in reference \cite{Asida2017}. Ono \emph{et al.} interpreted the nonzero geodesic curvature of light orbits with respect to Riemannian geometry $\alpha_{ij}$ as ``gravitomagnetism'' effects in stationary spacetimes. In our work, the same contribution can be naturally interpreted as the geodesic curvature contribution induced by the existence of the  non-Riemannian one-form $\beta$ in Randers geometry.

There is a subtlety that needs clarification for the geodesic curvature contribution $\kappa_{\beta}$ for prograde and retrograde photon orbits. The above derivation of expression (\ref{geodesic curvature contribution from beta}) is valid for the prograde motion (co-rotating motion) of lights. If the retrograde motion (counter-rotating motion) of light is considered, the tangent vector $T^{\text{OP}}$ changes to the opposite direction. We flip the normal vector into the opposite direction to make the frame $\{ e_{1}^{(\alpha)}, e_{2}^{(\alpha)}, e_{3}^{(\alpha)} \} = \{ T^{\alpha}, T^{\alpha}\times N^{\alpha}, N^{\alpha} \}$ maintain a left-handed system. Only the left-handedness of the frame can make the geodesic curvature properly expressed through equations (\ref{geodesic curvature alpha}) and (\ref{geodesic curvature alpha2}). Additionally, the nonzero ``covariant acceleration'' vector $a^{(\alpha)}$ given in equation (\ref{covariant acceleration}), which is proportional to the tangent vector $T^{(\alpha)} = \frac{dx}{dl}$ in the Riemannian geometry $dl^{2} = \alpha_{ij} dx^{i} dx^{j}$, also contributes to a minus sign for retrograde photon orbits. Therefore, the whole calculation procedure from equations (\ref{geodesic curvature alpha2})-(\ref{geodesic curvature contribution from beta}) eventually gives rise to an additional minus in the geodesic curvature $\kappa_{g}^{(\alpha)}$ (or $\kappa_{\beta}^{(\alpha)}$).
\begin{equation}
	\kappa_{g}^{(\alpha)} = - \kappa_{\beta}^{(\alpha)} 
	= + \frac{1}{\sqrt{\alpha_{rr}\alpha_{\phi\phi}}} 
	\frac{\partial \beta_{\phi}}{\partial r} 
	\ \ \ \text{for retrograde motions} 
	\label{geodesic curvature expression retrograde orbit}
\end{equation}

For retrograde motions of lights, the additional minus sign to the geodesic curvature contributions $\kappa_{g}^{(\alpha)}$ and $\kappa_{\beta}^{(\alpha)}$ in equation (\ref{geodesic curvature expression retrograde orbit}) can be understood in an alternative way. The additional minus sign reflects the irreversibility of the geodesics in Randers-Finsler geometry. Due to the presence of the non-Riemannian one-form $\beta=\beta_{i}dx^{i}$, the Randers-Finsler geometry defined using $F(x,y) = \sqrt{\alpha_{ij} \cdot y^{i}y^{j}} + \beta_{i}y^{i}$ generally does not preserve the reversal symmetry, because $F(x,y) \neq F(x,-y)$ usually holds for the same curve in Finsler geometry. When revering the tangent direction, the retrograde orbit is no longer a geodesic of Randers-Finsler geometry $F(x,y)$. Instead, the retrograde orbit follows the geodesic curve of its reversal Finsler geometry $F(x,-y)$. However, there does not mean that we should perform a detailed calculation using the reversal Finsler geometry $F(x,-y)$. Fortunately, for the optical geometry of axially symmetric spacetimes defined in (\ref{optical geometry rotational3}), the non-Riemannian part $\beta$ is determined by metric component $g_{t\phi}$, whose sign flips when the angular momentum of the gravitational source is reversed. A straightforward inspection of the optical geometry reveals that
\begin{equation}
	\beta_{i}(x,a) = - \beta_{i}(x,-a)  
\end{equation}
with $a=J/M$ to be the angular momentum parameter for the rotational spacetime. Consequently, the reversal Finsler geometry $F(x,-y)$ satisfies the following relation
\begin{subequations}
	\begin{eqnarray}
		F(x,-y) 
		& = & \tilde{F}(x,y), 
		\\
		\tilde{F}(x,y) 
		& = & \sqrt{ \alpha_{ij} \cdot y^{i}y^{j} } + \beta_{i}(x,-a) \cdot y^{i} \nonumber
		\\
		& = & \sqrt{ \alpha_{ij} \cdot y^{i}y^{j} } - \beta_{i}(x,a) \cdot y^{i}
	\end{eqnarray}
\end{subequations}
Thus, the retrograde photon motions can be analyzed simply by replacing the angular momentum $a$ with $-a$ (alternatively, by replacing $\beta_{i}$ with $-\beta_{i}$). Based on this analysis, if the retrograde motion of light is considered, we can apply the replacement $\beta_{i} \to -\beta_{i}$ to equivalently transform the retrograde orbits into the prograde orbits, while altering the non-Riemannian part to $-\beta$ in the optical geometry. This operation eventually leads to an additional minus sign in the calculation of geodesic curvature, which results in the same contribution as those presented in equation (\ref{geodesic curvature expression retrograde orbit}).

Furthermore, for practical calculations with specific spacetime metrics, it is convenient to adopt the convention where the angular momentum parameter is kept to be positive ($a>0$). This choice ensures that the prograde motion of lights corresponds to a positive angular velocity $\Omega = \frac{d\phi}{dt} > 0$, while the retrograde motion of lights corresponds to a negative angular velocity $\Omega = \frac{d\phi}{dt} < 0$. Under this convention, the geodesic curvatures for prograde and retrograde photon orbits can be expressed as follows 
\footnote{In Asida's work \cite{Asida2017}, the geodesic curvature is used to investigate the gravitational deflection angle of light orbits, via the Gauss-Bonnet theorem $\int_{D}\mathcal{K}dS+\int_{\partial D}\kappa_{g}dl + \sum_{i} \theta_{i} = \chi(D)$. To obtain the appropriate deflection angle, they enforce the regulation $dl>0$ for prograde motions of light and $dl<0$ for retrograde motions of light. In our convention, the plus and minus sign has been adapted into the geodesic curvature contribution $\kappa_{\beta}$ and $\kappa_{g}^{(\alpha)}$, and there is no need to impose the regulation $dl>0$ for prograde motions (or $dl<0$ for retrograde motions) artificially.}
\begin{widetext}
\begin{eqnarray}
	\kappa_{g}^{(\alpha)} 
	= - \kappa_{\beta}^{(\alpha)}
	= 
	\left\{
	\begin{aligned}
		- \frac{1}{\sqrt{\alpha_{rr}\alpha_{\phi\phi}}} 
		\frac{\partial \beta_{\phi}}{\partial r} & \ \ \ \text{for prograde motion's orbits with} \ \ \Omega>0 %\ \ (\text{assuming} \ a>0)
		\\
		+ \frac{1}{\sqrt{\alpha_{rr}\alpha_{\phi\phi}}} 
		\frac{\partial \beta_{\phi}}{\partial r} & \ \ \ \text{for retrograde motion's orbits with} \ \ \Omega<0 . %\ \ (\text{assuming} \ a>0) 
	\end{aligned}
	\right.
	\label{geodesic curvature expression final}
\end{eqnarray}
\end{widetext}
This expression can be written in a simpler form
\begin{equation}
	\kappa_{g}^{(\alpha)} 
	= - \kappa_{\beta}^{(\alpha)}
	= - \frac{\text{sign}(\Omega)}{\sqrt{\alpha_{rr}\alpha_{\phi\phi}}} \cdot
	\frac{\partial \beta_{\phi}}{\partial r} .
\end{equation}

In summary, we have analyzed two types of geodesic curvatures for the same continuous curve: one is geodesic curvature $\kappa_{g}^{(F)}$ defined using the Randers-Finsler optical geometry $dt= \sqrt{\alpha_{ij}dx^{i}dx^{j}} + \beta_{i}dx^{i}$, the other is the geodesic curvature $\kappa_{g}^{(\alpha)}$ defined using its Riemannian part $dl^{2}=\alpha_{ij}dx^{i}dx^{j}$. These two geodesic curvatures are interconnected through the ``distortion'' term $\kappa_{\beta}$, which is induced by the non-Riemannian part $\beta$. Particularly, for the Randers-Finsler optical geometry given in expression (\ref{optical geometry rotational3}), the additional contribution $\kappa_{\beta}$ simply depends solely on a radial derivative of the $\phi$ component $\kappa_{\beta} \propto \frac{\partial \beta_{\phi}}{\partial r} \propto \frac{\partial}{\partial r} \big( \frac{g_{t\phi}}{g_{tt}} \big)$, reflecting the axi-symmetry of rotating gravitational fields.   

\section{Flag Curvature in Randers-Finsler Optical Geometry \label{appendix3}}

In this appendix, we present an introduction to the intrinsic flag curvature, which is the natural generalization of Gaussian curvature into Finsler geometry. First, we introduce the mathematical definition of flag curvature. Subsequently, we present the explicit analytical expression of flag curvature in the equatorial plane of optical geometry. 

\begin{figure*}[t]
	\includegraphics[width=0.75\textwidth]{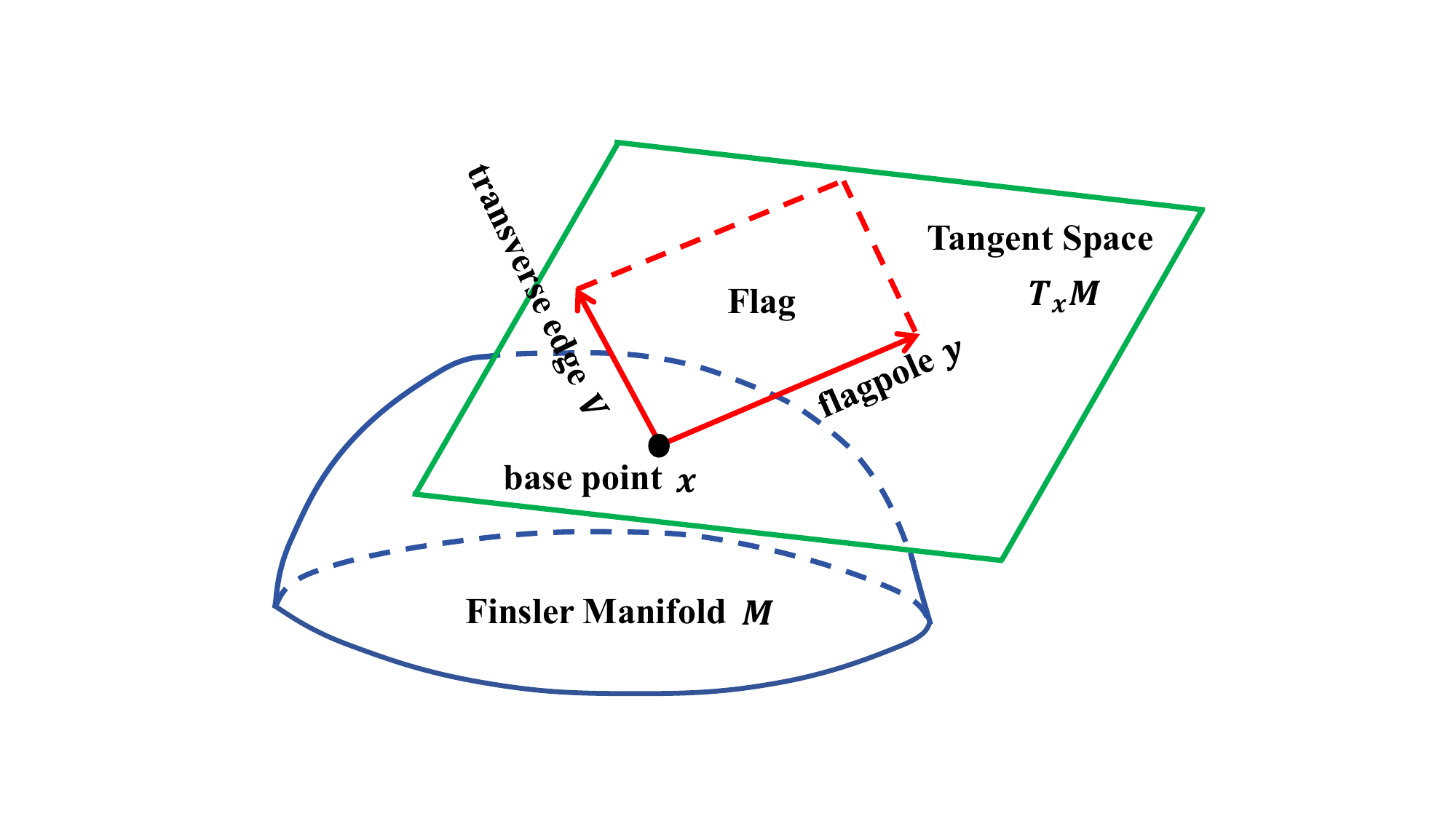}
	\caption{Illustration of the ingredients in the definition of flag curvature, including the base point $x$ in the Finsler manifold, as well as the flagpole $y$ and transverse edge $V$ in the tangent space $T_{x}M$.}
	\label{figure Flag Curvature}
\end{figure*}

Mathematically, every Finsler manifold is equipped with a Finsler function $F(x,y)$, which is a function generally defined in the tangent bundle $(x,y) \in TM$ such that the arc-length in the Finsler manifold is determined by $s_{AB} = \int_{\lambda_{A}}^{\lambda_{B}} F(x,T) d\lambda$, as shown in expression (\ref{Finsler length}). The Finsler function serves as the foundational quantity in Finsler geometry, from which all geometric quantities can be derived. In particular, a symmetric $(0,2)$-type fundamental tensor in this Finsler manifold can be induced through the second-order derivative of the Finsler function
\begin{equation}
	g^{(F)}_{ij}(x,y) \equiv \frac{\partial^{2}}{\partial y^{i} \partial y^{j}} \bigg[ \frac{F^{2}(x,y)}{2} \bigg] ,
\end{equation}
and a symmetric $(0,3)$-type Cartan tensor is defined through the third-order derivative
\begin{eqnarray} 
	A_{ijk}(x,y) & \equiv & \frac{F(x,y)}{4} \cdot \frac{\partial^{3}}{\partial y^{i} \partial y^{j} \partial y^{k}} \bigg[ F^{2}(x,y) \bigg] \nonumber
	\\
	& = & \frac{F(x,y)}{2} \cdot \frac{\partial g_{ij}(x,y)}{\partial y^{k}} .
\end{eqnarray} 
The fundamental tensor defines the inner product in tangent space, and it governs the raising/lowering of tensor indices in Finsler geometry. The Cartan tensor quantifies the deviation of Finsler geometry from Riemannian geometry, and it vanishes if and only if the Finsler manifold reduces to a Riemannian manifold (in this specific case, the Finsler function $F^{2}(x,y)$ exhibits a quadratic dependence on $y$).  
 
The flag curvature is a natural generalization of Gaussian curvature to Finsler geometry. To give a rigorous definition of flag curvature, it is necessary to construct a ``flag'' in the tangent space $T_{x}M$ of a base point $x$. A nonzero tangent vector $y \in T_{x}M$ serves as the ``flagpole'' and another vector $V=V^{i}\frac{\partial}{\partial x^{i}}$ (which is not collinear with flagpole $y$) is referred to as the transverse edge. The flagpole $y$ and transverse edge $V$ span a 2-dimensional subspace within the tangent space $T_{x}M$, which is usually called the ``flag'', as illustrated in Figure $\ref{figure Flag Curvature}$. Based on these elements, the intrinsic flag curvature in a Finsler manifold can be defined as
\begin{widetext}
\begin{equation}
	\mathcal{K}_{\text{flag}}(x,y,V) 
	\equiv 
	\frac{ V^{j}V^{k} \cdot y^{i}y^{l} \cdot R^{(F)}_{ijkl}(x,y) }{ <y,y>^{(F)}_{(x,y)} \cdot <V,V>^{(F)}_{(x,y)} - <y,V>^{(F)}_{(x,y)} \cdot <y,V>^{(F)}_{(x,y)} } .
	\label{flag curvature definition}
\end{equation}
\end{widetext}
The notation $<V,W>^{(F)}_{(x,y)}$ labels the inner product of two vector fields $V$ and $W$ in the Finsler manifold, which can be calculated through the fundamental tensor 
\begin{equation}
	<V,W>^{(F)}_{(x,y)} = g^{(F)}_{ij}(x,y) \cdot V^{i} W^{j} .
\end{equation}
Remarkably, it is worth noting that in Finsler geometry, the inner product of vector fields depends not only on the base point $x$, but also on the flagpole direction $y \in T_{x}M$. This feature exhibits a significant distinction from the situation in Riemannian geometry. In a Riemannian manifold, the inner product of vector fields is determined by the Riemannian metric, which is solely dependent on the base point $x$ and is independent of flagpole direction $y$. 
In the definition of flag curvature, the notation $R^{(F)}_{ijkl}$ and $R^{(F)}_{jk}$ represent the curvature tensors in Finsler geometry
\begin{widetext}
\begin{subequations}
\begin{eqnarray}
	(R^{(F)})^{i}_{jkl}(x,y) 
	& \equiv & \frac{ \delta \Gamma^{i}_{jl}(x,y) }{ \delta x^{k} } 
	- \frac{ \delta \Gamma^{i}_{jk}(x,y) }{ \delta x^{l} } 
	+ \Gamma^{i}_{sk}(x,y) \cdot \Gamma^{s}_{jl}(x,y) 
	- \Gamma^{i}_{sl}(x,y) \cdot \Gamma^{s}_{jk}(x,y) , \ \ \ 
	\\
	R^{(F)}_{ijkl}(x,y) & \equiv & g^{(F)}_{sj}(x,y) \cdot (R^{(F)})^{s}_{ikl}(x,y) ,
	\\
	R^{(F)}_{jk}(x,y) & \equiv & R^{(F)}_{ijkl}(x,y) \cdot y^{i} y^{l} .
\end{eqnarray}
\end{subequations}
In these expressions, $\Gamma^{i}_{jk}(x,y)$ is the coefficient of the renowned Chern-Rund connection in Finsler geometry. The Chern-Rund connection, which is torsion-free and almost metric-compatible, can be regarded as a generalization of the conventional Levi-Civita connection in Riemannian geometry to Finsler geometry \cite{ShenYB,ChernSS}. The explicit form of the Chern-Rund connection's coefficient is expressed as
\begin{eqnarray}
	\Gamma^{i}_{jk}(x,y) 
	& = & 
	\frac{g_{(F)}^{is}(x,y)}{2} \cdot 
	\bigg\{ 
	      \frac{ \delta }{ \delta x^{j} } \bigg[ g^{(F)}_{sk}(x,y) \bigg]
	    + \frac{ \delta }{ \delta x^{k} } \bigg[ g^{(F)}_{js}(x,y) \bigg]
	    - \frac{ \delta }{ \delta x^{s} } \bigg[ g^{(F)}_{ij}(x,y) \bigg]
	\bigg\} \nonumber
	\\
	& = & 
	\gamma^{i}_{jk}(x,y) 
	- g_{(F)}^{is}(x,y) \cdot 
	  \bigg[ 
	      A_{skl}(x,y) \cdot \frac{ N^{l}_{j}(x,y) }{ F(x,y) } 
	    + A_{jsl}(x,y) \cdot \frac{ N^{l}_{k}(x,y) }{ F(x,y) } 
	    - A_{jkl}(x,y) \cdot \frac{ N^{l}_{s}(x,y) }{ F(x,y) } 
	  \bigg] . 
\end{eqnarray}
Here, $\gamma^{i}_{jk}(x,y)$ is the formal Christoffel symbol in Finsler geometry
\begin{equation}
	\gamma^{i}_{jk}(x,y) 
	= \frac{g_{(F)}^{is}(x,y)}{2} \cdot 
	\bigg\{
	      \frac{ \partial }{ \partial x^{j} } \bigg[ g^{(F)}_{sk}(x,y) \bigg]
	    + \frac{ \partial }{ \partial x^{k} } \bigg[ g^{(F)}_{js}(x,y) \bigg]
	    - \frac{ \partial }{ \partial x^{s} } \bigg[ g^{(F)}_{jk}(x,y) \bigg]
	\bigg\} ,
\end{equation}
$g_{(F)}^{ij}(x,y)$ is the inverse of the fundamental tensor $g^{(F)}_{ij}(x,y)$, and the operator $\frac{\delta}{\delta x^{k}}$ is defined by
\footnote{In the Finsler geometry, it is more convenient to use $\big\{ \frac{\delta}{\delta x^{i}}, \frac{\partial}{\partial y^{i}} \big\}$ as the local frames in the tangent bundle $(x,y) \in TM$, instead of $\big\{ \frac{\partial}{\partial x^{i}}, \frac{\partial}{\partial y^{i}} \big\}$. This is because $\frac{\delta}{\delta x^{i}}$ follows a simpler transformation rule than $\frac{\partial}{\partial x^{i}}$ under coordinate transformations in the tangent bundle $TM$.}
\begin{equation}
	\frac{\delta}{\delta x^{k}} 
	\equiv 
	\frac{\partial}{\partial x^{k}} - N^{i}_{k}(x,y) \cdot \frac{\partial}{\partial y^{i}} .
	\label{delta operator}
\end{equation}
The nonlinear function $N$ in expression (\ref{delta operator}) can be defined through the fundamental tensor and Cartan tensor
\begin{equation}
	N^{i}_{j}(x,y) 
	\equiv 
	\gamma^{i}_{jk}(x,y) \cdot y^{k} 
	- \frac{ A^{i}_{jk}(x,y) }{ F(x,y) } \cdot \gamma^{k}_{ls}(x,y) \cdot y^{l} y^{s} .
\end{equation}
The $A^{i}_{jk}(x,y) = g_{(F)}^{is} A_{sjk}(x,y)$ is the contraction of Cartan tensor $A_{ijk}$ with the inverse fundamental tensor $g_{(F)}^{ij}(x,y)$. 

There are a number of useful formulas for calculating curvature tensors in Finsler geometry. Here, we present an elegant formula that can largely simplify the calculation of $(R^{(F)})^{i}_{j}$ \cite{ShenYB}
\begin{equation}
	(R^{(F)})^{i}_{j} (x,y)
	= 2 \cdot \frac{\partial G^{i}(x,y)}{\partial x^{j}} 
	- y^{k} \cdot \frac{\partial^{2}G^{i}(x,y)}{\partial x^{k} \partial y^{j}} 
	+ 2G^{k}(x,y) \cdot \frac{\partial^{2}G^{i}(x,y)}{\partial y^{k} \partial y^{j}}
	- \frac{\partial G^{i}(x,y)}{\partial y^{k}} \cdot \frac{\partial G^{k}(x,y)}{\partial y^{j}} ,
\end{equation}
where $G^{i}(x,y)$ is the geodesic spray coefficient in Finsler manifold
\begin{equation}
	G^{i}(x,y) = \frac{g_{(F)}^{ij}}{4}
	\cdot \bigg\{ y^{k} \cdot \frac{\partial^{2}}{\partial y^{j} \partial x^{k}} \bigg[ F^{2}(x,y) \bigg] - \frac{\partial}{\partial x^{j}} \bigg[ F^{2}(x,y) \bigg] \bigg\} .
\end{equation}

The Riemannian geometry emerges as a special case of Finsler geometry. In a general Finsler manifold, the Finsler function $F(x,y)$ is an arbitrary function in the tangent bundle $(x,y) \in TM$. However, when the square of the Finsler function takes a specific quadratic form in $y$
\begin{equation}
	F^{2}(x,y)= \alpha_{ij}(x) \cdot y^{i}y^{j} 
\end{equation}
the Finsler manifold reduces to a Riemannian manifold. In this case, the symmetric Cartan tensor vanishes, and the symmetric fundamental tensor in Finsler geometry reduces to a Riemannian metric tensor
\begin{subequations}
\begin{eqnarray}
	g^{(F)}_{ij}(x,y) & = & \frac{\partial^{2}}{\partial y^{i} \partial y^{j}} \bigg[ \frac{F^{2}(x,y)}{2} \bigg]
	= \alpha_{ij}(x) ,
	\\
	A_{ijk}(x,y) & = & \frac{F(x,y)}{4} \cdot \frac{\partial^{3}}{\partial y^{i} \partial y^{j} \partial y^{k}} \bigg[ F^{2}(x,y) \bigg] = 0 .
	\ \ \ \ \ \ \ \ \ \ \ 
\end{eqnarray}
\end{subequations}
In this context, the Chern-Rund connection coincides with the Levi-Civita connection of the Riemannian metric $\alpha$ \cite{ShenYB}
\begin{equation}
	\Gamma^{i}_{jk}(x,y) 
	= (\gamma^{(\alpha)})^{i}_{jk}(x) 
	= \frac{\alpha^{is}(x)}{2} 
	\cdot \bigg( \frac{\partial \alpha_{sk}}{\partial x^{j}} + 
	\frac{\partial \alpha_{js}}{\partial x^{k}} - \frac{\partial \alpha_{jk}}{\partial x^{s}} \bigg) .
\end{equation}
and the curvature tensor $(R^{(F)})^{i}_{jkl}$ in Finsler geometry reduces to the Riemannian curvature tensor $(R^{(\alpha)})^{i}_{jkl}$ \cite{ShenYB}
\begin{eqnarray}
	(R^{(F)})^{i}_{jkl} (x,y)
	& = & 
	(R^{(\alpha)})^{i}_{jkl} (x) \nonumber
	\\
	& = & 
	  \frac{ \partial }{ \partial x^{k} } \bigg[ (\gamma^{(\alpha)})^{i}_{jl} \bigg]
	- \frac{ \partial }{ \partial x^{l} } \bigg[ (\gamma^{(\alpha)})^{i}_{jk} \bigg]
	+ (\gamma^{(\alpha)})^{i}_{sk}(x) \cdot (\gamma^{(\alpha)})^{s}_{jl}(x) 
	- (\gamma^{(\alpha)})^{i}_{sl}(x) \cdot (\gamma^{(\alpha)})^{s}_{jk}(x) .
\end{eqnarray}
Particularly, in this 2-dimensional manifold, the flag curvature reproduces the Gaussian curvature in Riemannian geometry
\begin{equation}
	\mathcal{K}_{\text{flag}}(x,y,V) 
	= \frac{R_{1212}^{(\alpha)}}{\det(\alpha^{\text{2d}}_{ij})} 
	= \mathcal{K}^{(\alpha)}(x) .
\end{equation}

%\begin{widetext}
For axially symmetric spacetimes, the Finsler function in the equatorial plane of optical geometry is defined according to the arc-length given in expression (\ref{optical geometry rotational3}).
\begin{equation}
	F(x,y) = \sqrt{\alpha_{rr} \cdot (y^{r})^{2} + \alpha_{\phi\phi} \cdot (y^{\phi}})^{2} + \beta_{\phi} y^{\phi}
	= \sqrt{-\frac{g_{rr}}{g_{tt}} \cdot (y^{r})^{2} -\frac{g_{t\phi}^{2}-g_{tt}g_{\phi\phi}}{g_{tt}^{2}} \cdot (y^{\phi})^{2}}
	- \frac{g_{t\phi}}{g_{tt}} \cdot y^{\phi} .
\end{equation}
When studying light rings in the equatorial plane, it is convenient to choose the flagpole vector as the tangent vector of this light ring $y=T^{\text{OP}}$, with $y^{r}=0$, $y^{\phi}=\frac{d\phi}{dt}=\Omega$. Meanwhile, the transverse edge vector can be selected as the frame vector $V=\partial^{\text{OP}}_{r}$, with $V^{r}=1$ and $V^{\phi}=0$. In this scenario, the inner products of flagpole and transverse edge vectors in Randers-Finsler geometry become
\begin{subequations}
\begin{eqnarray}
	<y,y>_{(x,y)}^{(F)} 
	& = & 
	<T^{\text{OP}}, T^{\text{OP}}>_{(x,T^{\text{OP}})}^{(F)}
	\ = \ F^{2}(x,T^{\text{OP}}) ,
	\\
	<V,V>_{(x,y)}^{(F)} %= g_{ij}^{(F)}(x,y)V^{i}V^{j}
	& = &
	<\partial_{r}^{\text{OP}}, \partial_{r}^{\text{OP}}>_{(x,T^{\text{OP}})}^{(F)} 
	\ = \ g_{rr}^{(F)}(x,T^{\text{OP}}) \cdot 1 \ 
	= \alpha_{rr} \cdot \bigg( 1 + \frac{\beta_{\phi}}{\sqrt{\alpha_{\phi\phi}}} \bigg) ,
	\\
	<y,V>_{(x,y)}^{(F)} %= g_{ij}^{(F)}(x,y)y^{i}V^{j}
	& = &
	<T^{\text{OP}}, \partial_{r}^{\text{OP}}>_{(x,T^{\text{OP}})}^{(F)}
	\ = \ g_{r\phi}^{(F)}(x,T^{\text{OP}}) \cdot \Omega
	= 0 .
\end{eqnarray}
\end{subequations}
Using the aforementioned ``flag'' structure $\{y,V\}=\{T^{\text{OP}},\partial_{r}^{\text{OP}}\}$, the intrinsic flag curvature in the equatorial plane of Randers-Finsler optical geometry can be simplified as
\begin{eqnarray}
	\mathcal{K}_{\text{flag}}(x,y,V) 
	& = & \frac{ V^{j}V^{k} \cdot y^{i}y^{l} \cdot R^{(F)}_{ijkl}(x,y) }{ <y,y>^{(F)}_{(x,y)} \cdot <V,V>^{(F)}_{(x,y)} - <y,V>^{(F)}_{(x,y)} \cdot <y,V>^{(F)}_{(x,y)} } \nonumber
	\\
	\ \ \Rightarrow \ \ 
	\mathcal{K}_{\text{flag}}(r,T^{\text{OP}},\partial^{\text{OP}}_{r}) 
	& = & \frac{ \Omega^{2} \cdot R^{(F)}_{\phi rr\phi}(r,T^{\text{OP}}) }{ F^{2}(r,T^{\text{OP}}) \cdot \alpha_{rr} \big( 1+\frac{\beta_{\phi}}{\sqrt{\alpha_{\phi\phi}}} \big) }
	= \frac{ R^{(F)}_{rr}(r,T^{\text{OP}}) }{ \alpha_{rr} \big( 1+\frac{\beta_{\phi}}{\sqrt{\alpha_{\phi\phi}}} \big) } .
\end{eqnarray}
where we have used $<T^{\text{OP}},T^{\text{OP}}>^{(F)}_{(x,T^{\text{OP}})} = F^{2}(x,T^{\text{OP}}) = 1$ for unit tangent vector $y=T^{\text{OP}}$ along light rings. 
Through a detailed symbolic calculation via Wolfram Mathematica, the flag curvature in this Randers-Finsler optical geometry can be finally expressed as
\begin{eqnarray}
	\mathcal{K}_{\text{flag}}(r,T^{\text{OP}},\partial_{r}^{\text{OP}}) 
	& = & \frac{ R^{(F)}_{rr}(r,T^{\text{OP}}) }{ \alpha_{rr} \big( 1+\frac{\beta_{\phi}}{\sqrt{\alpha_{\phi\phi}}} \big) } \nonumber
	\\ 
	& = & 
	\frac{ 1 }{ \alpha_{rr} \big( 1 + \frac{\beta_{\phi}}{\sqrt{\alpha_{\phi\phi}}} \big) } 
	\times
	\bigg[ \ 
	\frac{ 3\Omega^{2} }{ 4\alpha_{rr} } \cdot \frac{d\alpha_{rr}}{dr}   \frac{d\alpha_{\phi\phi}}{dr}
	+ \frac{ 3\beta_{\phi}\Omega\sqrt{\alpha_{\phi\phi}\Omega^{2}} }{ 4\alpha_{rr}\alpha_{\phi\phi} } \cdot \frac{d\alpha_{rr}}{dr}  \frac{d\alpha_{\phi\phi}}{dr}
	+ \frac{ \beta_{\phi}\Omega \sqrt{\alpha_{\phi\phi}\Omega^{2}} }{ 4\alpha_{\phi\phi}^{2} } \cdot \bigg( \frac{d\alpha_{\phi\phi}}{dr} \bigg)^{2}
	\nonumber
	\\
	&   & 
	+ \frac{3\Omega \sqrt{\alpha_{\phi\phi}\Omega^{2}}}{2\alpha_{rr}} \cdot \frac{d\alpha_{rr}}{dr} \frac{d\beta_{\phi}}{dr}
	+ \frac{ 3\beta_{\phi}\Omega^{2} }{ 2\alpha_{rr} } \cdot \frac{d\alpha_{rr}}{dr} \frac{d\beta_{\phi}}{dr}
	- \frac{ \Omega \sqrt{\alpha_{\phi\phi}\Omega^{2}} }{ 4\alpha_{\phi\phi} } \cdot \frac{d\alpha_{\phi\phi}}{dr} \frac{d\beta_{\phi}}{dr}
	+ \frac{3 \beta_{\phi} \Omega^{2} }{ 4\alpha_{\phi\phi} } \cdot \frac{d\alpha_{\phi\phi}}{dr} \frac{d\beta_{\phi}}{dr}
	\nonumber
	\\
	&   & 
	+ \frac{\Omega^{2}}{2} \bigg( \frac{d\beta_{\phi}}{dr} \bigg)^{2} 
	+ \frac{ 3\beta_{\phi}\Omega \sqrt{\alpha_{\phi\phi}\Omega^{2}} }{ 2\alpha_{\phi\phi} } \bigg( \frac{d\beta_{\phi}}{dr} \bigg)^{2}
	- \frac{\Omega^{2}}{2} \cdot \frac{d^{2}\alpha_{\phi\phi}}{dr^{2}}
	- \frac{ \beta_{\phi}\Omega \sqrt{\alpha_{\phi\phi}\Omega^{2}} }{ 2\alpha_{\phi\phi} } \cdot \frac{d^{2}\alpha_{\phi\phi}}{dr^{2}} 
	\nonumber
	\\
	&   & 
	- \Omega \sqrt{\alpha_{\phi\phi}\Omega^{2}} \cdot \frac{d^{2}\beta_{\phi}}{dr^{2}}
	- \beta_{\phi} \Omega^{2} \cdot \frac{d^{2}\beta_{\phi}}{dr^{2}} 
	\ \bigg]
	\nonumber
	\\
	& = & 
	\frac{ 1 }{ \alpha_{rr} \big( 1+\frac{\beta_{\phi}}{\sqrt{\alpha_{\phi\phi}}} \big) } 
	\times
	\bigg\{ \
	\frac{ 3\Omega^{2} \sqrt{\alpha_{\phi\phi}} }{ 2\alpha_{rr} }  \frac{d\alpha_{rr}}{dr} 
	\cdot \bigg[ \frac{1}{2\sqrt{\alpha_{\phi\phi}}} \frac{d\alpha_{\phi\phi}}{dr} + \text{Sign}(\Omega) \frac{d\beta_{\phi}}{dr} \bigg]
	\nonumber
	\\
	&   &
	+ \frac{ 3\beta_{\phi}\Omega\sqrt{\Omega^{2}} }{ 2\alpha_{rr} } \frac{d\alpha_{rr}}{dr} \cdot \bigg[ \frac{1}{2\sqrt{\alpha_{\phi\phi}}} \frac{d\alpha_{\phi\phi}}{dr} + \text{Sign}(\Omega) \frac{d\beta_{\phi}}{dr} \bigg]
	\nonumber
	\\
	&   & 
	+ \frac{ \beta_{\phi}\Omega \sqrt{\alpha_{\phi\phi}\Omega^{2}} }{ \alpha_{\phi\phi} } 
	\cdot \bigg[ \frac{1}{2\alpha_{\phi\phi}} \frac{d\alpha_{\phi\phi}}{dr} + \text{Sign}(\Omega) \frac{d\beta_{\phi}}{dr} \bigg] 
	\cdot \bigg[ \frac{1}{2\alpha_{\phi\phi}} \frac{d\alpha_{\phi\phi}}{dr} - \text{Sign}(\Omega) \frac{d\beta_{\phi}}{dr} \bigg]
	\nonumber
	\\
	&   &
	- \frac{ \Omega \sqrt{\Omega^{2}} }{ 2 }  \frac{d\beta_{\phi}}{dr} \cdot \bigg[ \frac{1}{2\sqrt{\alpha_{\phi\phi}}} \frac{d\alpha_{\phi\phi}}{dr} + \text{Sign}(\Omega) \frac{d\beta_{\phi}}{dr} \bigg]
	+ \frac{ 3\beta_{\phi} \Omega^{2} }{ 2\sqrt{\alpha_{\phi\phi}} } \frac{d\beta_{\phi}}{dr} \cdot \bigg[ \frac{1}{2\sqrt{\alpha_{\phi\phi}}} \frac{d\alpha_{\phi\phi}}{dr} + \text{Sign}(\Omega) \frac{d\beta_{\phi}}{dr} \bigg]
	\nonumber
	\\
	&   & 
	+ \bigg( \sqrt{\alpha_{\phi\phi}\Omega^{2}} + \beta_{\phi}\Omega \bigg) 
	\cdot \bigg[ 
	\frac{ \sqrt{\alpha_{\phi\phi}\Omega^{2}} }{ \alpha_{\phi\phi} } 
	\cdot \bigg( \frac{d\beta_{\phi}}{dr} \bigg)^{2} 
	- \frac{ \sqrt{\alpha_{\phi\phi}\Omega^{2}} }{ 2\alpha_{\phi\phi} } 
	\cdot \frac{d^{2}\alpha_{\phi\phi}}{dr^{2}} 
	- \Omega \cdot \frac{d^{2}\beta_{\phi}}{dr^{2}} 
	\bigg]
	\ \bigg\} .
	\label{flag curvature expression}
\end{eqnarray} 
\end{widetext}

\begin{acknowledgments}
This work is partly supported by the National Science Foundation of China (Grants No. 12575048 and 12205013), the Scientific and Technology Research Program of Chongqing Municipal Education Commission (Grants No. KJZD-K202301110 and KJQN202201126), and the Natural Science Foundation of Chongqing Municipality (Grant No. CSTB2022NSCQ-MSX0932). Minyong Guo is also supported by the Open Fund of Key Laboratory of Multiscale Spin Physics (Ministry of Education), Beijing Normal University. 
\end{acknowledgments}

%\begin{acknowledgments}
%\end{acknowledgments}


%\bibliography{reference.bib}

\end{document}